\documentclass[11pt]{article}	
\pdfoutput=1 
\usepackage{cite}
\usepackage{color} 
\usepackage{amsmath}
\usepackage{amsfonts}
\usepackage{amssymb}
\usepackage{array,booktabs}
\usepackage{float}
\usepackage{caption}
\usepackage{subcaption}
\usepackage{graphicx}
\usepackage{slashed}            
\usepackage{bbm}                
\usepackage{xspace}				
\usepackage{collref}				
\usepackage{framed}				
\usepackage{placeins}
\usepackage{jheppub}							
\usepackage{mathrsfs}
\usepackage{hyperref}
\usepackage{tikz}
\usetikzlibrary{arrows,shapes}
\usetikzlibrary{trees}
\usetikzlibrary{matrix,arrows} 				
\usetikzlibrary{positioning}				
\usetikzlibrary{calc,through}				
\usetikzlibrary{decorations.pathreplacing}  
\usepackage{pgffor}							

\usetikzlibrary{decorations.pathmorphing}	
\usetikzlibrary{decorations.markings}
\usetikzlibrary{snakes}
\usepackage[normalem]{ulem}

\tikzset{
    vector/.style={decorate, decoration={snake}, draw},
	provector/.style={decorate, decoration={snake,amplitude=2.5pt}, draw},
	antivector/.style={decorate, decoration={snake,amplitude=-2.5pt}, draw},
    fermion/.style={draw, postaction={decorate},
        decoration={markings,mark=at position .55 with {\arrow[draw]{>}}}},
    fermionbar/.style={draw, postaction={decorate},
        decoration={markings,mark=at position .55 with {\arrow[draw=black]{<}}}},
    fermionnoarrow/.style={draw},
    gluon/.style={decorate, draw,decoration={coil,amplitude=4pt, segment length=6pt}, line width=1},
    scalar/.style={dashed,draw, postaction={decorate},
        decoration={markings,mark=at position .55 with {\arrow[draw]{>}}}},
    scalarbar/.style={dashed,draw, postaction={decorate},
        decoration={markings,mark=at position .55 with {\arrow[draw]{<}}}},
    scalarnoarrow/.style={dash pattern = on 6 pt off 3 pt,draw},
    electron/.style={draw, postaction={decorate},
        decoration={markings,mark=at position .55 with {\arrow[draw]{>}}}},
	bigvector/.style={decorate, decoration={snake,amplitude=4pt}, draw},
	vectorscalar/.style={loosely dotted,draw, postaction={decorate}},
}
\usepackage{relsize}
\usepackage{slashed}

\begin{document}

\def\lsim{\mathrel{\rlap{\lower4pt\hbox{\hskip1pt$\sim$}}
    \raise1pt\hbox{$<$}}}
\def\gsim{\mathrel{\rlap{\lower4pt\hbox{\hskip1pt$\sim$}}
    \raise1pt\hbox{$>$}}}
\newcommand{\vev}[1]{ \left\langle {#1} \right\rangle }
\newcommand{\bra}[1]{ \langle {#1} | }
\newcommand{\ket}[1]{ | {#1} \rangle }
\newcommand{\ev}{ {\rm eV} }
\newcommand{\kev}{{\rm keV}}
\newcommand{\mev}{{\rm MeV}}
\newcommand{\gev}{{\rm GeV}}
\newcommand{\tev}{{\rm TeV}}
\newcommand{\mpl}{$M_{Pl}$}
\newcommand{\mw}{$M_{W}$}
\newcommand{\Ft}{F_{T}}
\newcommand{\Zparity}{\mathbb{Z}_2}
\newcommand{\BLambda}{\boldsymbol{\lambda}}
\newcommand{\be}{\begin{eqnarray}}
\newcommand{\ee}{\end{eqnarray}}
\newcommand{\met}{\;\not\!\!\!{E}_T}

\newcommand{\sla}[1]{\setbox0=\hbox{$#1$}           
   \dimen0=\wd0                                     
   \setbox1=\hbox{/} \dimen1=\wd1                   
   \ifdim\dimen0>\dimen1                            
      \rlap{\hbox to \dimen0{\hfil/\hfil}}          
      #1                                            
   \else                                            
      \rlap{\hbox to \dimen1{\hfil$#1$\hfil}}       
      /                                             
   \fi} 

\newcommand{\Tab}[1]{Table~\ref{tab:#1}}
\newcommand{\tab}[1]{table~\ref{tab:#1}}
\newcommand{\tabl}[1]{\label{tab:#1}}
\newcommand{\Fig}[1]{figure~\ref{fig:#1}}
\newcommand{\Figl}[1]{\label{fig:#1}}
\newcommand{\draftnote}[1]{{\bf\color{red} #1}}
\def\YT#1{{\bf  \textcolor{red}{[YT: {#1}]}}}
\def\ES#1{{\bf  \textcolor{blue}{[ES: {#1}]}}}
\def\YTc#1{{  \textcolor{red}{{#1}}}}
\def\RP#1{{\bf  \textcolor{blue}{[RP: {#1}]}}}
\def\RPc#1{{  \textcolor{blue}{{#1}}}}

\title{The Dark Penguin Shines Light at Colliders}
\author{Reinard Primulando,$^1$ Ennio Salvioni$^{2}$ and Yuhsin Tsai$^{2}$}
\affiliation{$^1$Department of Physics and Astronomy, Johns Hopkins University, Baltimore, Maryland 21218\\ 
$^2$Department of Physics, University of California Davis, Davis, California 95616 }
\date{\today}
\abstract{Collider experiments are one of the most promising ways to constrain Dark Matter (DM) interactions. For several types of DM-Standard Model couplings, a meaningful interpretation of the results requires to go beyond effective field theory, considering simplified models with light mediators. This is especially important in the case of loop-mediated interactions. In this paper we perform the first simplified model study of the magnetic dipole interacting DM, by including the one-loop momentum-dependent form factors that mediate the coupling -- given by the Dark Penguin -- in collider processes. We compute bounds from the monojet, monophoton, and diphoton searches at the $8$ and $14$ TeV LHC, and compare the results to those of direct and indirect detection experiments. Future searches at the $100$ TeV hadron collider and at the ILC are also addressed. We find that the optimal search strategy requires loose cuts on the missing transverse energy, to capture the enhancement of the form factors near the threshold for on-shell production of the mediators. We consider both minimal models and models where an additional state beyond the DM is accessible. In the latter case, under the assumption of anarchic flavor structure in the dark sector, the LHC monophoton and diphoton searches will be able to set much stronger bounds than in the minimal scenario. A determination of the mass of the heavier dark fermion might be feasible using the $M_{T2}$ variable. In addition, if the Dark Penguin flavor structure is almost aligned with that of the DM mass, a displaced signal from the decay of the heavier dark fermion into the DM and photon can be observed. This allows us to set constraints on the mixings and couplings of the model from an existing search for non-pointing photons.}

\preprint{}

\maketitle

\section{Introduction}
The existence of Dark Matter (DM) is firmly established by a large number of astrophysical and cosmological observations. Despite the fact that it contributes a large component of the energy density of the universe, however, its precise properties remain almost completely mysterious. The common belief is that most of the DM is in the form of a stable particle, which is neutral or charged very weakly under the electric force, but interacts at least gravitationally with baryons. If such a particle carries non-gravitational interactions with the Standard Model (SM), the production of DM particles at high energy collider experiments offers one of the most promising opportunities to identify the nature of the DM interactions. Thanks to the capability to produce DM particles in a wide mass range and to the obvious independence from astrophysical uncertainties, collider searches provide complementary results to the direct and indirect detection experiments \cite{Birkedal:2004xn,Bai:2010hh,Goodman:2010yf,Cheung:2012gi,Arrenberg:2013rzp,Malik:2014ggr}. 

When setting collider constraints on DM interactions, the most straightforward way of parameterizing the DM-SM coupling is through an effective field theory (EFT) description. The non-observation of events with significant missing energy in excess of the SM background is then translated into upper bounds on the coefficients of the effective operators that couple the DM to the SM fields \cite{Cao:2009uw,Goodman:2010ku,Fox:2011fx,Fox:2011pm,Fox:2012ee,Bai:2012xg,Bell:2012rg,Carpenter:2012rg,Chen:2013gya,Aad:2013oja,Zhou:2013raa,Lin:2013sca,Dreiner:2013vla,Askew:2014kqa,Racco:2015dxa}. This simple method is independent of the details of the ultraviolet (UV) completion of the model. However, it is based on the assumption that the EFT gives a valid description of the collider process, i.e. that the mediators can be integrated out at the energy scale of the collision. Unfortunately, the sensitivity of the current LHC searches does not correspond to heavy enough mediators for many of the EFT couplings \cite{Buchmueller:2013dya,Busoni:2013lha}, and unitarity usually sets stronger bounds than the collider search \cite{Shoemaker:2011vi,Yamamoto:2014pfa}. Thus in many instances, to extract meaningful information from collider searches it is necessary to consider perturbative models with light mediators. This `simplified model' approach has been applied to several DM-SM couplings whose UV completions feature the tree-level exchange of mediators \cite{Goodman:2011jq,An:2012ue,An:2013xka,Bai:2013iqa,DiFranzo:2013vra,Chang:2013oia,Cheung:2013dua,Busoni:2014sya,Papucci:2014iwa,deSimone:2014pda,Berlin:2014cfa,Buchmueller:2014yoa,Abdallah:2014hon,Curtin:2014pda}.

The inadequacy of the EFT approach is manifestly even more dramatic in the case of loop-mediated couplings. Among the loop-induced DM-SM interactions, an especially important status is held by the dipole and Rayleigh operators \cite{Sigurdson:2004zp,Weiner:2012cb}, which play important roles in DM model building. Collider constraints on these couplings within the EFT approach have been explored in several previous studies \cite{Fortin:2011hv,Barger:2012pf,Nelson:2013pqa,Cotta:2012nj,Crivellin:2015wva}, but the resulting bounds translate into very weak constraints on the mediator masses, thus calling for a simplified model description. 

In this work we perform the first systematic collider study of a loop-induced DM-SM coupling, by considering a perturbative UV completion with light mediators and including the momentum-dependent form factors in the description of collider processes.\footnote{A first brief discussion of the loop form factors for dipole-interacting DM production at colliders was presented in Ref.~\cite{Tamarit:2013nna}. See also Ref.~\cite{Altmannshofer:2014cla}, where the loop contribution of the dark sector to dilepton production was considered, although in a different model.} In our study we focus on a generic UV completion of the magnetic and electric dipole operators, the \emph{Dark Penguin}. We emphasize, however, that the method introduced in this work can be applied to any other loop-induced DM-SM coupling.

If the DM is a singlet under the SM gauge symmetry, the magnetic and electric dipole operators $\mu_M\,\bar{\chi}\sigma^{\mu\nu}\chi F_{\mu\nu}+\mu_E\,\bar{\chi}\sigma^{\mu\nu}\chi \tilde{F}_{\mu\nu}$, generated by the dark penguin diagrams, give the lowest order interactions of the DM with the SM gauge fields. As a consequence, these operators play important roles in the possible explanation of various $\gamma$-ray excesses observed at indirect detection experiments \cite{Weiner:2012gm,Rajaraman:2012fu,Dudas:2014ixa,Geng:2014zqa,Cheung:2014tha,Lee:2014koa,Pierce:2014spa}. Moreover, being chirality-flipping, these operators can possess a non-trivial flavor structure when more than one species of dark fermions $\chi_i$ is present. In this case the photon dipole couplings in the mass basis can connect the light and excited DM state, and give interesting inelastic scattering signals at direct detection experiments \cite{Chang:2010en,Banks:2010eh,Kumar:2011iy,Pospelov:2013nea,Frandsen:2013bfa,DelNobile:2014eta}. The typical momentum exchange in the direct and indirect detection experiments is very small, therefore it is reasonable to use the EFT description even when the mediators are light with respect to collider energies. 

In order to allow a comparison to the results of direct and indirect detection experiments, the collider searches need to provide bounds on both the DM-mediator coupling and the mediator mass. The search for mediator decays in various UV completions of the dipole and Raleigh operators has been studied in \cite{Liu:2013gba}. However, since the decay of the mediators only depends on their branching ratios, the information about the DM-mediator coupling is lost in these searches. Thus the study of the dark penguin process would be crucial even if the mediators were directly discovered first.  

The dark penguin serves as a good example of a simplified model that can be constrained by different collider searches: depending on the `dark flavor' structure -- the flavor structure of dark fermions, -- it can provide signals in the monojet, monophoton, diphoton, and even non-pointing photon searches. Besides allowing us to set meaningful constraints, the inclusion of light mediators helps us to identify the optimal cuts to be used in the collider searches, which are different from those commonly employed in the study of the EFT couplings. When the mediator mass is much smaller than the typical momentum exchange in the dipole, the missing transverse energy (MET) distribution of the DM signal is much softer than the one obtained assuming EFT couplings. Therefore the intuition of setting harder cuts to increase the signal excess no longer applies: on the contrary, the best strategy is to keep the cuts as low as possible, to include the enhancement of the form factors corresponding to the mediators being produced on-shell. It follows that in the search for the dark penguin with light mediators, lowering the background is more important than increasing the collider energy. Looking ahead towards future experiments, this implies that the International Linear Collider (ILC) would have the capability to set much stronger constraints than a very high energy ($100$ TeV) hadron collider, as will be shown in our analysis.  

The organization of the paper is as follows. In Sec.~\ref{sec:model} we present the simplified model used in the analysis, and address the possibility of having displaced photon signals from an aligned flavor structure. We begin Sec.~\ref{sec:searches} by explaining our method for including the loop-mediated dipole couplings in the simulation of collider processes. Then we discuss in detail the missing energy searches used to set constraints on the simplified model, paying particular attention to the estimation of the systematic uncertainties in the projection to the $14$ TeV LHC. In Sec.~\ref{sec:NF1} we present the constraints from LEP, the $8$ and $14$ TeV LHC, and the future $100$ TeV collider and ILC on the dark penguin parameter space, assuming a single flavor of dark fermions. Furthermore, we point out that the sensitivity to light mediators can be improved by choosing MET cuts weaker than those used in the search for effective couplings. To give an example of the case with more than one dark flavors, we analyze the model with two dark fermions in Sec.~\ref{sec:NF2}, setting bounds from LEP and the monophoton and diphoton searches at the LHC. For the diphoton channel, the possible application of the $M_{T2}$ variable to determine the mass of the heavier dark fermion is discussed. In Sec.~\ref{sec:displacedphoton} we consider the dark penguin with a flavor structure almost aligned to the dark fermion masses, in which case displaced photon signals can be observed at the LHC. We use the $8$ TeV ATLAS search for non-pointing photons to compute the bound on the dark mixing angle and coupling. In Sec.~\ref{sec:compare} the collider constraints on the dark penguin are compared to the current results from direct and indirect DM detection experiments, by showing the reach of the different searches on the magnetic dipole moment and the annihilation cross section of $\chi\bar{\chi}\to\gamma\gamma$, respectively. We conclude by summarizing our result in Sec.~\ref{sec:conclude}. Finally, App.~\ref{append:formfactor} contains general formulas for the dark penguin form factors, whereas App.~\ref{append:statistics} collects the basic statistics we used for setting limits.

It is worth pointing out that Sec.~\ref{sec:searches} is somewhat technical, therefore the reader mainly interested in the results of our work might prefer, after having become familiar with the dark penguin in Sec.~\ref{sec:model}, to move directly to Sec.~\ref{sec:NF1}.

\section{A simplified model of dark penguin}\label{sec:model}
%
Here we describe a simple UV completion of the magnetic and electric dipole operators, which will be employed throughout the paper. The model is similar to the one discussed in \cite{Weiner:2012gm,Liu:2013gba} and contains dark Dirac fermions $\chi_i$ with flavor index $i=1,...,N_{\chi}$, a fermion mediator $\psi$, and a scalar mediator $\phi$. Both mediators carry some hypercharge $Y$, and the $\chi_i$ are SM singlets. A specific assignment of dark charge which stabilizes the DM particle may affect the decay of mediators, but not the dark penguin process we are interested in. The Lagrangian in the mass basis is written as
\begin{equation}\label{eq:model}
L\supset \bar{\chi}_i (i\slashed\partial-m_{i})\chi_i+\bar{\psi}\,i\slashed D\psi-M_f\bar{\psi}\psi+|D_{\mu}\phi|^2-M_s^2\left |\phi\right|^2+\left(\lambda^R_i\bar{\psi}\,P_R\,\chi_i\,\phi + \lambda^L_{i}\bar{\psi}P_L\,\chi_i\,\phi + \mathrm{h.c.}\right),
\end{equation}
where $D_{\mu}=\partial_{\mu}-i\,Yg'B_{\mu}$. Several simplifications will be made on the Lagrangian: we assume the fermion and scalar mediators to have the same mass, $M_f=M_s=M$,\footnote{Notice, however, that in App.~\ref{append:formfactor} we present general results for the dark penguin with $M_f\neq M_s$.} and also assume the Yukawa-type couplings $\lambda_i = \lambda^L_i = \lambda^R_i$ to be real. This in particular implies that no electric dipole moment operator is generated from Eq.~\eqref{eq:model}. Depending on the details of a specific model, one of the mediators can decay into SM particles, while the mediator that carries the dark charge stabilizing the DM can decay into the DM and SM particles. A general analysis of the phenomenology of the mediators can be found in \cite{Liu:2013gba}. Their results show that the discovery prospects at the LHC depend strongly on the charge assignment. The most challenging scenario corresponds to $SU(2)_L$-singlet mediators with $Y=-1$, and $\psi$ mixed with the right-handed tau lepton. Even at 14 TeV with $300$ fb$^{-1}$, the LHC will have no sensitivity to this model \cite{Liu:2013gba}. However, different hypercharge assignments can change dramatically the decays of the mediators, leading to significantly better prospects. In our analysis of the dark penguin we wish to be independent from the details of the model building, therefore we treat $Y$ and $M$ as free parameters. Precision electroweak measurements do not set relevant constraints on mediators carrying only hypercharge. Even if the mediators are doublets under $SU(2)_{L}$, as considered for example in Ref.~\cite{Weiner:2012gm}, no contribution to the $S$ and $T$ parameters arises at one loop.

\begin{figure}
\begin{center}
\begin{tikzpicture}[line width=1.5 pt, scale=1.0] 
			\draw[color=white] (-1.7,-1.7) rectangle (1.7,1.7);
			\draw[vector] (-1.2,0) -- (0,0);
			\draw[scalar] (1,1) arc (90:180:1);
			\draw[scalar] (-0,0) arc (180:270:1);
			\draw[fermion,line width=2.0] (1,-1) -- (1,1);
			\draw[fermionnoarrow,line width=1.5] (1,1) -- (2,1.5);
			\draw[fermionnoarrow,line width=1] (2,-1.5) -- (1,-1);
			\draw[->, line width =1] (-1.1,-.4) -- (-.5,-.4);
			\draw[->, line width =1] (1,-1.4) -- (1.5,-1.65);
			\draw[->, line width =1] (1,1.4) -- (1.5,1.65);
						  \node at (-0.8,.5) {$B^\mu$};
			  \node at (-1.3,-.4) {$q$};
			  \node at (1.75,-1.75) {$p_j$};
			   \node at (1.8,1.75) {$p_i$};
						          \node at (-0.2,-.8) {$\phi$};
						           \node at (2.35,1.5) {$\chi_i$};
						           \node at (1.4,0) {$\psi$};
						           \node at (2.35,-1.5) {$\bar{\chi}_{j}$};
						                                  		\end{tikzpicture}
								\qquad
\begin{tikzpicture}[line width=1.5 pt, scale=1.0] 
			\draw[color=white] (-1.7,-1.7) rectangle (1.7,1.7);
			\draw[vector] (-1.2,0) -- (0,0);
			\draw[fermion,line width=2.0] (1,1) arc (90:180:1);
			\draw[fermion,line width=2.0] (0,0) arc (180:270:1);
			\draw[scalar] (1,-1) -- (1,1);
			\draw[fermionnoarrow,line width=1.5] (1,1) -- (2,1.5);
			\draw[fermionnoarrow,line width=1] (2,-1.5) -- (1,-1);
			\draw[->, line width =1] (-1.1,-.4) -- (-.5,-.4);
			  \draw[->, line width =1] (1,-1.4) -- (1.5,-1.65);
			\draw[->, line width =1] (1,1.4) -- (1.5,1.65);
			  \node at (-0.8,.5) {$B^\mu$};
			  \node at (-1.3,-.4) {$q$};
			   \node at (1.75,-1.75) {$p_j$};
			   \node at (1.8,1.75) {$p_i$};
						          \node at (-0.2,-.8) {$\psi$};
						           \node at (2.35,1.5) {$\chi_i$};
\node at (1.4,0) {$\phi$};
						           \node at (2.35,-1.5) {$\bar{\chi}_{j}$};
						                                  		\end{tikzpicture}
\end{center}
\caption{Diagrams of the dark penguin. The corresponding amplitude is given in Eq.~(\ref{eq:loopresult}).}
\label{fig:IDMloopFeynDiag}
\end{figure}
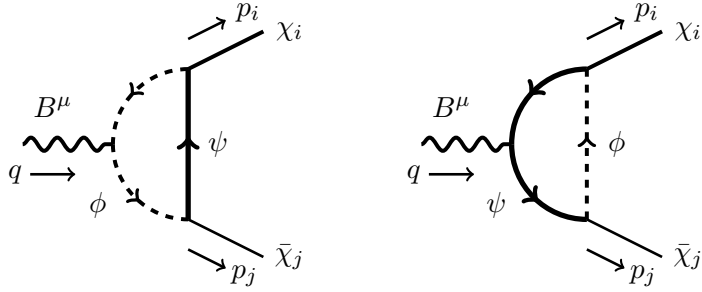

The model in Eq.~\eqref{eq:model} generates a magnetic dipole operator through the dark penguin process in Fig.~\ref{fig:IDMloopFeynDiag}. The amplitude for $B^\mu \to \chi_i \bar{\chi}_j$ is written into a gauge invariant form
\begin{equation}\label{eq:loopresult}
i\mathcal{M}^{\mu}_{\text{penguin}}=\frac{i \lambda_i \lambda_j g'YN}{32\pi^2}\bar{u}(p_i) \times\left[\left(q^2\gamma^{\mu}-(m_{\chi_i}-m_{\chi_j})q^{\mu}\right)F_q-i\sigma^{\mu\nu}q_{\nu}F_{\sigma}\right]v(p_j)\,,
\end{equation}
In this paper, we use various collider searches to set an upper bound on\footnote{Our analysis is insensitive to the sign of the hypercharge of the mediators. For simplicity, from now on we denote with $Y$ the absolute value of the hypercharge.} 
\begin{equation}
\lambda\sqrt{YN}\,.
\end{equation}
where the factor $\lambda_i \lambda_j$ encodes the dark flavor structure. The factor $N$ is the multiplicity of the mediators ($N > 1$ can arise, for example, if the mediators transform non-trivially under $SU(2)_{L}$). The form factors $F_{q,\sigma}$ as functions of the masses $(m_{\chi_i},m_{\chi_j},M_f,M_s)$ and momentum $q^2$ (defined in Fig.~\ref{fig:IDMloopFeynDiag}) are given in App.~\ref{append:formfactor}. It is important to note that the model discussed here can also generate other loops, such as those for the Rayleigh operator, $\bar{\chi}_i\chi_j B_{\mu\nu}B^{\mu\nu}$. However, the contributions of this operator to the collider processes considered in this paper carry either extra gauge couplings or phase space suppressions, and thus only give sub-leading effects, except in a few cases on which we will comment in what follows. We leave the detailed study of the Rayleigh operators for future work.

In presence of a non-trivial flavor structure of the dark fermions $\chi_i$, one important difference between collider searches and (in)direct detection experiments is that the collider processes can generically involve more than one dark flavor. Indeed, as we will show, the current and upcoming missing energy searches at colliders are strikingly more sensitive to the case where more than one dark flavors are within kinematic reach. In this case, some assumptions on the flavor structure are necessary in order to compare results between the different experiments. In this paper, we assume the $\lambda_i$ couplings have a \emph{totally anarchic structure}, with no unnatural hierarchies between different flavors. The assumption of anarchic structure permits a direct comparison of the collider bounds among themselves and to the direct and indirect detection experiments, which are typically sensitive only to the couplings of the lightest dark fermion (the DM particle). To be more precise, in the following we assume $\lambda_i=\lambda\,(1+\delta_i)$ with $|\delta_i| \ll 1$. For models with different flavor structures, one can rescale our bounds from the different searches to set proper constraints.

According to naive dimensional analysis (NDA), the perturbative bound is $\lambda\sqrt{N}\lsim 4\pi$. When showing our results, we allow $\lambda\sqrt{YN}$ to be as large as $8\pi$, which corresponds to the NDA perturbative bound for $Y=4$ (corresponding to $g'Y\simeq 1$). In a different UV completion of the dipole interaction, such as for example a model where the scalar mediator is replaced with a gauge boson, the quantitative result for the dark penguin would be modified. However, the fact that gauge invariance constrains the structure of the amplitude forbids a qualitative alteration of our analysis.
\subsection{Displaced signals and the aligned flavor structure}\label{sec:alignment}
Another interesting scenario to explore is the case where the flavor structures of the DM mass matrix and the dark penguin are nearly aligned. This means that the fermion mass matrix $m_{ij}\bar{\chi}_i\chi_j$ is almost proportional to the dark penguin, whose flavor structure is determined by $\lambda_i\lambda_j$
\begin{equation}\label{eq:flavormisalign}
m_{ij}\propto\lambda_{i}\lambda_j(1+\epsilon_{ij}) + \hat{m}\delta_{ij},\qquad|\epsilon_{ij}|\ll1\,,
\end{equation}
where the term proportional to the identity gives the lightest dark fermion a mass $\sim \hat{m}$. When rotating the dark fermion fields into the mass basis, the dark penguin between different mass eigenstates carries an extra $\mathcal{O}(\epsilon)$ suppression that can result into a displaced decay of a heavy dark fermion $\chi_h$ to the light DM $\chi_l$ and a photon, with a width (assuming $M \gg m_{\chi_{h}}\gg m_{\chi_{l}}$)
\begin{equation}\label{eq:displaceddecay}
\Gamma(\chi_h\to\chi_l\gamma)\simeq\frac{e^{2}\lambda^4 Y^2 N^2 m_{\chi_h}^3}{8\pi(32\pi^2)^2M^2}\times\epsilon^2.
\end{equation}
Contrarily to the case where the decay is displaced because of phase space suppression \cite{Bai:2011jg}, here the photon is hard, therefore the process is accessible in LHC searches.

The aligned flavor structure can be generated if the dark sector has a single flavor breaking spurion, and both the DM masses and dipole interactions are generated by the same loop-level mediation. Since the gauge coupling is flavor blind, the flavor structures of the two operators are identical
\begin{equation}
\sim\frac{\lambda^2\,\hat{y}_{ij}\,e Q_{\psi}}{16\pi^2M}\,\bar{\chi}_i\sigma_{\mu\nu}F^{\mu\nu}\chi_j+\frac{\lambda^2\hat{y}_{ij}}{16\pi^2}\,M\bar{\chi}_i\chi_j\,,
\end{equation}
implying that no heavy-light dipole coupling is present in the mass basis for dark fermions. However, if there exists a small chiral symmetry breaking in the infrared, or an extra flavor symmetry breaking from an even higher order mediation is present, the misalignment between the newly generated mass and the dark penguin leads to a small heavy-light dipole coupling and thus a suppressed decay rate as in Eq.~(\ref{eq:displaceddecay}). With the assumption of anarchic couplings $\lambda_i$, the search of non-pointing photons$+$MET can set interesting bounds on the $(\lambda,\epsilon)$ plane. The result is discussed in Sec.~\ref{sec:displacedphoton}.

\section{DM searches at colliders}\label{sec:searches}
In this section we describe the collider analyses used in this paper to set bounds on the DM parameter space. We begin by outlining our procedure for taking into account the full loop dark penguin form factors in the collider simulations. Then we move on to describe in detail the relevant searches, including the monojet, monophoton and diphoton final states at the $8$ and $14$ TeV LHC. Then we turn to the $8$ TeV search for non-pointing photons. Finally, the projections to a $100$ TeV collider and to the ILC are studied for some of the searches.  

In our analysis of $8$ TeV searches, we reproduce the shape of each SM background using our MonteCarlo (MC) simulations, and compute the additional global rescaling factor needed to obtain exact agreement with the distributions reported in the experimental papers. We then apply these rescaling factors in the $14$ TeV projections. In the $14$ TeV projection of monojet and monophoton searches, we estimate the improvement of systematic uncertainties by separating them into two parts. For uncertainties that relate to the normalization of the SM background, we follow the data-driven analysis at $8$ TeV by simulating the control region sample at $14$ TeV. For other uncertainties, including those on the PDFs and the acceptance of the detector, we assume an improvement proportional to the square-root of the luminosity. We do not attempt to simulate the QCD background at $14$ TeV in this work. In the $8$ TeV search, the pure QCD processes contribute less than $\sim 1\%$ ($10\%$) of the monojet (monophoton) background, and we make the reasonable assumption that the background can be kept subdominant at $14$ TeV by setting a harder MET (photon $p_T$) cut. On the other hand, the QCD background does play an important role in the $8$ TeV diphoton$+$MET search. We include additional jet and lepton vetoes to suppress it at $14$ TeV.

We generated both the signals and backgrounds using MadGraph5 \cite{Alwall:2011uj} and showered the parton level events using Pythia 6 \cite{Sjostrand:2006za}. We used PGS~4 for the detector simulation and cross-checked the results using Delphes~3 \cite{deFavereau:2013fsa}. To compute the QCD $K$-factors for some SM backgrounds, we used MCFM \cite{Campbell:2010ff} and VBFNLO \cite{Baglio:2014uba}.

\subsection{Including loop form factors in a collider process}\label{sec:formfactors}
To properly describe loop-mediated processes at colliders, we simulate the DM production using EFT operators and reweight the events by employing the expressions of the form factors given in App.~\ref{append:formfactor}. 
As described in Eq.~(\ref{eq:loopresult}), the amplitude of the dark penguin contains three distinct Lorentz structures
\begin{equation}\label{eq:eftcouplings}
F_{\sigma}(q^2)\,\bar{u}_{\chi_i}\,\sigma^{\mu\nu}v_{\chi_j}q_\nu B_{\mu},\quad iF_{q}(q^2)\,\bar{u}_{\chi_i} \gamma^{\mu}v_{\chi_j}q^2 B_{\mu},\quad iF_{q}(q^2)\,\bar{u}_{\chi_i}v_{\chi_j}(m_{\chi_j}-m_{\chi_i})q^{\mu}B_{\mu}\,.
\end{equation}
The coefficient of the last operator $\sim q^\mu$ vanishes in the $N_{\chi}=1$ case, since $m_{\chi_i}=m_{\chi_j}$. For $N_{\chi}>1$ the contribution of this operator to the amplitude for pair production of DM particles $f\bar{f}\to \chi_i \bar{\chi}_j$ is proportional to $\bar{v}_f (\slashed{p}_f + \slashed{p}_{\bar{f}}) u_f$ and is thus strongly suppressed by the light SM fermion masses, after using the equations of motion. Therefore we neglect the operator $\sim q^\mu$ altogether and only consider the $\sigma^{\mu\nu}q_{\nu}$ and $q^2\gamma^{\mu}$ operators.

For a given process, we first generate the events in MadGraph5 using a linear combination of the two relevant effective operators, then we reweight the events using the ratio of the partonic matrix element squared computed retaining the form factors to the value obtained using the EFT.   

To give a concrete example, we consider the $N_{\chi}=2$ monophoton analysis. We first simulate the signal process in Fig.~\ref{fig:IDMloop} using two operators implemented through FeynRules \cite{Alloul:2013bka}:\footnote{The sum of the second and third term in Eq.~\eqref{eq:eftcouplings} actually corresponds to the gauge invariant dimension-$6$ operator $\bar{\chi}_i\gamma^{\mu}\chi_j \partial^{\nu}B_{\mu\nu}/\Lambda^2 + \mathrm{h.c.}$ (see also App.~\ref{append:formfactor}). To simplify the event generation we use $\mathcal{O}_V$ instead, and include the $q^2$ factor from the derivatives in the reweighting of the events.}
\begin{equation} \label{effops}
\mathcal{O}_\sigma \sim \frac{1}{\Lambda}\bar{\chi}_h\,\sigma^{\mu\nu}\chi_l B_{\mu\nu} + \mathrm{h.c.},\qquad \mathcal{O}_{V}\sim \bar{\chi}_h\gamma^{\mu}\chi_l B_{\mu} + \mathrm{h.c.}.
\end{equation}
After performing the detector simulation, we apply the cuts for the monophoton search and obtain a list of signal events. For each event, we obtain the value of $q^2$ from the corresponding parton-level four momenta and use it to compute the form factors $F_{\sigma,q}(q^2)$. We then reweight the event using the expression of the matrix element squared, schematically
\begin{equation}\label{eq:rescale}
\frac{d\sigma_{\text{dark penguin}}}{dq^2}\simeq\frac{\left|\mathcal{M}_{F_\sigma}(q^2)+\mathcal{M}_{F_q}(q^2)\right|^2}{\left|\mathcal{M}_{\mathcal{O}_{\sigma}}+\mathcal{M}_{\mathcal{O}_V}\right|^2}\frac{d\sigma_{\text{EFT}}}{dq^2}\,,
\end{equation}
where $\mathcal{M}_{F_{\sigma,q}}(q^2)$ are the amplitudes corresponding to the first and second term in Eq.~\eqref{eq:eftcouplings}, respectively, whereas $\mathcal{M}_{\mathcal{O}_{\sigma, V}}$ are the amplitudes corresponding to the effective operators. 

The procedure described above fully accounts for the interference between the two relevant Lorentz structures, and was applied in the study of the $N_{\chi}=2$ scenario, where both monophoton and diphoton signals are given by $2\to 2$ scatterings followed by the decay of $\chi_h$, as shown in Figs.~\ref{fig:IDMloop} and \ref{fig:diphotonloop}. On the other hand, for $N_\chi = 1$ the main constraint comes from the monojet process in Fig.~\ref{fig:colliderdigm}, which is genuinely a $2\to 3$ scattering. For the sake of simplicity, in this case we neglect the interference term in Eq.~(\ref{eq:rescale}), which carries an additional $\sim m_{\chi}/\sqrt{q^2}$ suppression due to the different chirality structure between the two operators. In the $N_{\chi}=2$ case, the inclusion of the interference term gives at $14$ TeV an increase of the cross section of less than $40\%$, which translates into a correction of less than $10\%$ to the constraint on $\lambda\sqrt{YN}$. Such a deviation is acceptable when compared to the possible uncertainties in our projection. Thus we simulate the monojet events using the incoherent sum of the two operators in Eq.~(\ref{effops}) to obtain the dark penguin result.

\begin{figure}
\begin{center}
\includegraphics[width=0.45\textwidth]{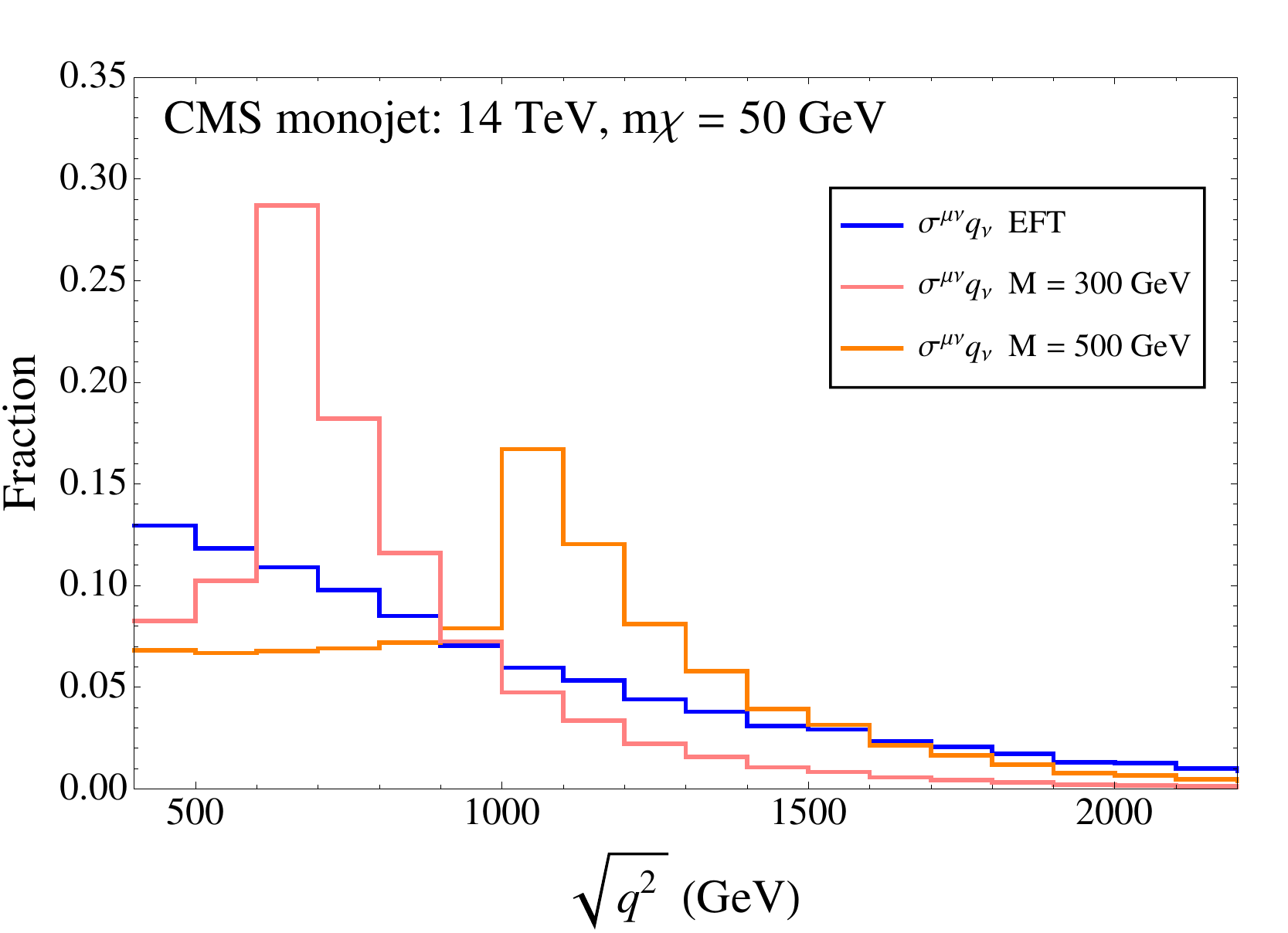}\qquad\includegraphics[width=0.45\textwidth]{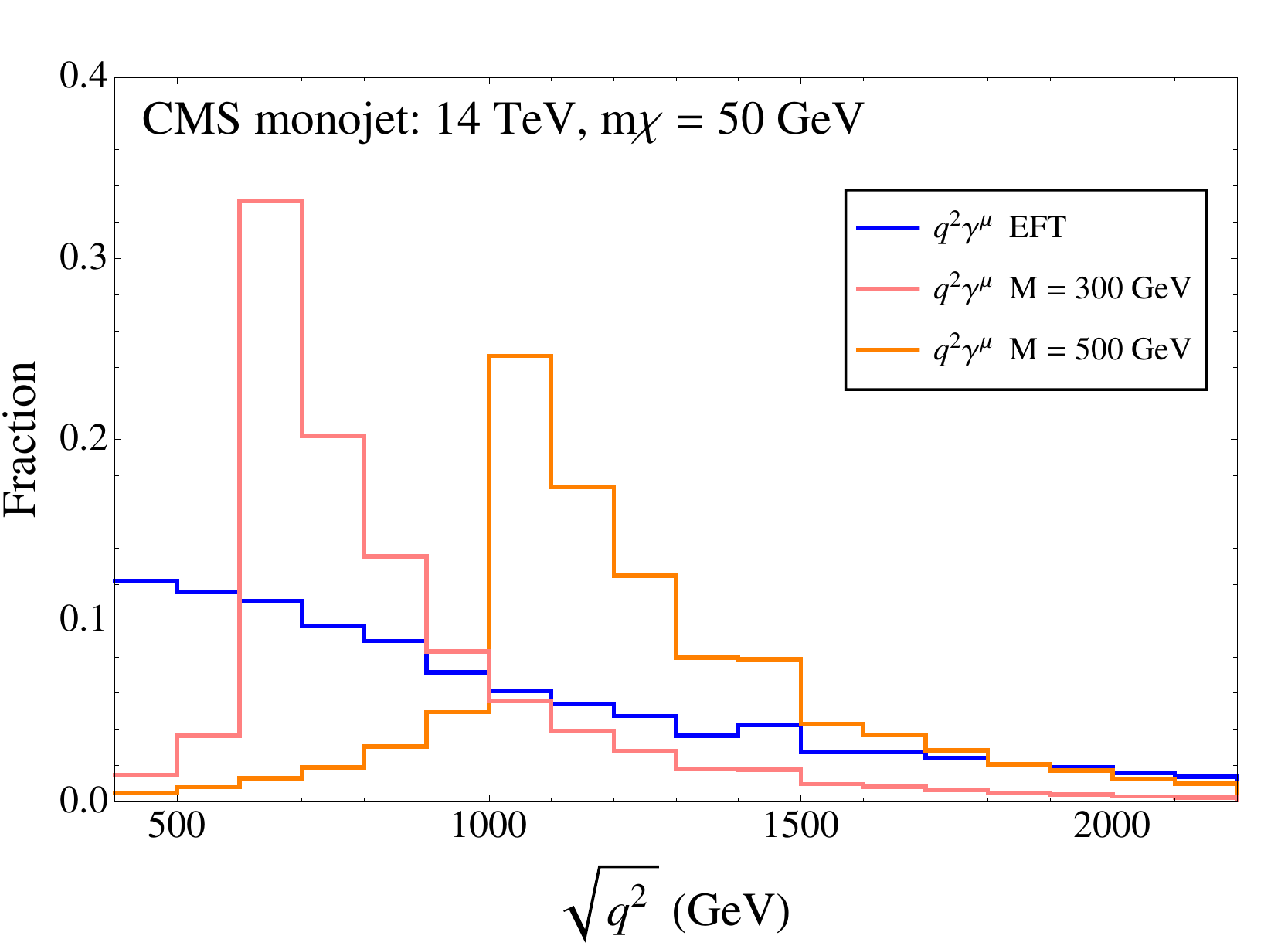}
\end{center}
\caption{The normalized $\sqrt{q^2}$ distribution of the DM signal in the monojet search. The blue histograms correspond to the simulation of the effective couplings $\bar{\chi}\sigma^{\mu\nu}\chi B_{\mu\nu}/\Lambda$ (left) and $\bar{\chi}\gamma^{\mu}\chi\partial^{\nu}B_{\mu\nu}/\Lambda^2$ (right), while the pink (orange) histograms correspond to the full dark penguin form factors with mediator mass $M=300\; (500)$ GeV. The form factors peak at $\sqrt{q^2}\simeq2M$. The cuts applied are described in Sec.~\ref{sec:monojet} (the cut on the MET is set to $550$ GeV).}
\label{fig:rescalingexample}
\end{figure}

To demonstrate the effect of the form factors, we plot in Fig.~\ref{fig:rescalingexample} the $\sqrt{q^2}$ distribution of the monojet events in both the dark penguin and EFT cases. Notice that due to the kinematics of the process, the MET is given by $\sqrt{q^2}$ multiplied by some angular factors. The distribution of the dark penguin is remarkably different from the one of the EFT: the dark penguin exhibits an enhancement around $\sqrt{q^2}\simeq 2M$, corresponding to the threshold for the production of on-shell mediators \cite{Tamarit:2013nna}, and it is softer than the EFT at high energy. As a consequence, contrarily to the strategy used for effective DM couplings, where a strong cut on the MET is generically preferred, in the search for the dark penguin with light mediators a softer cut is favored. We will return to this point when discussing the monojet result, see Fig.~\ref{fig:METcuts}. The preference for soft MET cuts applies also to the monophoton and diphoton searches.

\subsection{LHC monojet}\label{sec:monojet}

\begin{figure}
\begin{center}
\includegraphics[width=0.455\textwidth]{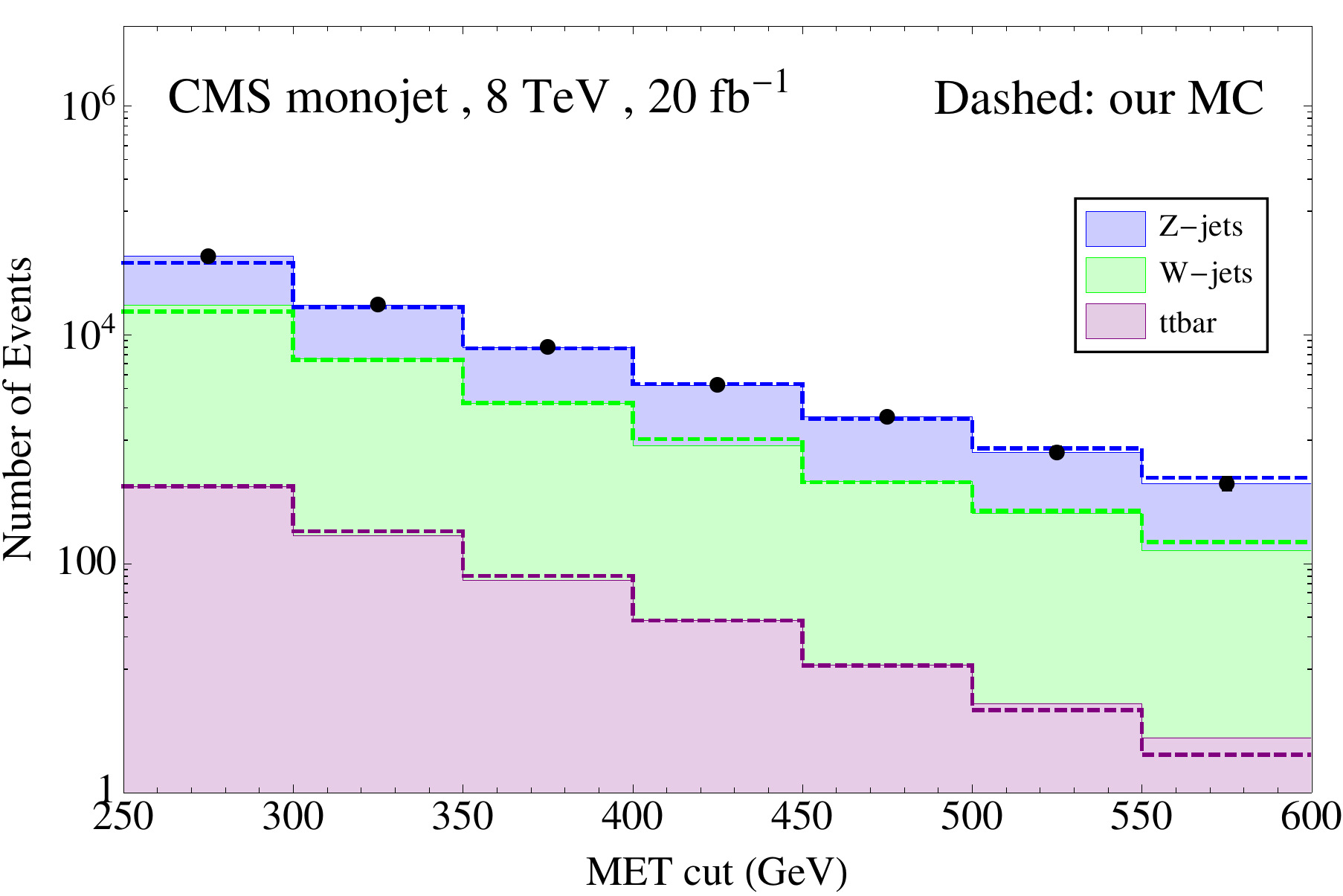}\qquad\includegraphics[width=0.47\textwidth]{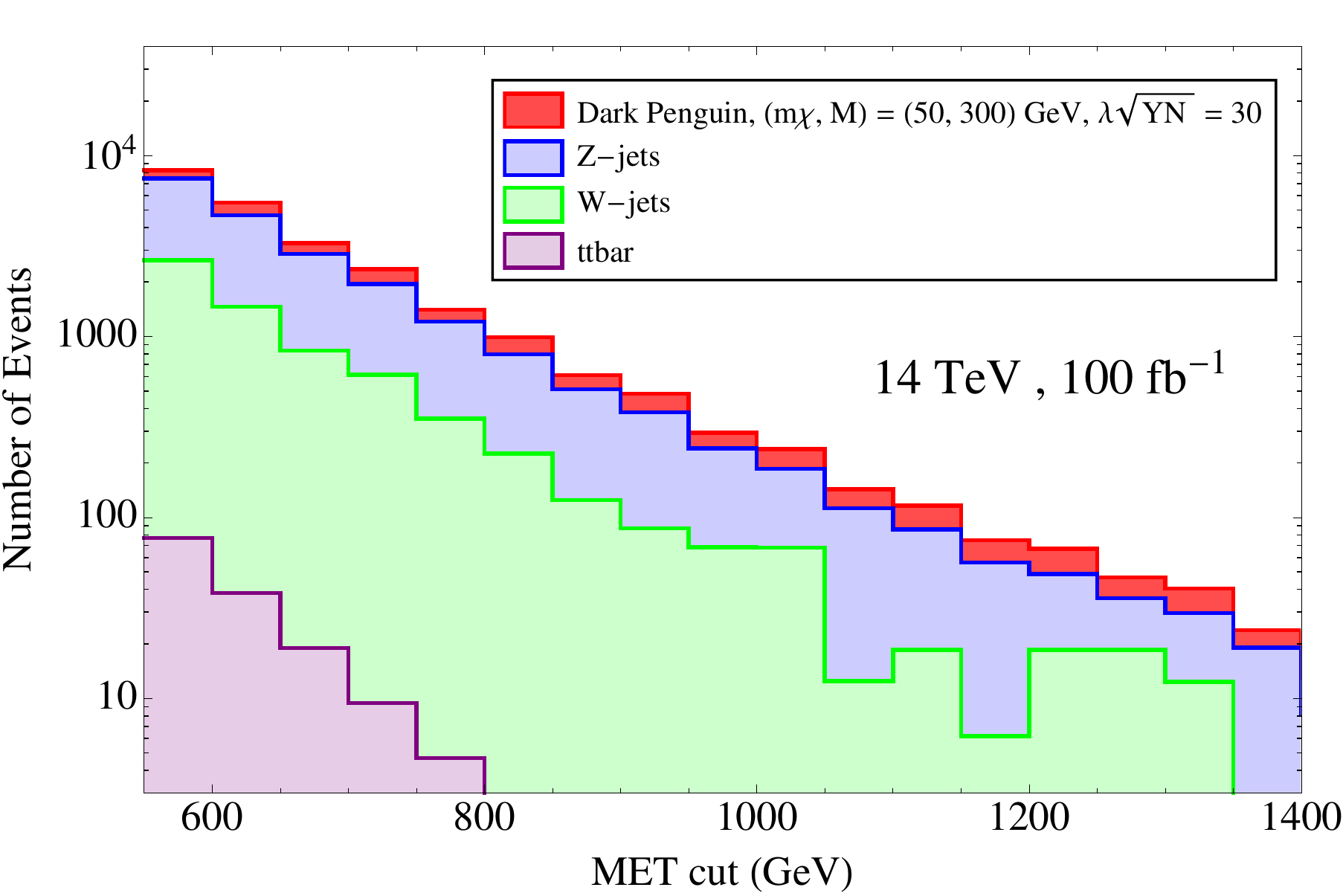}
\end{center}
\caption{Left panel: the dominant SM backgrounds at 8 TeV from the CMS study in \cite{CMS:rwa} (shaded regions) and from our MC simulation (dashed lines). Right panel: distribution of the dominant backgrounds at 14 TeV. Here we take a large DM coupling, $\lambda\sqrt{YN}=30$, for the visualization of the signal distribution.}
\label{fig:monojetbg}
\end{figure}

For the monojet channel, we follow and extend the CMS analysis in \cite{CMS:rwa}. The event selection requires one jet with $p_T(j_1) > 110$ GeV and $|\eta (j_1)| < 2.4$. A second jet with $p_T(j_2) > 30$ GeV and $|\eta(j_2)| < 4.5$ is allowed, as long as the $\Delta \phi (j_1,j_2) < 2.5$. Events containing a third jet satisfying $p_T(j_3) > 30$ GeV and $|\eta(j_3)| < 4.5$ are vetoed. To reduce the backgrounds from $Z$ and $W$ production, events containing electrons with $p_T > 10$ GeV and $|\eta| < 2.5$, muons with $p_T > 10$ GeV and $|\eta| < 2.1$, or hadronic taus with $p_T > 20$ GeV and $|\eta| < 2.3$ are also vetoed. The counting experiments in \cite{CMS:rwa} are performed in 7 signal regions, with MET cut from $250$ to $550$ GeV with a step of $50$ GeV. In the left panel of Fig.~\ref{fig:monojetbg} we show the comparison of our simulated backgrounds with those reported by CMS. For each of the three main backgrounds $Z+$jets, $W+$jets and $t\bar{t}$, we use MCFM to obtain the QCD $K$-factor and apply an additional rescaling factor to match the CMS result. The rescaling factors are $0.84$, $0.95$ and $0.60$, respectively. The rescaling factor for $Z+$jets is also applied to the DM signal events.
 
\begin{figure}
\begin{center}
\includegraphics[width=0.46\textwidth]{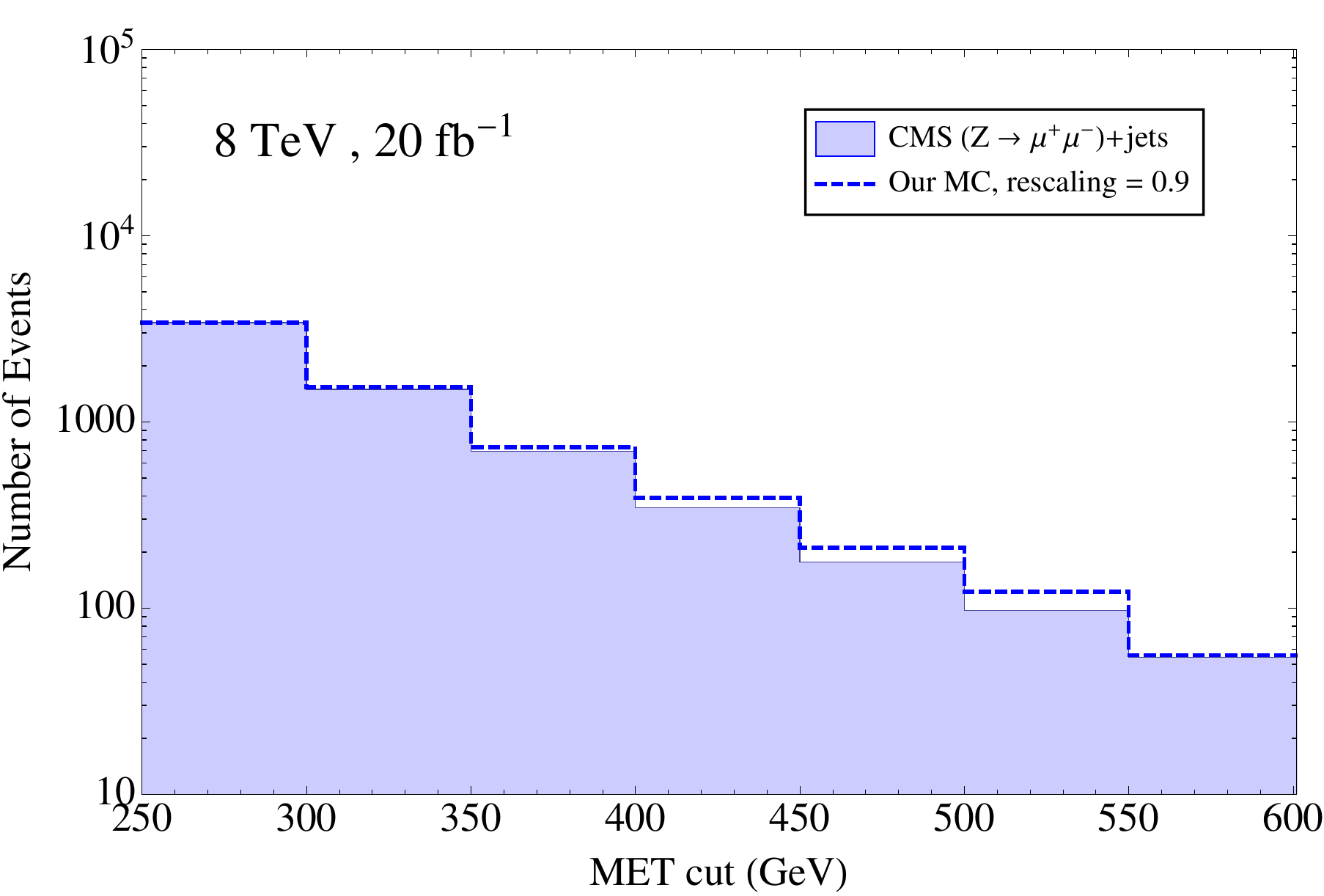}\qquad
\includegraphics[width=0.46\textwidth]{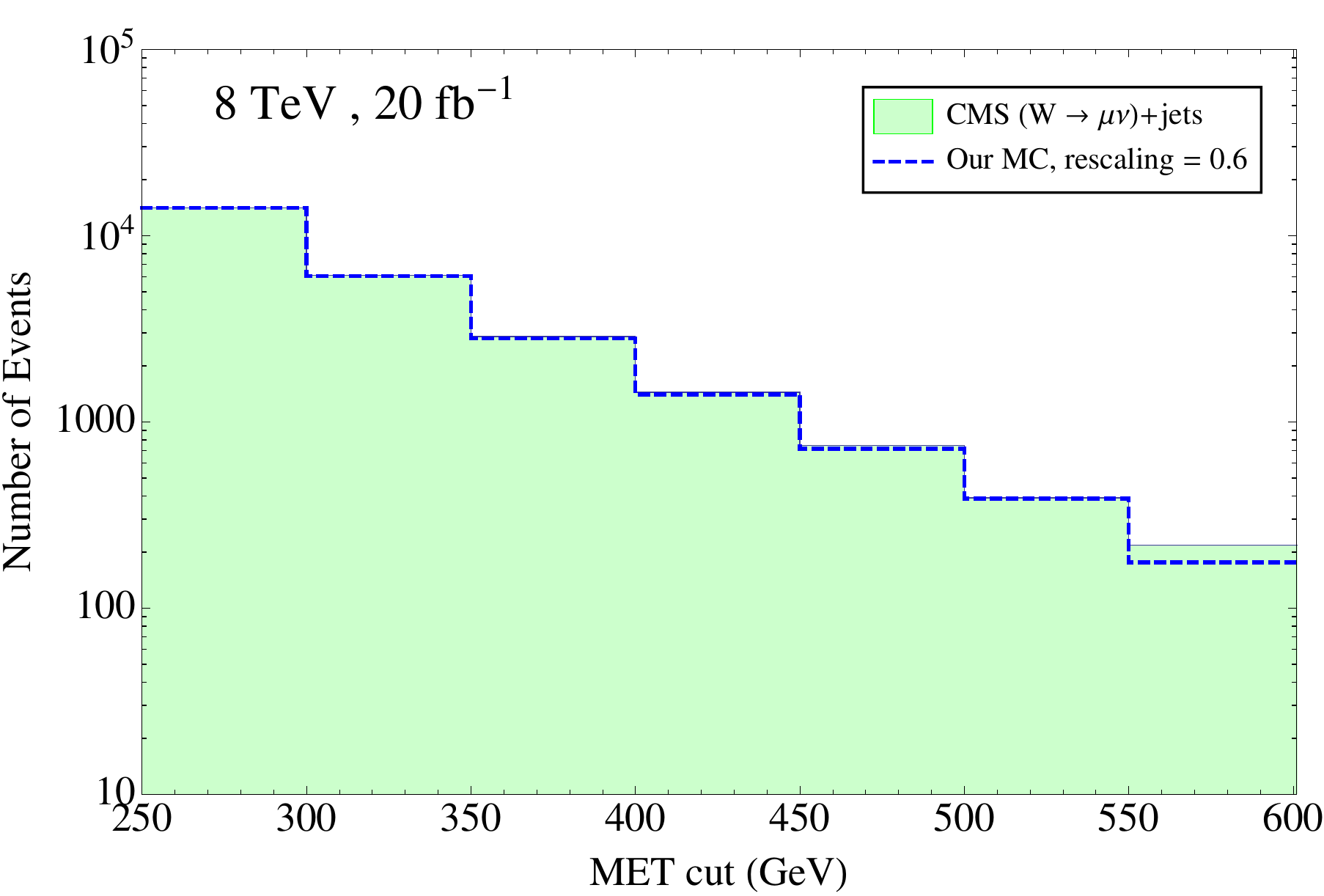}
\end{center}
\caption{Event distributions of $Z(\mu^+\mu^-)+$jets and $W(\mu\nu)+$jets in the control region.}
\label{fig:monojetvalidation}
\end{figure}

The CMS search uses a data driven analysis based on a $\mu+$jets control sample to determine the normalization of the $jZ$ and $jW$ backgrounds. The corresponding events in the control sample are $Z(\mu^+\mu^-)+$jets and $W(\mu\nu)+$jets, respectively. The cuts for the control sample are the same as for the monojet search, except the lepton vetoes are not applied. The $Z(\mu^+\mu^-)+$jets sample requires two muons with $p_T > 20$ GeV and $|\eta| < 2.1$, with at least one of the muons passing isolation requirements, and the invariant mass of the muon pair between $60$ and $120$ GeV. Similarly, the $W(\mu\nu)+$jets sample requires one isolated muon with $p_T > 20$ GeV and $|\eta| < 2.1$, and the transverse mass of the muon plus neutrino system in the range $50$ GeV $< M_T < $ $100$ GeV. Fig.~\ref{fig:monojetvalidation} shows the comparison between our MC simulation of the control regions and the CMS results. For $Z(\mu^+\mu^-)+$jets, the MET is defined as the sum of the muon transverse momenta, while for $W(\mu\nu)+$jets the MET is given by the neutrino transverse momentum. The good agreement with the CMS results allows us to simulate the data driven analysis in the $14$ TeV study.

\begin{figure}
\begin{center}
\includegraphics[width=0.39\textwidth]{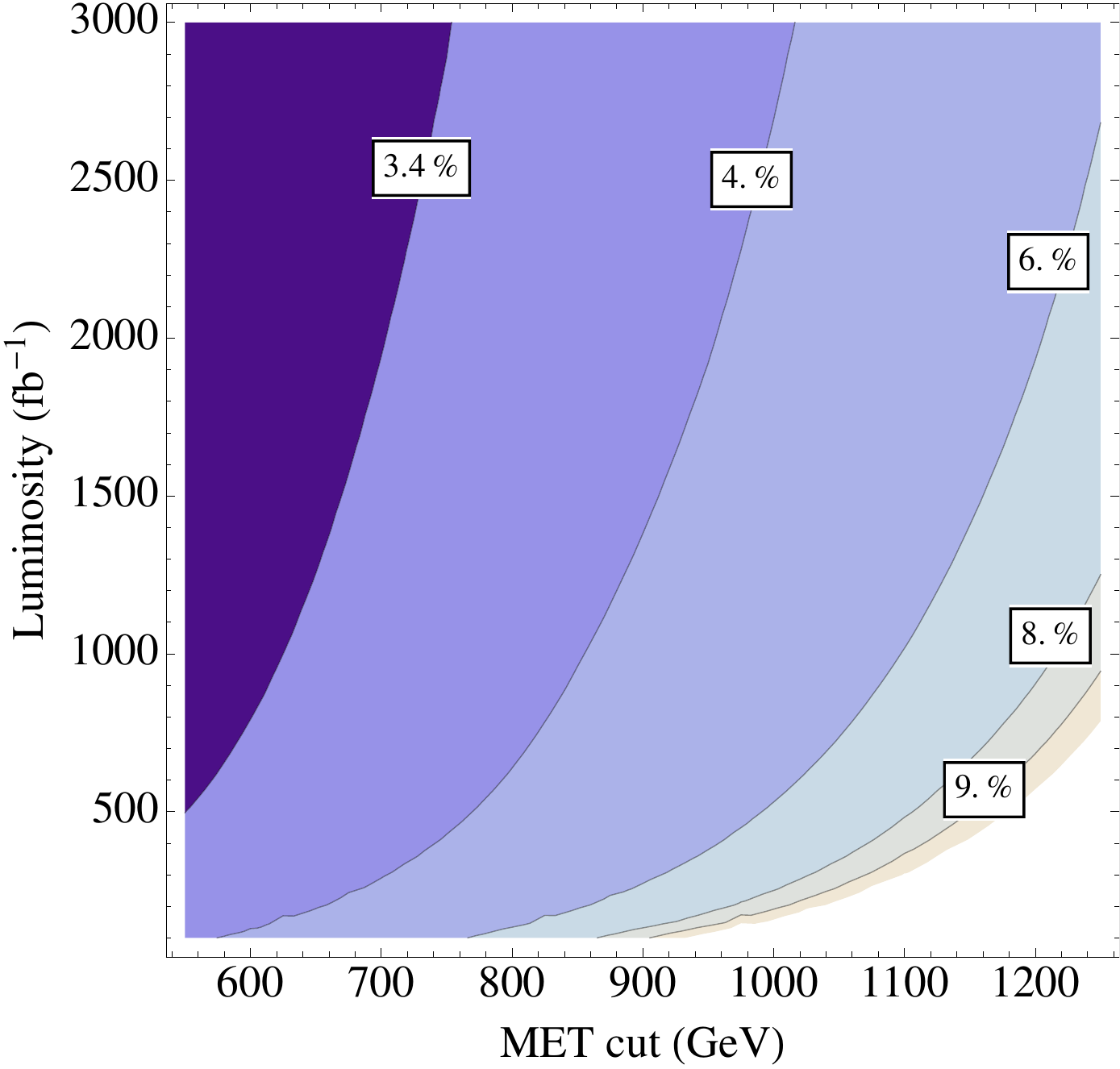}
\qquad\includegraphics[width=0.4\textwidth]{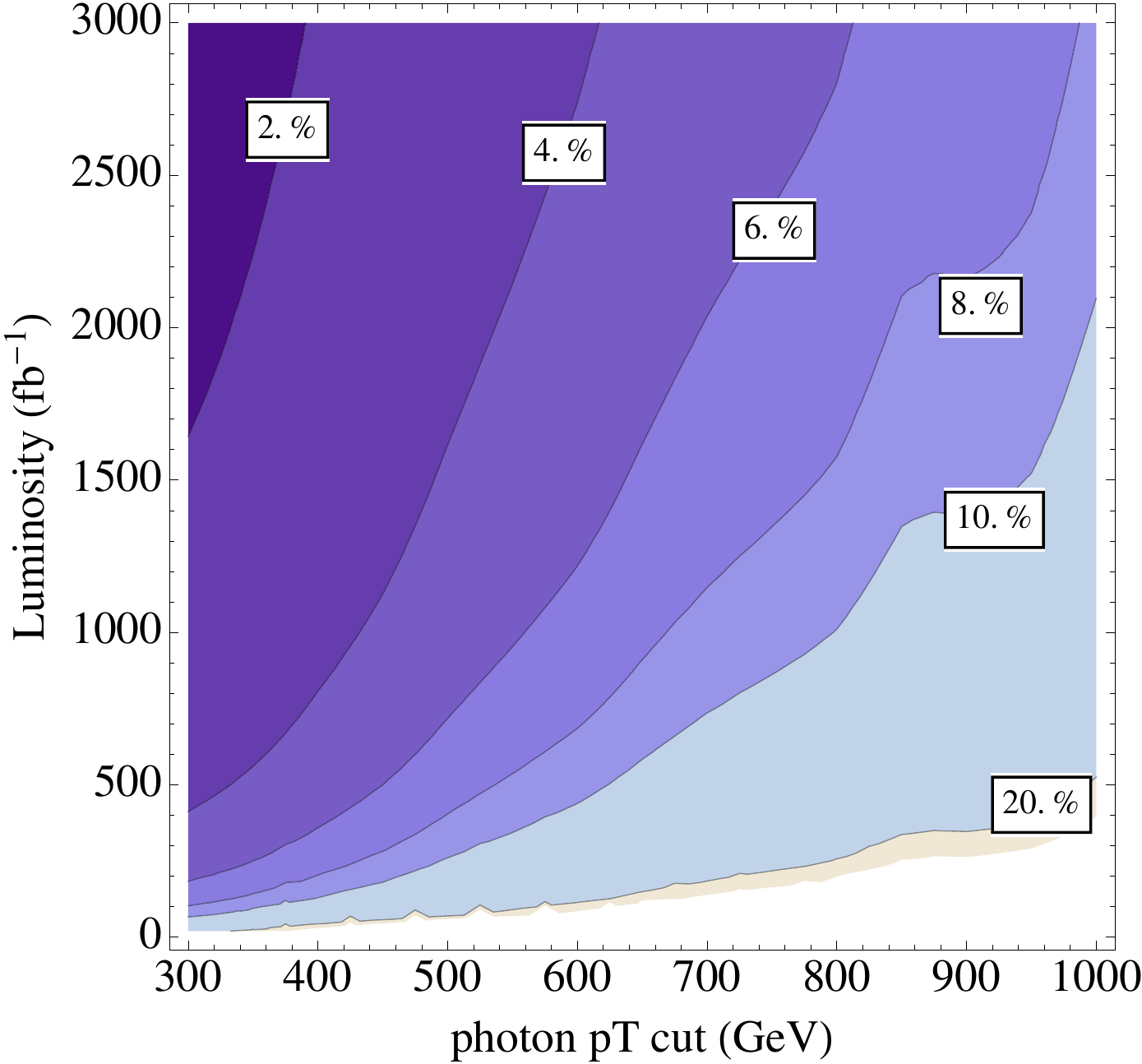}
\end{center}
\caption{Monojet and monophoton systematic uncertainties as functions of luminosity and MET cut, following the assumptions in Tables~\ref{Tab:channels} and \ref{Tab:channelsphoton}. The number shown on each contour is the total systematic uncertainty divided by the number of background events.}
\label{fig:MonoXerror}
\end{figure}

For the $14$ TeV projection we follow the same cuts in the CMS $8$ TeV analysis, apart from varying the MET cut from $550$ to $2250$ GeV with a step of $100$ GeV. The dominant systematic uncertainty for each MET cut choice is obtained from the control sample simulation. For the remaining uncertainties, we take the values quoted in the CMS analysis for a MET cut of $550$ GeV, and assume they will decrease with the square root of luminosity, see Table~\ref{Tab:channels}. Since at $8$ TeV these additional uncertainties do not depend strongly on the MET cut, it is a reasonable guess to use their value also at $14$ TeV. The right panel of Fig.~\ref{fig:monojetbg} shows the projected monojet backgrounds studied in this work, together with the dark penguin signal. The total systematic uncertainty is shown in Fig.~\ref{fig:MonoXerror} as a function of luminosity and MET cut. To provide some figure of merit, with $3$ ab$^{-1}$ of data and MET cut of $550$ GeV the background is $\sim 10^6$ events, which requires $\sim 10^4$ DM signal events for a few-$\sigma$ excess, given an uncertainty of order $1\%$. The corresponding $\lambda\sqrt{YN}$ is $\sim 20$ when the mediator mass is $M\simeq 500$ GeV.

\begin{table}
   \begin{center}
   \begin{tabular}{|c|c|c|} 
   \hline
    & CR statistics & Other systematics \\
   \hline\hline
   \vspace{0mm}
   $Z+$jets & $N_{CR_{Zj}}^{-1/2}$ & $3.9\,\hat{\mathcal{L}}^{-1/2}$ \\
   \hline    
   \vspace{0mm}
   $W+$jets & $N_{CR_{Wj}}^{-1/2}$ & $4.6\,\hat{\mathcal{L}}^{-1/2}$ \\      
   \hline
   \end{tabular}      
   \end{center}
   \caption{Summary of the contributions (in $\%$) to the uncertainty used for the $14$ TeV monojet study, following the analysis in \cite{CMS:rwa}. The dominant uncertainty comes from the limited number of control sample events. $\hat{\mathcal{L}}\equiv\mathcal{L}/(20\;\mathrm{fb}^{-1})$, and we assume the other sources of uncertainty will be improved with the increase of data.}
   \label{Tab:channels}
\end{table}

\subsection{LHC monophoton}

\begin{figure}
\begin{center}
\includegraphics[width=0.48\textwidth]{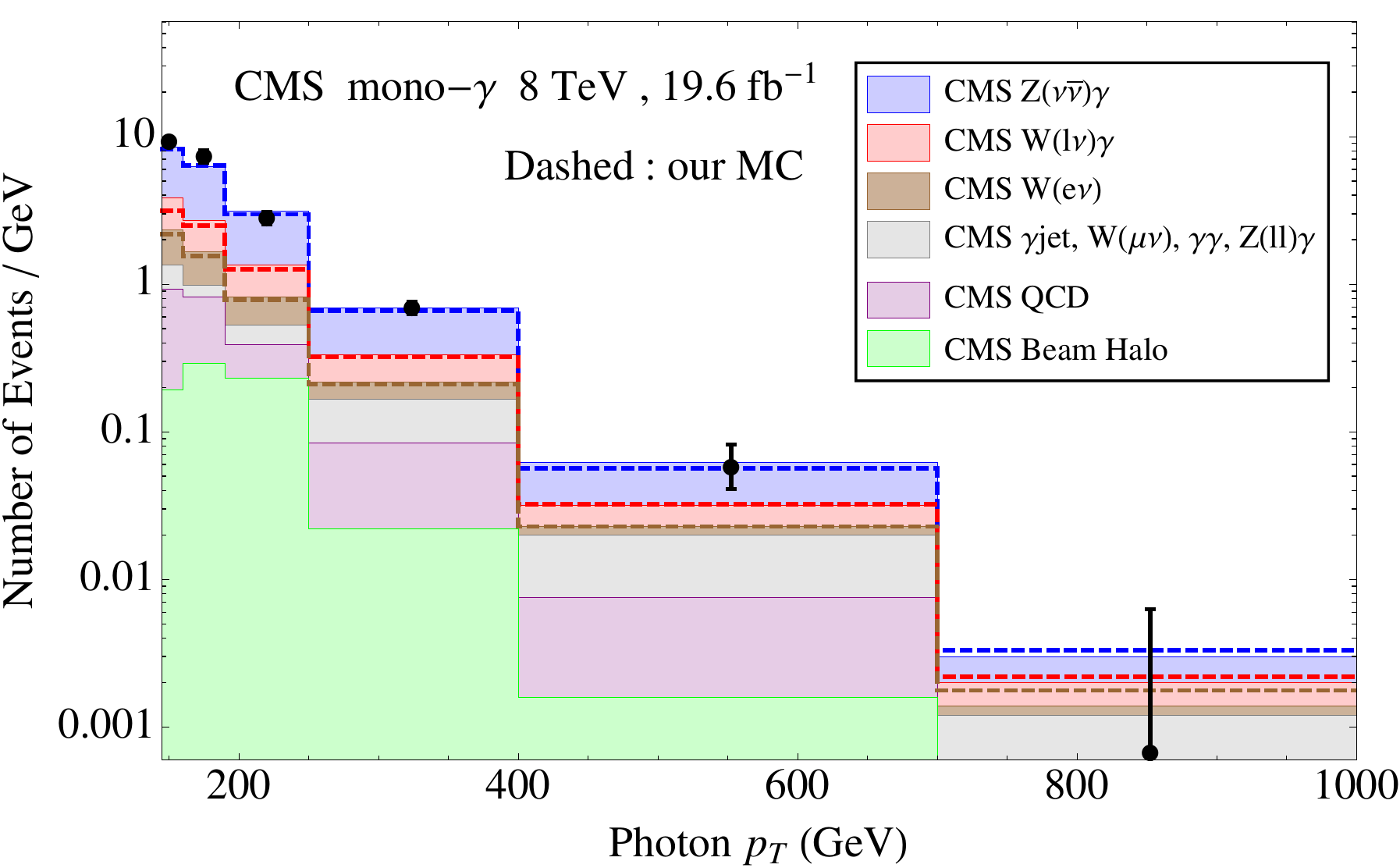}\qquad\includegraphics[width=0.462\textwidth]{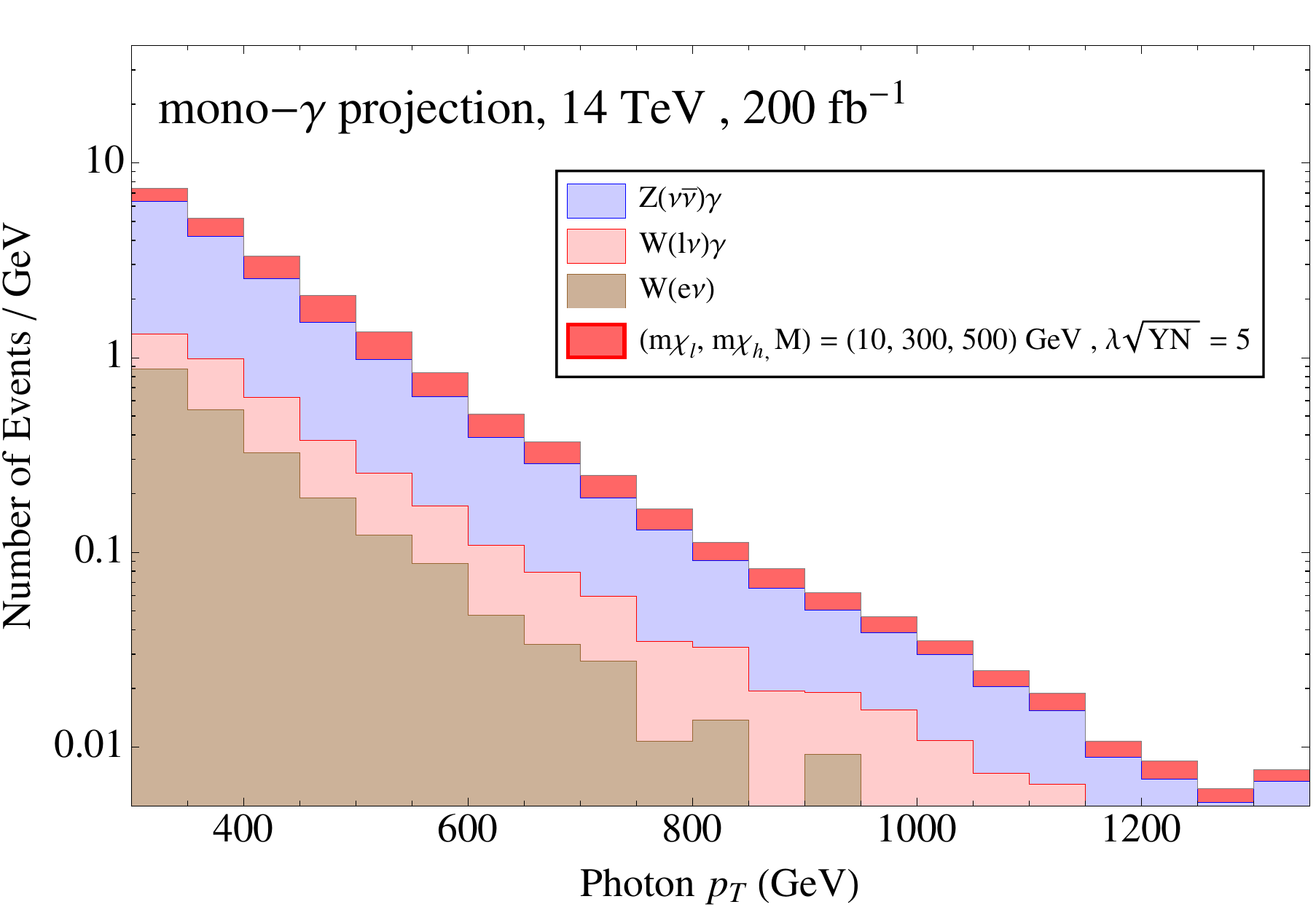}
\end{center}
\caption{Left panel: comparison of the $8$ TeV monophoton backgrounds from our MC simulations and from the CMS analysis. Our results are shown as dashed histograms, stacked on top of the sum of the subleading backgrounds ($W(\mu\nu)$, $Z(ll)\gamma$, $\gamma\,j$, $\gamma\gamma$, QCD and beam halo) as given by CMS \cite{CMS-PAS-EXO-12-047}, in order to facilitate the comparison. Right panel: the projected $14$ TeV background used in the analysis.}
\label{fig:Monophotonbackground8TeV}
\end{figure}

For the monophoton channel, we follow the 8 TeV CMS search in \cite{CMS-PAS-EXO-12-047}. The event selection requires one photon with $p_T^\gamma > 145$ GeV and $|\eta^{\gamma}|<1.4442$, and in addition $\slashed E_T > 140$ GeV with $\Delta\phi(\slashed E_T,\gamma)>2$. Events containing electrons or muons with $p_T > 10$ GeV, $|\eta| < 2.5$, and $\Delta R_{\ell\gamma}>0.5$ are vetoed, as well as events containing more than one reconstructed jet with $p_T > 30$ GeV,  $|\eta| < 2.5$ and $\Delta R_{j\,\gamma}>0.5$. The comparison of our simulation of the photon $p_T$ distribution to the one obtained in the CMS analysis is given in Fig.~\ref{fig:Monophotonbackground8TeV}. For each background we apply a $p_T$-dependent $K$-factor obtained from the NLO MCFM calculation \cite{Campbell:2010ff}.\footnote{The $K$-factor is very large for $W(\ell\nu)\gamma$, $K\gtrsim 4.5$ in the region considered $p_T^\gamma > 145\;\mathrm{GeV}$ and growing with increasing $p_T^\gamma$. This is a consequence of the presence of a `radiation zero' \cite{Brodsky:1982sh,Brown:1982xx} that strongly suppresses the LO amplitude. The NLO QCD corrections (in particular those associated with the $qg$ partonic channel, which is not suppressed by the radiation zero) therefore contribute a large fraction of the total cross section, leading to the big $K$-factor \cite{Ohnemus:1992jn}.} In addition, we apply an overall rescaling in order to match the normalization provided by CMS: the rescaling factor is equal to $1.1$ for $Z(\bar{\nu}\nu)\gamma$, $1.2$ for $W(e\nu)$ and $1.8$ for $W(\ell\nu)\gamma$.\footnote{We do not apply to the DM signal the rescaling factor $1.1$ obtained from the $Z(\nu\bar{\nu})\gamma$ background, since it amounts to a negligible correction.} We do not attempt to simulate the backgrounds given by the misidentification of leptons ($W(\mu\nu)$, $Z(ll)\gamma$), jets ($\gamma\,j$), $\gamma\gamma$, beam halo, and QCD, but these backgrounds are subdominant as is shown in the left plot of Fig.~\ref{fig:Monophotonbackground8TeV}. Only the three dominant backgrounds $Z(\bar{\nu}\nu)\gamma,\,W(e\nu),\,W(\ell\nu)\gamma$ are included in the $14$ TeV analysis.
\begin{figure}
\begin{center}
\includegraphics[width=0.47\textwidth]{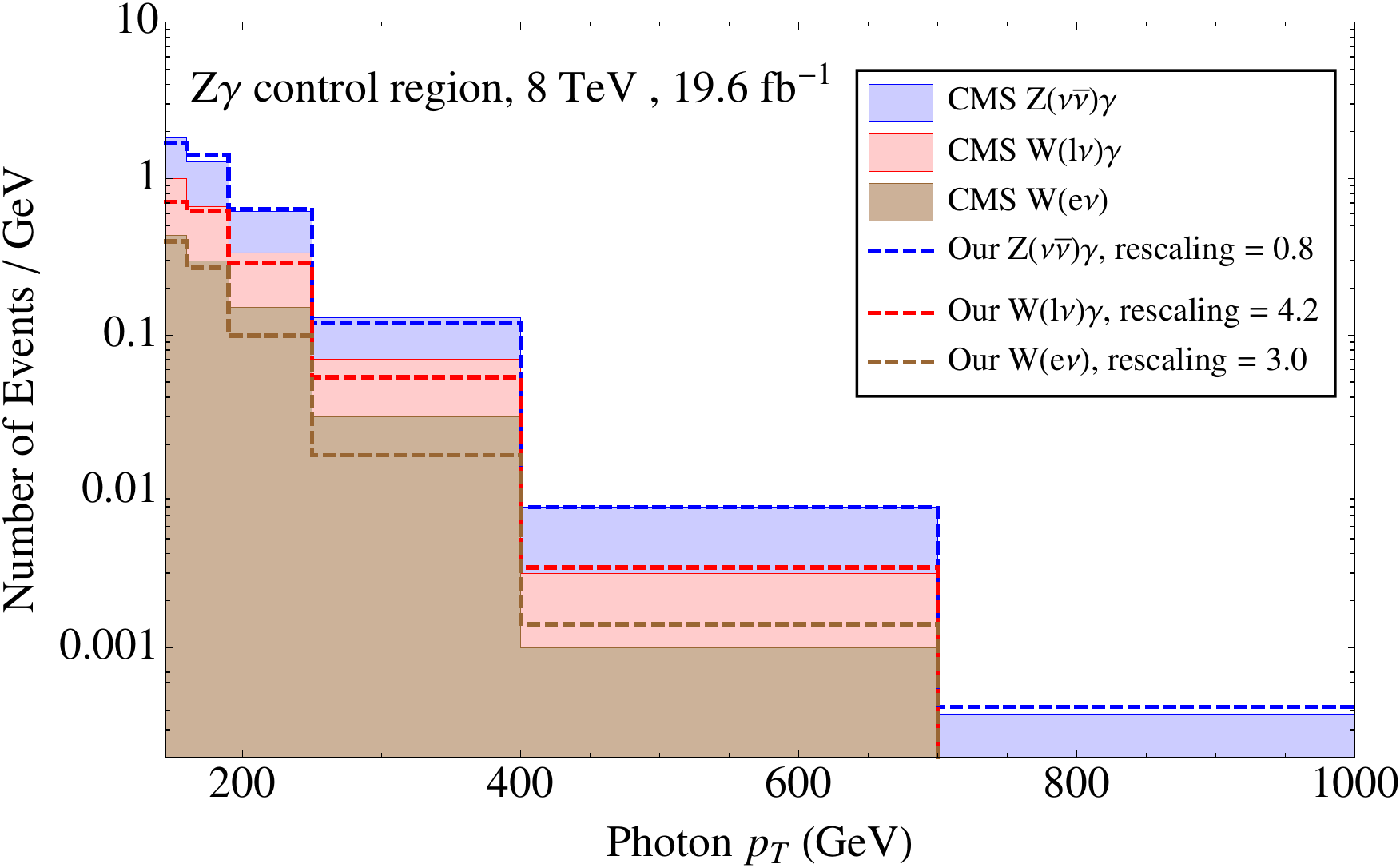}\qquad\includegraphics[width=0.47\textwidth]{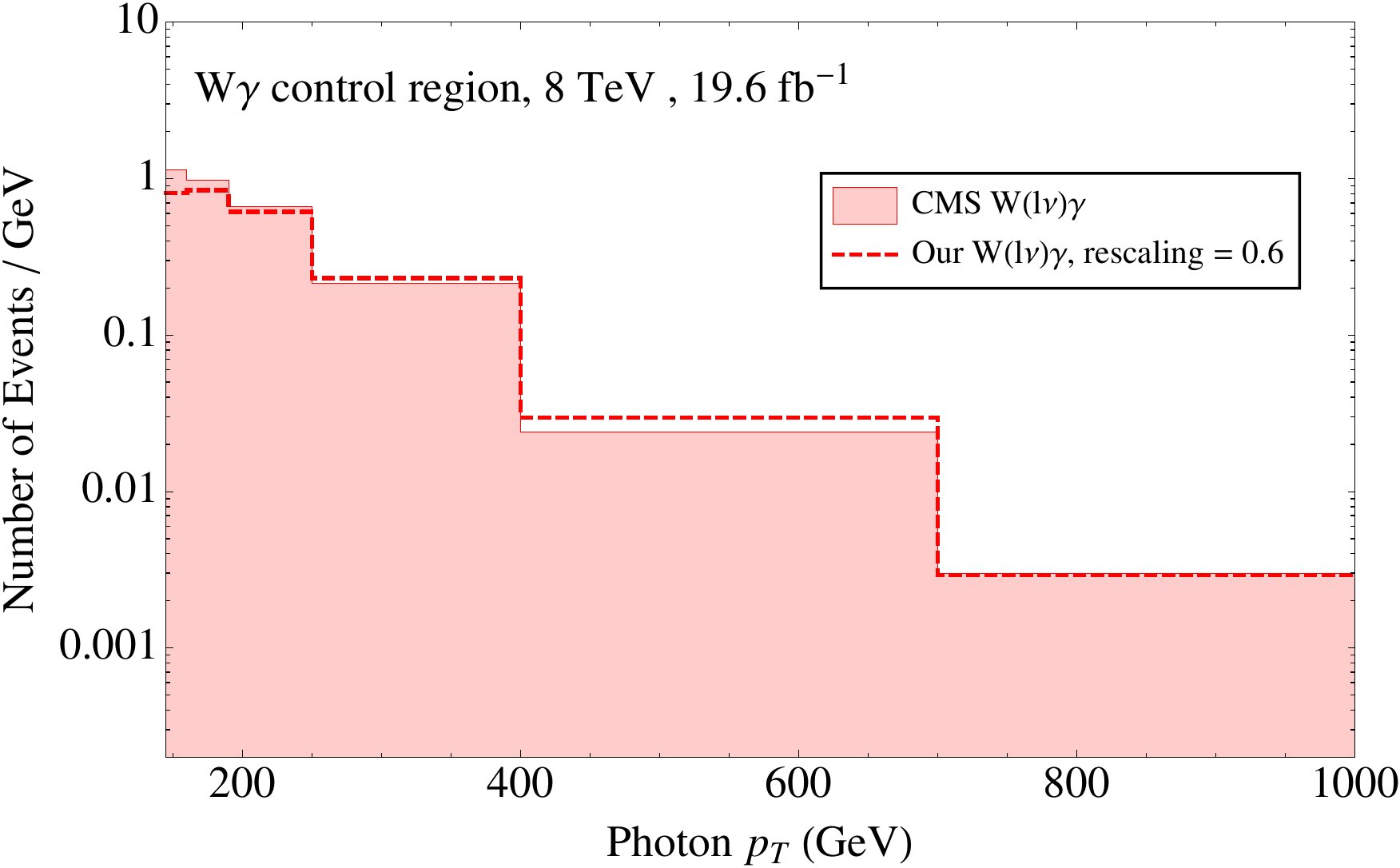}
\end{center}
\caption{Control regions for the $Z\gamma$ (left) and $W\gamma$ (right) backgrounds, with the inverted angular cut $\Delta\phi(\slashed E_T,\gamma)<2.9$ for the $Z(\nu\bar{\nu})\gamma$ events and the inverted lepton veto for the $W(\ell\nu)\gamma$ events.}
\label{fig:MonophotonCR}
\end{figure}

In the CMS analysis \cite{CMS-PAS-EXO-12-047}, the main source of systematic uncertainty is given by the higher-order QCD corrections to the $Z(\bar{\nu}\nu)\gamma$ and $W(\ell\nu)\gamma$ backgrounds. Although the collaboration performs a data-driven analysis for each of the two backgrounds and compares the results to those obtained from MCFM, the limited number of control sample events at $8$ TeV does not help to reduce the uncertainty. Therefore, in our 8 TeV analysis we made directly use of the uncertainties quoted by CMS. However, following the same strategy of the monojet case, we are going to assume that the data-driven analysis will play an important role in the future. With the increase of luminosity, the systematic uncertainty will be based on the statistics of the control sample. In this case, using too stringent kinematic cuts in the search can increase the size of the systematic uncertainty, and a proper choice of the cut is required to optimize the signal significance. 

To perform a data-driven analysis at $14$ TeV, we compare our $8$ TeV simulation of the two control region samples to the CMS results, and then apply the same analysis for the $14$ TeV case. The $Z(\nu \bar{\nu})\gamma$ control region is defined by the inverted angular cut $\Delta\phi(\slashed E_T,\gamma)<2.9$, while the $W(\ell\nu)\gamma$ control region is defined by the inverted lepton veto. In both cases we obtain a result in good agreement with CMS, see Fig.~\ref{fig:MonophotonCR}.\footnote{Notice that we need to apply large rescaling factors for the $W(\ell\nu)\gamma$ and $W(e\nu)$ backgrounds in the $Z\gamma$ control region, $4.2$ and $3.0$, respectively. However, these backgrounds are subdominant to $Z(\nu\bar{\nu})\gamma$.} This allows us to simulate the control region sample at $14$ TeV and derive the relative systematic uncertainty shown in Fig.~\ref{fig:MonoXerror}. As given in Table~\ref{Tab:channelsphoton}, the total uncertainty also contains various systematic uncertainties in the control region analysis for $Z\gamma$ and $W\gamma$, as well as the uncertainty on the probability for an electron to be misidentified as photon for the $W(e\nu)$ background. We take the values of these uncertainties from the $8$ TeV analysis \cite{CMS-PAS-EXO-12-047}, and assume they will be improved with the increased luminosity as $\mathcal{L}^{-1/2}$. This is based on the fact that various data-driven analyses have been used to determine these uncertainties at $8$ TeV.

For the $14$ TeV monophoton projection, we follow the same event selection used in the $8$ TeV CMS search, except for tighter cuts on the photon $p_T$ and MET: both are required to be larger than $300$ GeV, to suppress the QCD and beam halo backgrounds, which are not included in the $14$ TeV estimate. As discussed in Sec.~\ref{sec:formfactors}, choosing a tighter cut does not improve the sensitivity to the dark penguin.
 
\begin{table}
   \begin{center}
   \begin{tabular}{|c|c|} 
   \hline
    Background & Relative systematic uncertainty\\
   \hline\hline
   $Z(\bar{\nu}\nu)\gamma $ & $\sqrt{N_{CR_{tot}}+(0.10\,\hat{\mathcal{L}}^{-1/2}N_{CR_{Z\gamma}})^2+(0.16\,\hat{\mathcal{L}}^{-1/2}N_{CR_{W(\ell\nu)\gamma}})^2}/N_{CR_{tot}}$\\
   \hline
   $W(\ell\nu)\gamma$ & $\sqrt{N_{CR_{tot}}+(0.22\,\hat{\mathcal{L}}^{-1/2}N_{CR_{W(\ell\nu)\gamma}})^2}/N_{CR_{tot}}$\\
   \hline
   $W(e\nu)$ & $0.10\,\hat{\mathcal{L}}^{-1/2}$\\
      \hline     
   \end{tabular}      
   \end{center}
   \caption{Summary of the contributions to the uncertainty used for the $14$ TeV monophoton search. $\hat{\mathcal{L}}=\mathcal{L}/(20\,{\rm fb}^{-1})$, and the number of control sample events depends on the photon $p_T$ cut. The relative uncertainties are taken from the $8$ TeV CMS analysis \cite{CMS-PAS-EXO-12-047}.}
   \label{Tab:channelsphoton}
\end{table}
\subsection{LHC diphoton$+$MET: prompt}
If there is more than one flavor of $\chi_i$ ($N_{\chi}>1$), a heavier $\chi_h$ can decay to the lightest $\chi_l$ and a photon: $\chi_h \rightarrow \chi_l \gamma$. If $\chi_h$ is pair produced at the collider, diphoton$+$MET can be important in probing the multi-flavor scenario. For this channel, we follow the diphoton search at 8 TeV of energy and 20.3 $\textrm{fb}^{-1}$ of data performed by the ATLAS collaboration \cite{ATLAS-CONF-2014-001}. The search is aimed to constrain the parameter space of gauge mediated supersymmetry breaking models. The event selection applied in this analysis is as follows. At least two photons with $p_{T}^\gamma > 75$ GeV and $|\eta_\gamma| < 2.37$ are required. Jets are reconstructed using the anti-$k_t$ algorithm with a radius parameter of 0.4. The jets are required to have $p_{T}^j > 30$ GeV and $|\eta_j| < 2.8$. No vetoes are applied on the number of leptons and jets. Several new variables are introduced as follows. An angular separation variable, $\phi^{\text{min}}_\gamma$, is defined as the minimum azimuthal angle between $E_T^\text{miss}$ and the two selected photons. In presence of jets, a variable $\phi^{\text{min}}_\text{jet}$ is introduced and defined as the minimum azimuthal angle between $E_T^\text{miss}$ and the two highest reconstructed jets, where the jets are required to have $p_{T}^j > 75$ GeV. The total visible energy, $H_T$, is defined as the scalar sum of the transverse momenta of photons, jets and leptons.

\begin{table}
   \begin{center}
   \begin{tabular}{|c|c|c|c||c|} 
   \hline
     & WP1 & WP2 & MIS & 14 TeV\\
   \hline\hline
 	 $\phi^{\text{min}}_\gamma$ & 0.5 & 0.0 & 0.0 & 0.0\\	
	 $\phi^{\text{min}}_\text{jet}$ & 0.5 & 0.5 & 0.5 & 0.5\\
	 $H_T > $ (GeV) & 400 & 600 & 0 & 0\\
	  $E_T^{miss} > $ (GeV) & 200 & 150 & 250 & 350 \\
	 jet veto & no veto & no veto & no veto &$p_T > 50$ GeV, $|\eta| < 2.8$ \\
	  lepton veto & no veto & no veto & no veto & $p_T >$ 25 GeV, $|\eta| < 2.5$ \\	
      \hline     
       Predicted & $1.01\pm0.36$ & $2.38\pm0.69$ & $1.59\pm0.58$ & 2.90 \\
       Observed & 1 & 5 & 2 & -- \\
       \hline
   \end{tabular}      
   \end{center}
   \caption{Cuts, number of predicted events and number of observed events for the signal regions used in the $8$ TeV ATLAS diphoton$+$MET analysis and in our 14 TeV projection. For the latter a luminosity of 300 fb$^{-1}$ was assumed.}
   \label{Tab:diphotonATLASsignalreg}
\end{table}

There are three relevant signal regions defined in the first three columns of Table \ref{Tab:diphotonATLASsignalreg}. The region WP1 is preferred for a larger mass splitting between $m_{\chi_h}$ and $m_{\chi_l}$, because the cut $H_T > 400$ GeV can only be satisfied with high $p_T$ photons. In this signal region the SM background is smaller than in the other two regions, therefore a stricter bound can be achieved. For a lower mass splitting, the region MIS is preferred because it does not have any requirement on the $H_T$ value. At 8 TeV, the region WP2 provides a weaker bound compared with the other regions, because of an upward fluctuation in the number of observed events. 

Note that none of the signal regions veto on the leptons and the jets. In the benchmark model used in the ATLAS analysis, the NLSP are produced in the decay chain of either gluino or chargino, and either jets or leptons are always present in the final states. At the partonic level, the final states of our model contain only a pair of photons and missing energy from the DM particles $\chi_l$. Hence vetoing jets and leptons increases the sensitivity of the search to our model. 

\begin{figure}
\begin{center}
\includegraphics[width=0.7\textwidth]{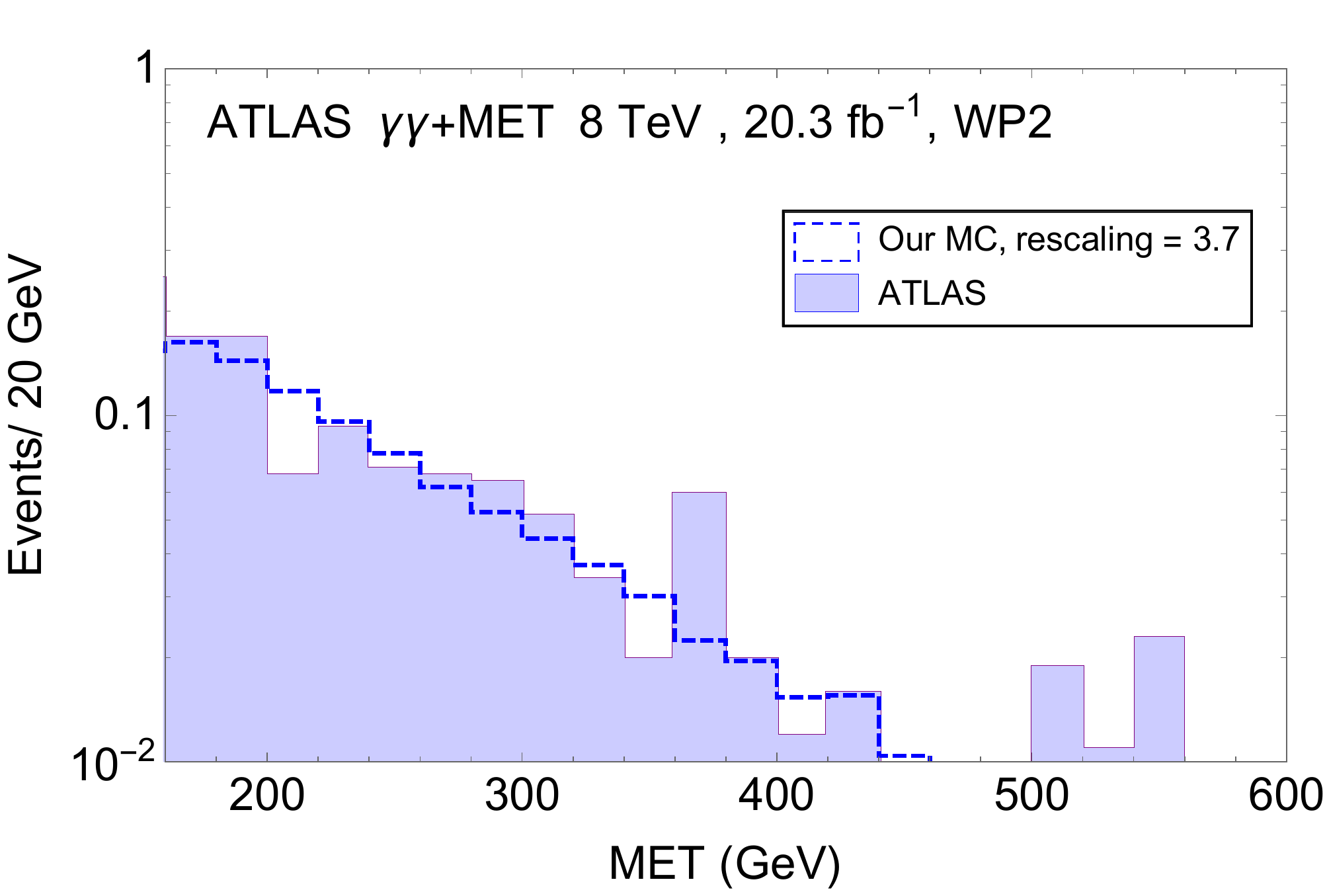}
\end{center}
\caption{Comparison between the ATLAS result and our MC simulation for the sum of $W\gamma\gamma$ and $Z\gamma\gamma$ in the signal region WP2 (described in Table~\ref{Tab:diphotonATLASsignalreg}).}
\label{fig:Diphotonbackground8TeV}
\end{figure}

The main backgrounds for these regions are $W\gamma\gamma$, $Z\gamma\gamma$ and ``QCD''. In this case, ``QCD'' is defined as the sum of multi-jet, $\gamma$ + jets and $\gamma\gamma$ + jets processes. When estimating the background, we multiplied the LO cross section of the $W\gamma\gamma$ ($Z\gamma\gamma$) background obtained from MadGraph5 with a $K$-factor of 8.1 (1.8) obtained from VBFNLO \cite{Baglio:2014uba}. This result still needs to be multiplied by a factor of 3.7 (1.1) to match the distribution reported by ATLAS.\footnote{In analogy with the $W\gamma$ process, the very large $K$-factor for $W\gamma\gamma$ can be understood as due to the presence of a radiation zero in the LO amplitude when the two photons are collinear \cite{Bozzi:2011wwa}. We cannot explain the large rescaling factor of $3.7$ needed to match the ATLAS $W\gamma\gamma$ result, which was obtained by means of a data driven analysis.} The sum of the $W\gamma\gamma$ and $Z\gamma\gamma$ missing energy distributions in the WP2 signal region is shown in Fig.~\ref{fig:Diphotonbackground8TeV}. In this region the $W\gamma\gamma$ background dominates $Z\gamma\gamma$ by an order of magnitude.
  


For the $14$ TeV simulation, we use cuts inspired by the MIS region of the 8 TeV ATLAS analysis, see Table \ref{Tab:diphotonATLASsignalreg}. The signal region is defined to have a slightly tighter missing tranverse energy than the 8 TeV analysis. Additionally, in order to increase the sensitivity of the search, jet and lepton vetoes are applied. In order to take into account the low $p_T$ jet veto correctly, we generated jet matched samples for both the signal and backgrounds. Contrarily to the $8$ TeV case, at $14$ TeV $Z\gamma\gamma$ is the leading background, as a consequence of the lepton veto: the expected background for the $14$ TeV signal region is of 2.64 events from $Z\gamma\gamma$ and 0.26 events from $W\gamma\gamma$, assuming 300 fb$^{-1}$ of luminosity. In the 8 TeV ATLAS analysis, the $Z\gamma\gamma$ background was estimated using MC, with an associated systematic uncertainty of $50\%$. Similarly to the previous subsections, we assume that the systematics will improve with luminosity as $\mathcal L^{-1/2}$. The systematic error is thus estimated to be of 0.69 events at 300 fb$^{-1}$ and 2.2 events at 3 ab$^{-1}$. We neglect the systematics for the subdominant $W\gamma\gamma$ background, since the number of events expected from $W\gamma\gamma$ is smaller than the uncertainty on $Z\gamma\gamma$. Finally we did not simulate the ``QCD''  backgrounds, but they are expected to be small due to the jet and lepton vetoes together with a harder MET cut.


\subsection{LHC diphoton$+$MET: displaced}\label{sec:displacedphotonTech}
\begin{figure}
\begin{center}
\includegraphics[width=0.45\textwidth]{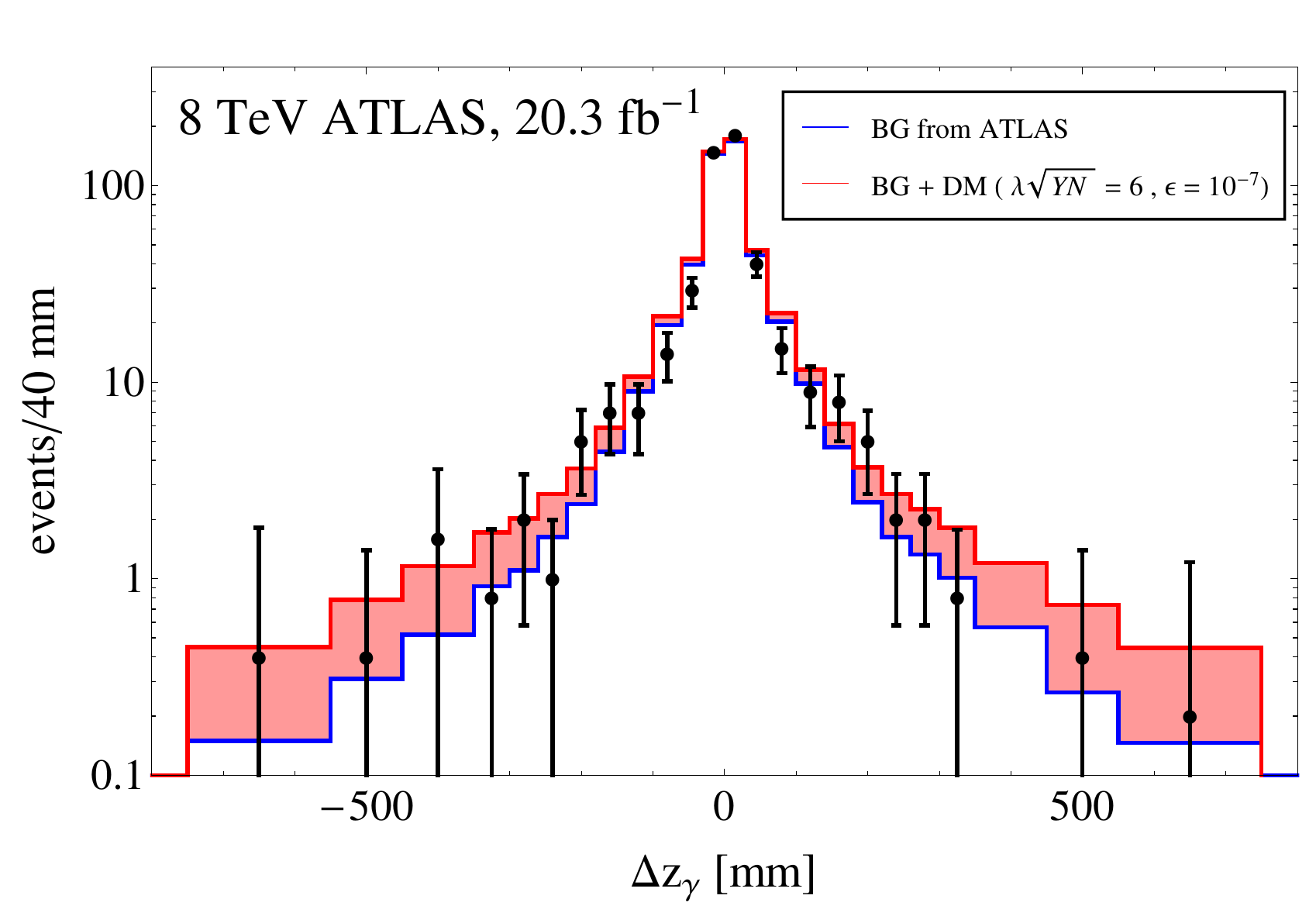}\qquad\includegraphics[width=0.5\textwidth]{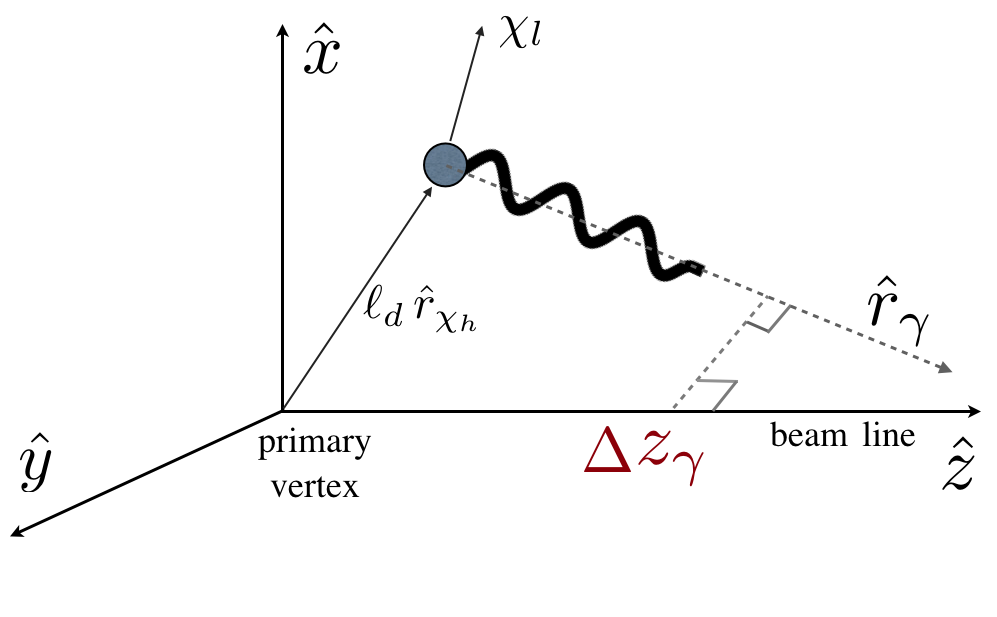}
\end{center}
\caption{Left panel: the $\Delta z_{\gamma}$ distribution of the non-pointing photon signals measured by ATLAS. The background reported by ATLAS (blue histogram) was obtained from a data driven analysis, using diphoton events with $\slashed E_T<20$ GeV. Also shown, stacked on top of the background (red histogram), is the signal distribution for $(m_{\chi_h},\,m_{\chi_l},\,M)=(300,\,10,\,300)$ GeV, $\lambda\sqrt{NY}=6$, and $\varepsilon=10^{-7}$. Right panel: the geometry of the displaced signals.}
\label{fig:nonpointing}
\end{figure}

For the search of the displaced photon signal plus missing energy, we follow the nonpointing photon analysis in \cite{Aad:2014gfa}, performed by the ATLAS collaboration on about $20$ fb$^{-1}$ of $8$ TeV data. The full search also uses the delayed photon measurement, however, due to the complication of modeling the time of flight of the photon from the displaced vertex to the electromagnetic calorimeter (ECAL), we only focus on the measurement of $\Delta z_{\gamma}$ of nonpointing photons (see Fig.~\ref{fig:nonpointing}). For DM signals given by the long-lived $\chi_h\to\chi_l\gamma$ decay, $\Delta z_{\gamma}$ can be related to the $\chi_h$ decay length $\ell_{d}$ in the lab frame:
\begin{equation}\label{eq:Zgamma}
\Delta z_{\gamma}=\ell_d\left(\hat{r}_{\chi_h,z}-\frac{\hat{r}_{\chi_h,T}\cdot\hat{r}_{\gamma,T}}{1-(\hat{r}_{\gamma,z})^2}\hat{r}_{\gamma,z}\right) = \ell_d \Big[\cos\theta_{\chi_h} -\cos (\phi_{\chi_h} - \phi_\gamma)\mathrm{cot}\,\theta_\gamma \sin \theta_{\chi_h} \Big]
\end{equation}
where $\hat{r}_{T,z}$ represent the transverse and longitudinal components of the unit vector $\hat{r}$, respectively, as shown in Fig.~\ref{fig:nonpointing}. To obtain the $\Delta z_{\gamma}$ distribution of the DM decay, we first simulate the prompt process, $p\,p \to \chi_h\bar{\chi}_h, \chi_h\to\chi_l\gamma, \bar{\chi}_h\to\bar{\chi}_l\gamma$ in MadGraph5, apply the cuts performed in the ATLAS analysis, and reweight the events using the dark penguin form factors. Then we calculate the proper lifetime of $\chi_h$ and boost it to the lab frame using the momenta of each parton-level event. The angular information of the photon and $\chi_h$ allow us to calculate $\Delta z_{\gamma}$ in Eq.~\eqref{eq:Zgamma} as a function of the decay length. Using this, each simulated MC event contributes to the differential cross section in $\Delta z_{\gamma}$ as
\begin{equation} \label{eq:displdistr}
\frac{d\sigma_{\text{displaced}}}{d\Delta z_{\gamma}} =\sigma_{\text{prompt}}\frac{d\,P}{d\Delta z_{\gamma}}=\sigma_{\text{prompt}}\frac{|\mu|}{2} e^{-\mu\,\Delta z_{\gamma}},
\end{equation}
where the $\mu$ characterizing the probability distribution $dP/d\Delta z_\gamma$ of the decay is defined as
\begin{equation}
\mu\equiv\frac{\Gamma_{\chi_h}m_{\chi_h}}{p_{\chi_h}}\left(\hat{r}_{\chi_h,z}-\frac{\hat{r}_{\chi_h,T}\cdot\hat{r}_{\gamma,T}}{1-(\hat{r}_{\gamma,z})^2}\hat{r}_{\gamma,z}\right)^{-1}.
\end{equation}
Summing the distributions derived from all the simulated events we obtain the differential cross section in $\Delta z_\gamma$, see Fig.~\ref{fig:nonpointing}.

The ATLAS search requires at least two loose photons with $|\eta|<2.37$ and $E_T>50$ GeV. At least one photon is required to be in the barrel region $|\eta|<1.37$. To avoid collisions due to satellite bunches, both photons are required to have an arrival time at the ECAL $t_{\gamma}$ smaller than $4$ ns, with zero defined as the expected time of arrival for a prompt photon from the primary vertex. We do not attempt to fully simulate $t_\gamma$, which would require a more complex detector description, but rather we approximate $t_\gamma$ with the time of flight of the $\chi_h$, requiring it to be smaller than $4$ ns.
In our estimation we do not include the detailed isolation cuts on the photon. We also neglect the effect of the displaced decay on the angular acceptance of the photons, simply imposing the requirements on $|\eta|$ at the level of the prompt event.
The signal region also requires $\slashed E_T>75$ GeV. 
Finally, to simplify the discussion we assume that every event has a reconstructed primary vertex in the geometrical center of the detector.

For events where only one photon satisfies $|\eta|<1.37$ (i.e. it is in the barrel calorimeter), this photon is used for the measurement of $\Delta z_{\gamma}$. For events where both photons are in the barrel, the photon with larger $t_{\gamma}$ is used. We approximate this timing condition by taking the photon emitted by the more boosted $\chi_h$, in which case the average decay is more delayed. In Fig.~\ref{fig:nonpointing} the generated $\Delta z_{\gamma}$ signal distribution is shown, on top of the expected background. The latter is taken from Fig.~4 of the ATLAS paper \cite{Aad:2014gfa}. Because we are focusing on the non-pointing photon signals, to set constraints on the DM couplings we remove events with $|\Delta z_{\gamma}|< 30$ mm. In our exploratory analysis we only consider the statistical uncertainty on the background, neglecting the effect of systematics.

\subsection{$100$ TeV collider and ILC}\label{sec:futurecollider}
To give an idea of how much higher energy colliders can improve the sensitivity to the dark penguin, we estimate the bound from the mono-jet search at a $100$ TeV collider. Following the discussion in \cite{Cirelli:2014dsa}, we simulate the DM signal and backgrounds with $\sqrt{s}=100$ TeV and impose the cuts $p_T>2.5$ TeV, $|\eta|<2.2$ on the leading jet, together with $\slashed E_T>3$ TeV. A second jet with $p_T>100$ GeV is allowed as long as it has $|\eta|<4.5$ and the azimuthal separation from the leading jet is $\Delta\phi<2.2$. Events with leptons (taus) are vetoed if the lepton satisfies $|\eta|<2.5$ and $p_T>20$ (40) GeV. We assume a $2\%$ systematic uncertainty when calculating bounds with $3$ ab$^{-1}$ of data. 

To show the importance of lowering the SM background for the dark penguin search, we also estimate the monophoton bound from the ILC-500P and ILC-1000P scenarios in Sec.~\ref{sec:NF1}. Our analysis follows the one in \cite{Chae:2012bq} by assuming a $500$ GeV ($1$ TeV) ILC with $250$ ($500$) fb$^{-1}$ of data, and polarizations equal to $P_{-}= +\, 0.8$ and $P_{+}= \, 0.5$ for the electron and positron beams, respectively. The $Z$-related SM background $e^+e^-\to  Z(\nu\bar{\nu})\gamma$ can be eliminated by cutting away the $Z$ pole, but the background process involving a $t$-channel $W$ cannot be reduced by simple kinematic cuts. The polarization of the ILC beams plays an important role in reducing the latter background, since the $W$ only couples to the left-handed electron. The proposed search requires $E_{\gamma}>8$ GeV, $\left|\cos\theta_{\gamma}\right|<0.995$, and imposes a veto on events with photon energy $238<E_{\gamma}<245$ ($490<E_{\gamma}<495$) GeV at ILC-500P (-1000P) in order to suppress the $Z(\nu \bar{\nu})\gamma$ background. Notice that in our simulation of the dark penguin monophoton signals at the ILC, we only considered the case where the photon is emitted by ISR. However, in principle the contribution from the Rayleigh operator $\bar{\chi}\chi B_{\mu\nu} B^{\mu\nu}$ is of the same order and should be consistently included. This could lead to quantitative changes in our results, although the qualitative features would remain the same. We leave the inclusion of the Rayleigh operators for future work.


\section{Collider phenomenology of the $N_{\chi}=1$ case}\label{sec:NF1}
In this section we present the collider reach on the scenario where only a single species of dark fermion $\chi$ is accessible at colliders. In this case, signals in the monojet and monophoton plus missing energy channels can be produced through initial state radiation. In addition, when the DM mass is less than half of the $Z$ mass, the measurement of the invisible $Z$ decay at the $Z$-pole also sets a strong constraint. 
\subsection{Bound from the $Z$ invisible width}
If the DM is lighter than $m_Z/2$, the decay $Z\to \bar{\chi} \chi$ is kinematically allowed and constrained by the measurement at LEP/SLD of the invisible decay width of the $Z$. For $m_Z \ll M$, the decay $Z\to \bar{\chi} \chi$ is well described by an effective dipole interaction $\bar{\chi}\sigma^{\mu\nu}\chi B_{\mu\nu}$ (see Eq.~\eqref{eq:eftcoupling}), leading to the following approximate expression for the width
\begin{equation}\label{eq:Zdecay}
\Gamma(Z\to \bar{\chi}\chi) \simeq \frac{\lambda^4 Y^2 N^2 g^{\prime\,2} s_w^2}{24576\, \pi^5}\frac{m_Z^3}{M^2}\left(1+8\frac{m_{\chi}^2}{m_Z^2}\right)\sqrt{1-\frac{4m_{\chi}^2}{m_Z^2}}\,,
\end{equation}
The uncertainty on $\Gamma(Z\to \mathrm{invisible})$ is of $1.5$ MeV \cite{Beringer:1900zz}, therefore the $95\%$ bound reads
\begin{equation}
\mathrm{BR}(Z\to  \bar{\chi} \chi) \lesssim 1\times 10^{-3}\,.
\end{equation}
This bound is shown as an orange curve in the right panel of Fig.~\ref{fig:monojetresult}. 
\subsection{LHC and future colliders}
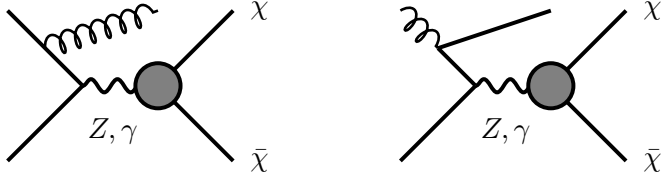
\begin{figure}
\begin{center}
\begin{tikzpicture}[line width=1.5 pt, scale=1] 
			\draw[color=white] (-1.7,-1.7) rectangle (1.7,1.7);
			\draw[fermionnoarrow] (-1,1) -- (0,0);
			\draw[fermionnoarrow] (0,0) -- (-1,-1);
			\draw[vector] (0,0) -- (1,0);
			\draw[fermionnoarrow] (2,1) -- (1,0);
			\draw[fermionnoarrow] (1,0) -- (2,-1);
			\draw[gluon] (-.5,.5) -- (1,1);
			\draw[fill=gray] (1,0) circle (.3cm);
									           \node at (2.35,1.0) {$\chi$};
						          \node at (0.4,-.6) {$Z,\gamma$};
						           \node at (2.35,-1.0) {$\bar{\chi}$};
						                                  		\end{tikzpicture} \qquad\begin{tikzpicture}[line width=1.5 pt, scale=1] 
			\draw[color=white] (-1.7,-1.7) rectangle (1.7,1.7);
			\draw[gluon] (-1,1) -- (-0.5,0.5);
			\draw[fermionnoarrow] (0,0) -- (-1,-1);
			\draw[vector] (0,0) -- (1,0);
			\draw[fermionnoarrow] (2,1) -- (1,0);
			\draw[fermionnoarrow] (1,0) -- (2,-1);
			\draw[fermionnoarrow] (-.5,.5) -- (1,1);
			\draw[fermionnoarrow] (-.5,.5) -- (0,0);
			\draw[fill=gray] (1,0) circle (.3cm);
									           \node at (2.35,1.0) {$\chi$};
						          \node at (0.4,-.6) {$Z,\gamma$};
						           \node at (2.35,-1.0) {$\bar{\chi}$};
						                                  		\end{tikzpicture}								\end{center}
\caption{Diagrams for the monojet process. The grey circle indicates the dark penguin.}\label{fig:colliderdigm}
\end{figure}

\begin{figure}
\begin{center}
\includegraphics[width=0.46\textwidth]{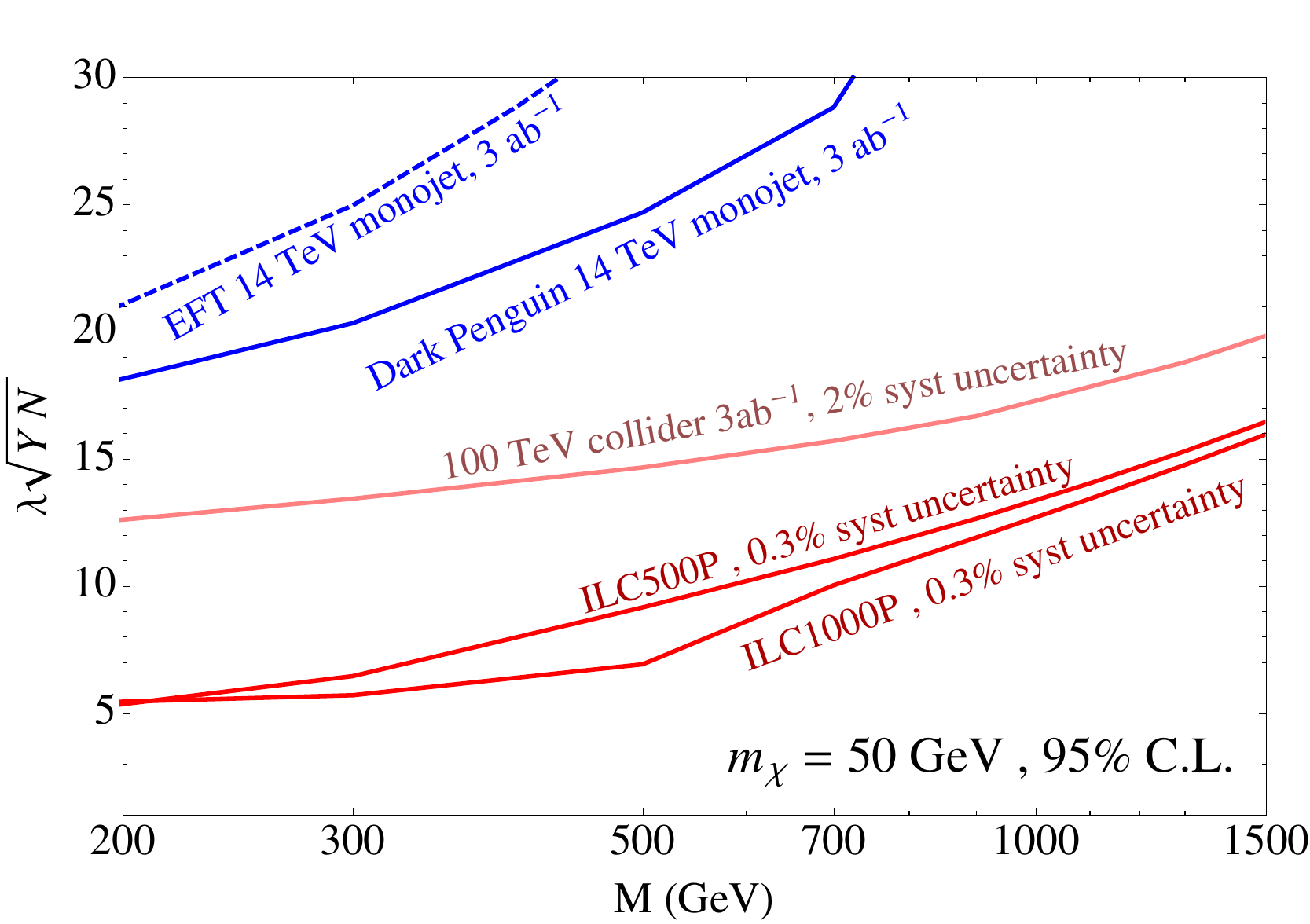}\quad\includegraphics[width=0.46\textwidth]{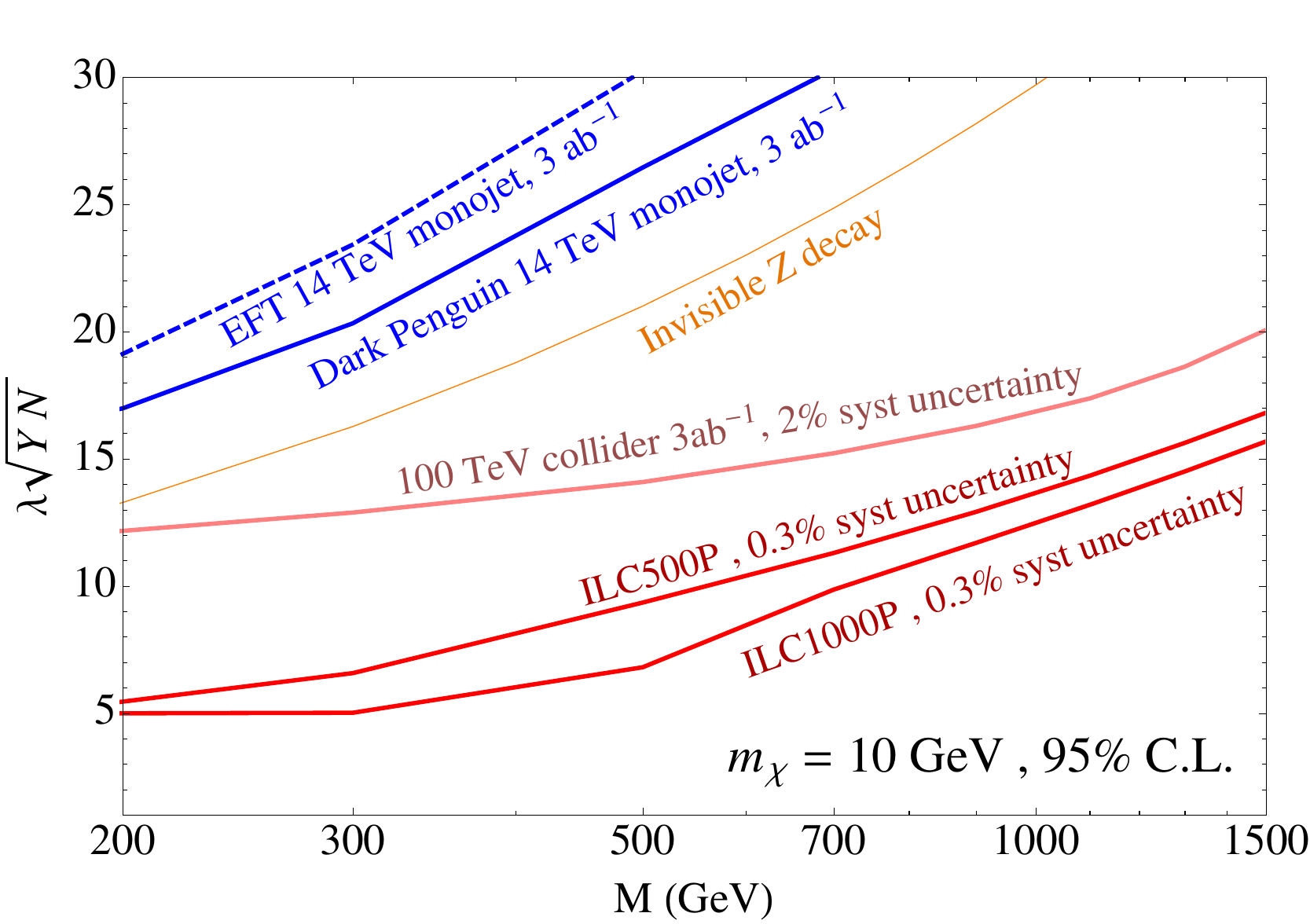}
\end{center}
\caption{Upper bounds on the DM-mediators coupling for $N_\chi = 1$, as a function of the mediator mass $M$. The dark matter mass is assumed to be $m_{\chi}=50$ GeV (left) and $10$ GeV (right). Here we optimize the LHC bound by choosing the best MET cut for each mediator mass. The blue dashed curve shows the bound from the effective dipole interaction, $\bar{\chi}\sigma^{\mu\nu}\chi B_{\mu\nu}$ in Eq.~(\ref{eq:eftcoupling}). The monophoton constraint from the ILC500-P (ILC1000-P) assumes $250$ ($500$) fb$^{-1}$ of data with $500$ GeV ($1$ TeV) center of mass energy and a polarization $P_{-}= +\, 0.8$ and $P_{+}= \, 0.5$. The orange curve shows the constraint from the current invisible $Z$ decay measurement. The perturbative bound on $\lambda\sqrt{YN}$ depends on the hypercharge of the mediator, $\lambda\sqrt{YN}\lsim 4\pi \sqrt{Y}$.} 
\label{fig:monojetresult}
\end{figure}

At the LHC, the most promising search for the dark penguin with $N_{\chi}=1$ is monojet, in which a high $p_T$ jet is produced through the QCD ISR process, see Fig.~\ref{fig:colliderdigm}. The details of the analysis were given in Sec.~\ref{sec:monojet}. The signal and background yields and the systematic uncertainty on the background all depend on the MET cut and luminosity. The projected $14$ TeV constraint is computed varying the MET cut from $550$ to $2250$ GeV with a step of $100$ GeV, and taking the strongest bound. The results are shown as solid blue curves in Fig.~\ref{fig:monojetresult} for $m_{\chi}=50$ GeV (left panel) and $10$ GeV (right panel). The LHC reach is limited: the $14$ TeV run can only cover part of the perturbative parameter space for relatively large $Y\simeq 4$, in which case the perturbative bound is $\lambda\sqrt{YN}\lsim 8\pi$.

\begin{figure}
\begin{center}
\includegraphics[width=0.33\textwidth]{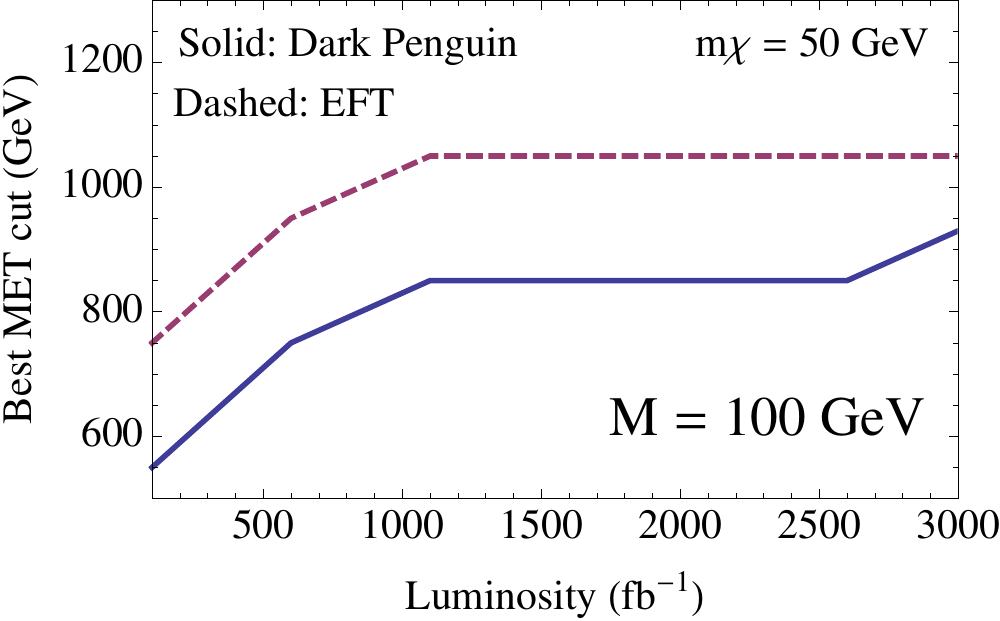}
\includegraphics[width=0.33\textwidth]{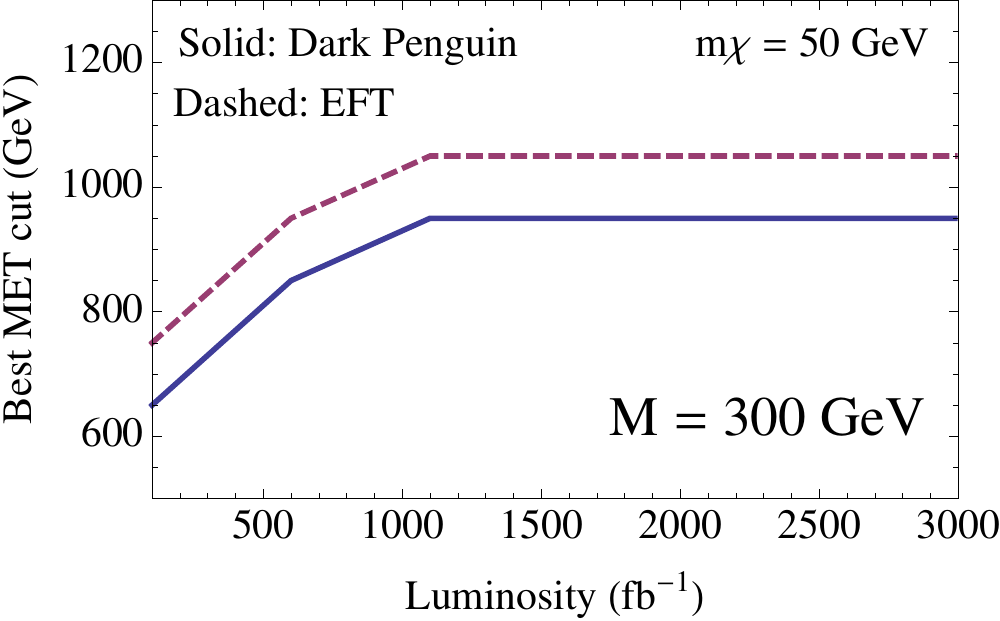}\includegraphics[width=0.33\textwidth]{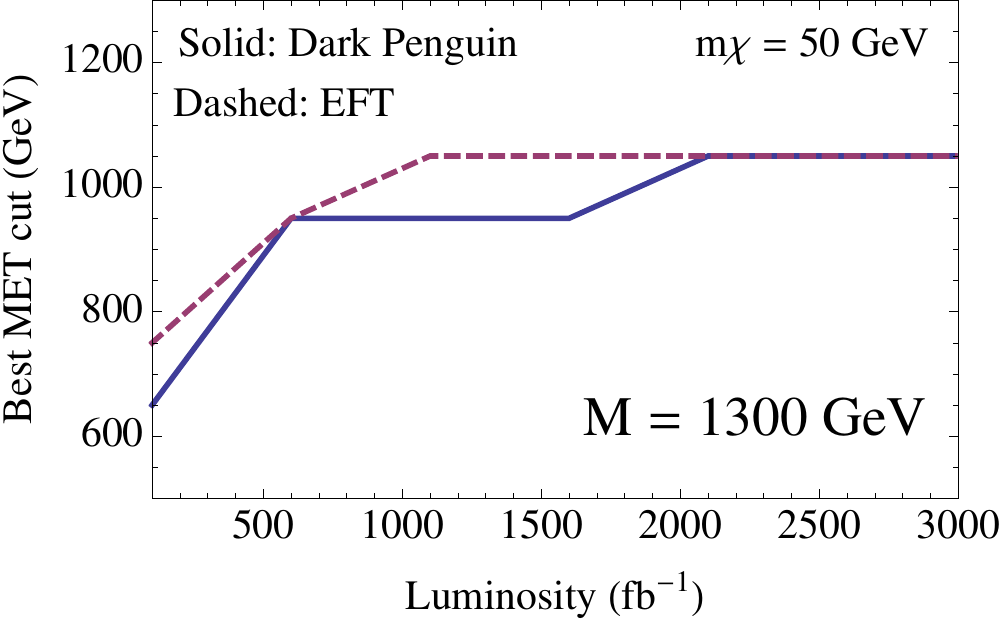}
\end{center}
\caption{The MET cut of the projected $14$ TeV search that optimizes the DM signal significance as a function of luminosity. As can be seen in the plots, dark penguins with light mediators prefer lower MET cuts. When the mediators are heavier the MET distribution becomes harder, similar to the EFT case, and a tighter cut is preferred. The optimal MET cut increases with luminosity, because in our data-driven analysis the systematic uncertainties are mainly controlled by the statistics of the control samples.}
\label{fig:METcuts}
\end{figure}

To compare the dark penguin to the EFT, we also compute the bounds by parameterizing the DM-SM coupling with the effective dipole interaction. The resulting constraint, shown by the dashed blue lines in Fig.~\ref{fig:monojetresult}, is weaker than the dark penguin one, since the dipole EFT neglects the enhancement of the form factor $F_\sigma$ for $\sqrt{q^2}\simeq 2M$, as well as the sizable contribution from the $q^2\gamma^{\mu}$ term in Eq.~\eqref{eq:loopresult} when $q^2$ is large (in the EFT the $q^2\gamma^{\mu}$ piece would correspond to a dimension-$6$ operator, see the first term in Eq.~\eqref{eq:eftcoupling}). Furthermore, the optimal MET cut for the dark penguin signal is typically softer than for the EFT coupling. In Fig.~\ref{fig:METcuts} we show a comparison of the best MET cut for the dark penguin and the effective dipole coupling, as a function of the mediator mass. When the mediators are light, the best MET cut for the dark penguin is lower than for the effective coupling. Notice that in both cases the optimal MET cut increases at larger luminosity, because in our data-driven analysis the systematic uncertainties are mainly determined by the statistics of the control samples.

The LHC monophoton search gives a much weaker bound compared to monojet, and will not be discussed in detail.

Given the limited sensitivity of the LHC to the $N_\chi = 1$ scenario, it is important to estimate how the bounds could be improved at future colliders. The study of DM production at very high energy machines, such as a $100$ TeV $pp$ collider, mandates the full inclusion of the loop form factors, since the typical partonic energy is much larger than the masses of the mediators. Differently than in the search for heavy particles, however, the significance of the dark penguin signal will not be appreciably improved at a higher energy machine, unless the systematic uncertainty can be greatly reduced. In fact, when the mediators have a sub-TeV mass, both the SM background and DM production at $100$ TeV are mediated by effectively nearly-massless particles. This results in similar shapes of the MET distribution for the signal and background, and as long as the background is systematics-dominated, the ratio between the signal and background does not vary significantly when increasing the collider energy. The pink curves in Fig.~\ref{fig:monojetresult} give an estimate of the $100$ TeV reach (the search is described in detail in Sec.~\ref{sec:futurecollider}). The result is an improvement of the bound on $\lambda\sqrt{YN}$ by only a factor $\sim 2$ with respect to the projected $14$ TeV LHC constraint. The sensitivity of the $100$ TeV collider ameliorates if the MET cut can be kept very weak, below the TeV scale, as in this case the signal region includes the enhancement of the form factors for $\sqrt{q^2}\sim 2M$, thus increasing the signal yield. However, it is not clear whether such a low MET cut would be achievable, and if so, whether the systematic uncertainties would be increased beyond the $2\%$ that we are using as benchmark in our estimate. More detailed studies are necessary to precisely assess the reach of a $100$ TeV machine. 

In comparison to hadron colliders, the monophoton search at the ILC appears to be more promising for the dark penguin signals. As discussed in Sec.~\ref{sec:futurecollider}, thanks to the beam polarization and full reconstruction of the missing invariant mass, the SM background can be efficiently reduced. As is shown in Fig.~\ref{fig:monojetresult}, the ILC500-P with $500$ GeV center of mass energy, $250$ fb$^{-1}$ of data and polarized beams can improve greatly the sensitivity to the dark penguin.

\section{Collider phenomenology of the $N_{\chi}=2$ case with anarchic DM couplings}\label{sec:NF2}
Differently from the direct and indirect detection experiments, colliders are capable of probing extended dark sectors, through the production of the additional states that accompany the DM. In this section we study the monophoton, diphoton, and rare $Z$ decay constraints on the $N_\chi = 2$ dark penguin model, where one heavy dark fermion $\chi_h$ is present in addition to the light DM particle $\chi_l$. In this scenario, large photon signals can be generated through the pair production of dark fermions, followed by the decay $\chi_h\to\chi_l\gamma$. Under the assumption of anarchic couplings between the dark fermions and the mediators, we can compare the bounds from different searches and provide complementary information to the direct and indirect detection experiments. In the diphoton search, the kinematic variable $M_{T2}$ can be used to determine the mass of $\chi_h$.

The constraint from the invisible $Z$ decay width applies also in the $N_{\chi} = 2$ case, as long as $m_{\chi_l} < m_Z/2$, as we assume. In addition, when also $m_{\chi_h}< m_Z/2$, the search at LEP1 for events containing two photons plus missing energy sets a much stronger constraint on the DM couplings. Therefore we focus first on the case $m_{\chi_h} > m_Z/2$, and discuss separately the more constrained possibility of a lighter $\chi_h$, in Sec.~\ref{sec:Zdecay}.
\subsection{LHC monophoton}
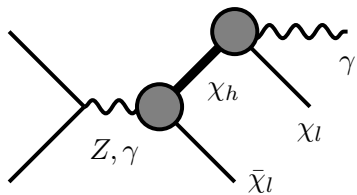
\begin{figure}
\begin{center}
\begin{tikzpicture}[line width=1.5 pt, scale=1] 
			\draw[color=white] (-1.7,-1.7) rectangle (1.7,1.7);
			\draw[fermionnoarrow] (-1,1) -- (0,0);
			\draw[fermionnoarrow] (0,0) -- (-1,-1);
			\draw[vector] (0,0) -- (1,0);
			\draw[fermionnoarrow, line width=3.5pt] (2,1) -- (1,0);
			\draw[fermionnoarrow] (1,0) -- (2,-1);
			\draw[fermionnoarrow] (2,1) -- (3,0);
			\draw[vector] (2,1) -- (3.5,1);
			\draw[fill=gray] (1,0) circle (.3cm);
			\draw[fill=gray] (2,1) circle (.3cm);
									           \node at (1.85,0.2) {$\chi_h$};
									           \node at (3.5,0.55) {$\gamma$};
						          \node at (0.4,-.6) {$Z,\gamma$};
						           \node at (2.35,-1.0) {$\bar{\chi}_l$};
						           \node at (3,-0.35) {$\chi_l$};						                                  		\end{tikzpicture}
\end{center}
\caption{Dominant process for the monophoton signal. The grey circle indicates the dark penguin.}
\label{fig:IDMloop}
\end{figure}

\begin{figure}
\begin{center}
\includegraphics[width=0.45\textwidth]{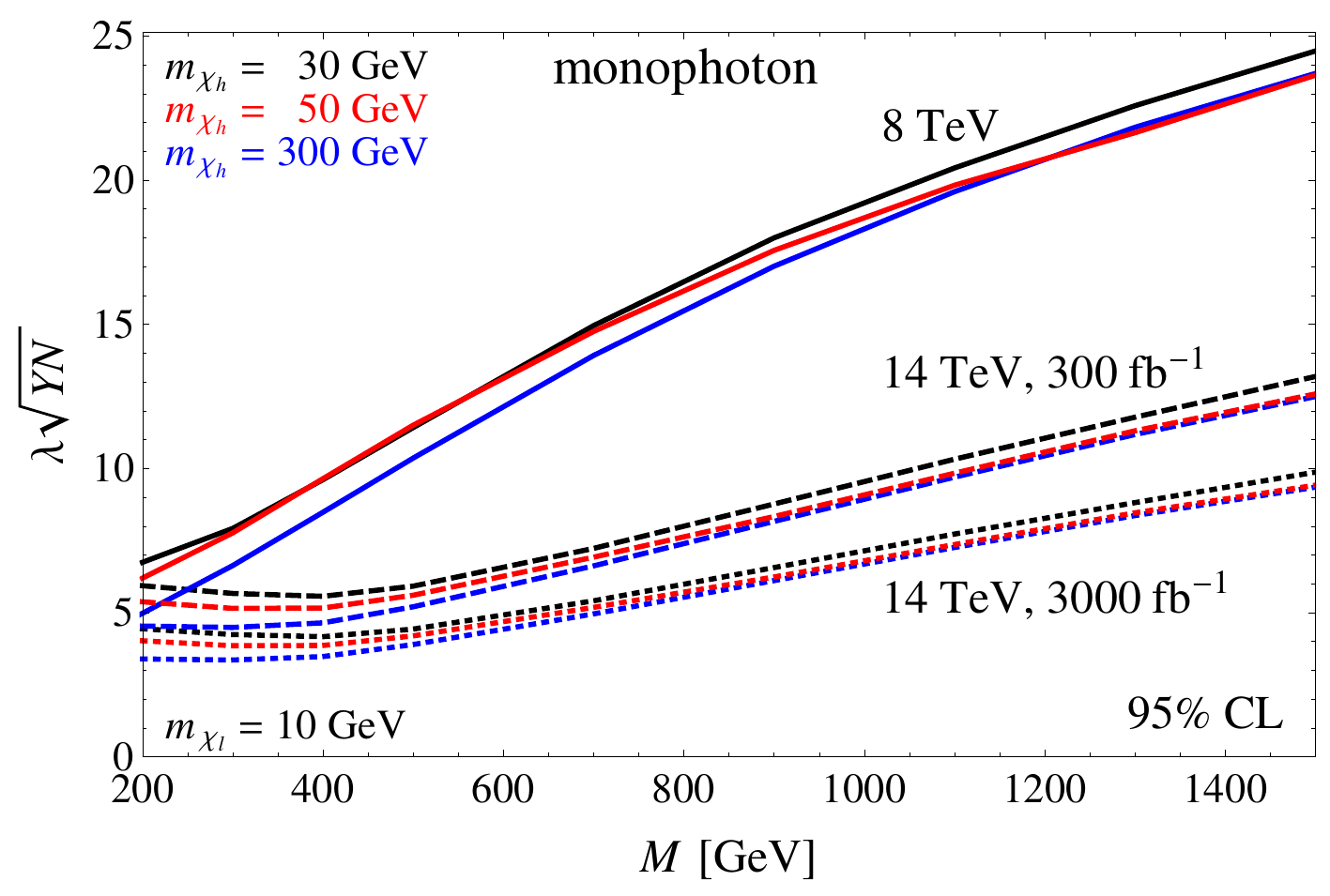}
\qquad\includegraphics[width=0.465\textwidth]{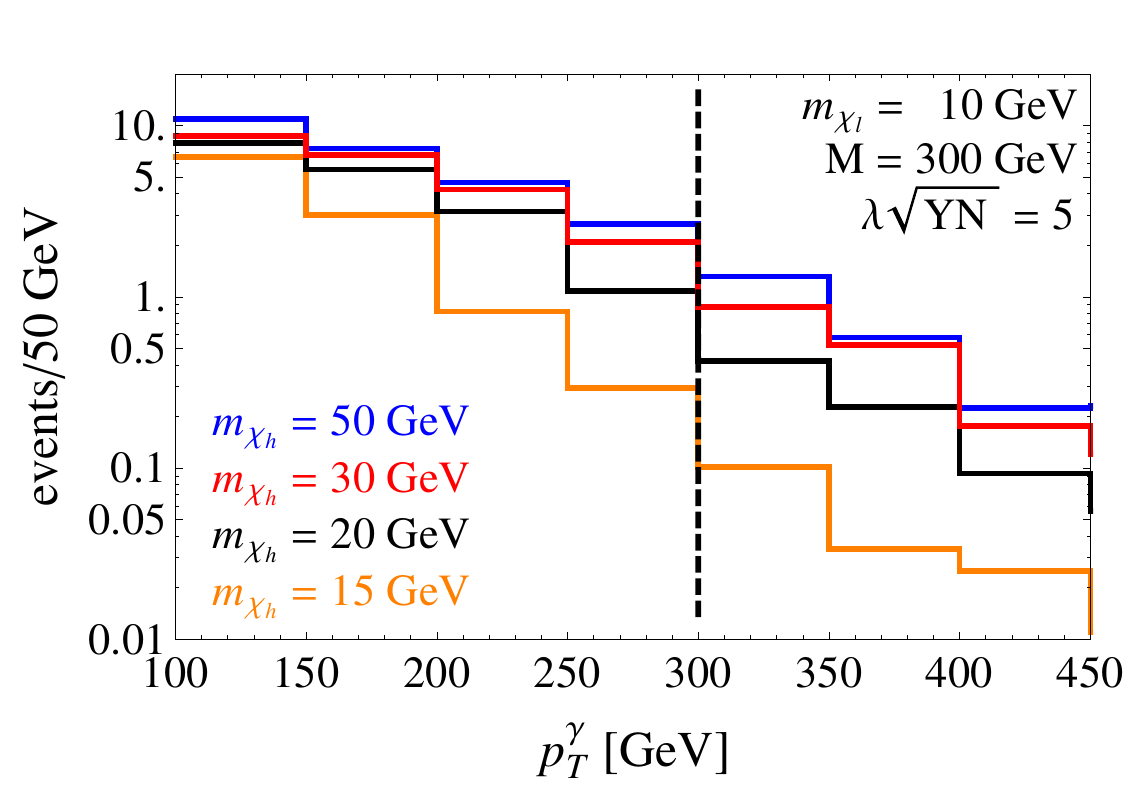}
\end{center}
\caption{Left panel: LHC monophoton bounds for $N_\chi = 2$. Right panel: dependence of the photon $p_T$ distribution at $14$ TeV on the mass of $\chi_h$. To generate the distribution, the cuts on the photon transverse momentum and MET were relaxed to $p_{T}^\gamma, \slashed{E}_T  > 100\;\mathrm{GeV}$. The luminosity is fixed to $300$ fb$^{-1}$. The vertical dashed line indicates the actual cut applied in the analysis, $p_T^\gamma > 300\;\mathrm{GeV}$. The perturbative bound on $\lambda\sqrt{YN}$ depends on the hypercharge of the mediator, $\lambda\sqrt{YN}\lsim 4\pi \sqrt{Y}$.}
\label{fig:monophotonresult}
\end{figure}

The monophoton signal arises through the production of $\chi_h\chi_l$, followed by the decay $\chi_h \to \chi_l \gamma$, see Fig.~\ref{fig:IDMloop}. The branching ratio for the decay is either $\sim c_w^2 \simeq 0.8$ or unity, depending on whether the $\chi_l Z$ channel is kinematically open or closed. Therefore the cross section is much larger than for the ISR process discussed in the case $N_\chi = 1$, whose amplitude is comparatively suppressed by the three-body phase space and by the additional electromagnetic coupling. In this section we calculate the bound on $\lambda\sqrt{YN}$ from the $8$ and $14$ TeV monophoton searches, and study the dependence of the bound on the mass splitting between $\chi_h$ and $\chi_l$. The result is shown in the left panel of Fig.~\ref{fig:monophotonresult}.
For light mediators, the slope of the bound on $\lambda\sqrt{Y N}$ decreases. This can be understood by noticing that the requirement of a minimum photon transverse momentum $p_T^{\gamma}$ imposes $\sqrt{q^2}\gtrsim 2p_T^{\gamma}$, while the dark penguin peaks at $\sqrt{q^2}\sim 2M$. Therefore, for $M\lesssim p_T^\gamma$ a significant fraction of the signal does not pass the event selection. The effect is more evident in the $14$ TeV case, where the cut on the photon $p_T$ is stronger ($300$ GeV).   

To gain some insight about how the mass difference between $\chi_h$ and $\chi_l$ affects the monophoton signal, we fix the DM mass to $m_{\chi_l}=10\;\mathrm{GeV}$ and show the photon $p_T$ distribution for various choices of $m_{\chi_h}$ in the right plot of Fig.~\ref{fig:monophotonresult}. The coupling $\lambda\sqrt{YN}$ is fixed to the value corresponding approximately to the reach of the $14$ TeV LHC. We see that the signal yield decreases significantly only for $m_{\chi_h}\lesssim 20\;\mathrm{GeV}$. Thus the monophoton search has sensitivity even for fairly degenerate dark fermions, with mass splitting $m_{\chi_h}-m_{\chi_l}$ much smaller than the required photon $p_T$, thanks to the boost of $\chi_h$ from the production process.

\begin{figure}
\begin{center}
\includegraphics[width=0.5\textwidth]{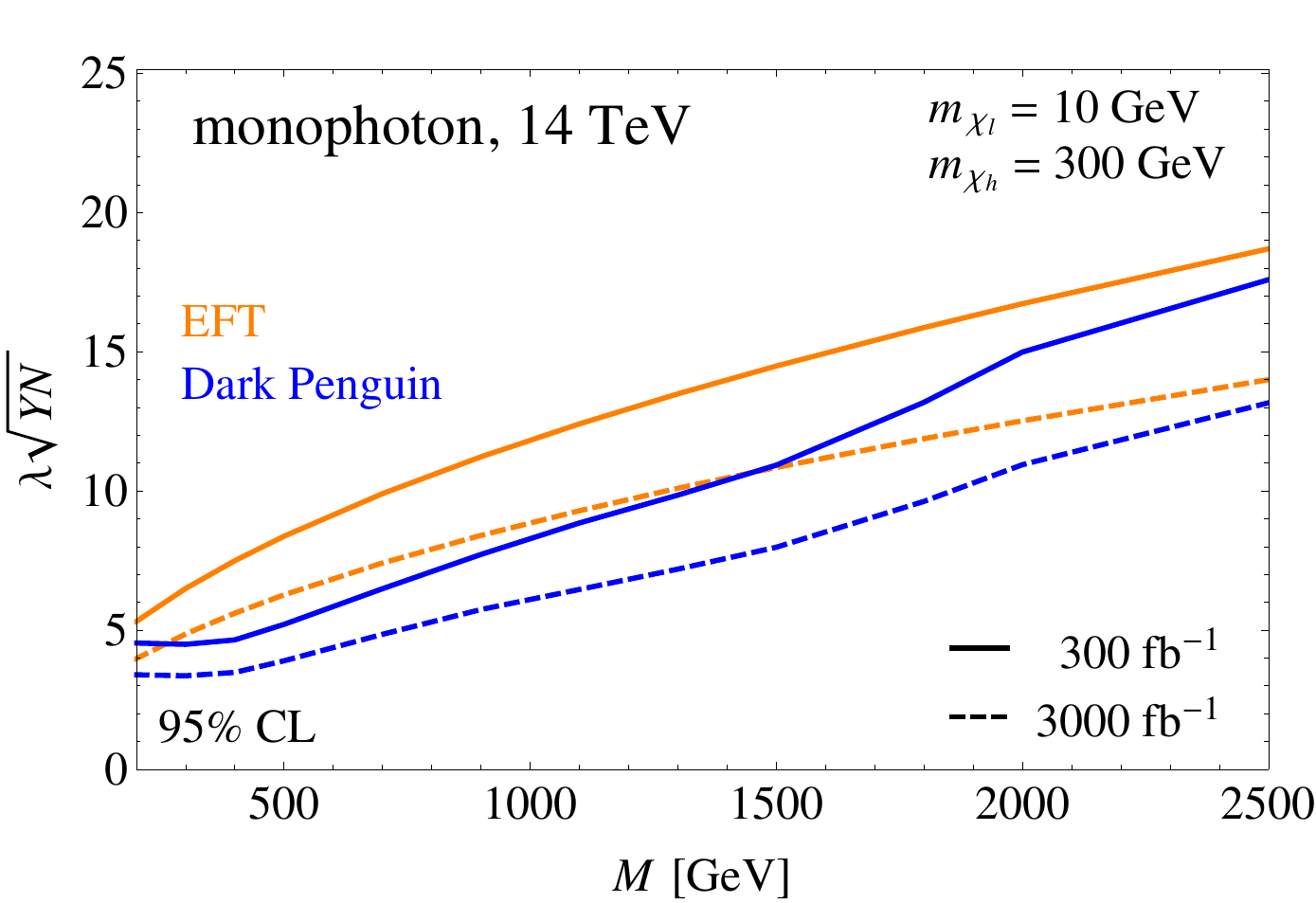}
\end{center}
\caption{A comparison between the monophoton bounds computed using the dark penguin amplitude (blue) and the effective dipole interaction $\bar{\chi}_h\sigma^{\mu\nu}\chi_lB_{\mu\nu} + \mathrm{h.c.}$  (orange). For each mediator mass, the bounds are optimized by choosing the best photon $p_T$ cut between $300$ and $1000$ GeV.}
\label{fig:monophotonEFTvsDP}
\end{figure}

In Fig.~\ref{fig:monophotonEFTvsDP} we show a comparison of the bound obtained using the dark penguin and the one computed using the effective dipole coupling $\bar{\chi}_h\sigma^{\mu\nu}\chi_lB_{\mu\nu} + \mathrm{h.c.}$ (see Eq.~(\ref{eq:eftcoupling})). For each mediator mass we vary the photon $p_T$ cut between $300$ and $1000$ GeV in $100$ GeV steps, and choose the value that gives the strongest constraint (therefore the dark penguin bound shown in Fig.~\ref{fig:monophotonEFTvsDP} is slightly better than the one in Fig.~\ref{fig:monophotonresult}, which was obtained with a fixed cut $p_T^{\gamma}>300$ GeV for all mediator masses). Similarly to the monojet case, the bounds computed by taking into account the full dark penguin are stronger than the EFT results, due to the form factor enhancement at $\sqrt{q^2}\sim 2M$ and the contribution of the $q^2\gamma^{\mu}$ term in Eq.~(\ref{eq:loopresult}) for large $q^2$. 

\subsection{LHC diphoton$+$MET}
\begin{figure}
\begin{center}
\begin{tikzpicture}[line width=1.5 pt, scale=1] 
			\draw[color=white] (-1.7,-1.7) rectangle (1.7,1.7);
			\draw[fermionnoarrow] (-1,1) -- (0,0);
			\draw[fermionnoarrow] (0,0) -- (-1,-1);
			\draw[vector] (0,0) -- (1,0);
			\draw[fermionnoarrow, line width=3.5pt] (2,1) -- (1,0);
			\draw[fermionnoarrow, line width=3.5pt] (1,0) -- (2,-1);
			\draw[fermionnoarrow] (2,1) -- (3,1);
			\draw[fermionnoarrow] (2,-1) -- (3,-1);
			\draw[vector] (2,1) -- (2.9,1.9);
			\draw[vector] (2,-1) -- (2.9,-1.9);
			\draw[fill=gray] (1,0) circle (.3cm);
			\draw[fill=gray] (2,1) circle (.3cm);
			\draw[fill=gray] (2,-1) circle (.3cm);
									           \node at (1.85,0.3) {$\chi_h$};
									           \node at (1.85,-0.3) {$\bar{\chi}_h$};
						                 \node at (0.45,-.6) {$Z,\gamma$};	
						          	     \node at (3,0.6) {$\chi_l$};
						                 \node at (3,-0.6) {$\bar{\chi}_l$};
						                 \node at (3,1.6) {$\gamma$};
						                 \node at (3,-1.6) {$\gamma$};
						         						                                  		\end{tikzpicture}
\end{center}
\caption{Dominant process for the diphoton signal. The grey circle indicates the dark penguin.}
\label{fig:diphotonloop}
\end{figure}
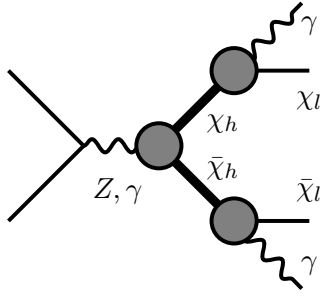
If the $\chi_h$ is light enough to be pair produced with sizable rate, the diphoton$+$MET signal can be generated through the process shown in Fig.~\ref{fig:diphotonloop}. Differently from the monophoton case, there can be a partial cancellation of the transverse momenta of the two $\chi_l$ particles, leading to a suppression of the MET. Nevertheless, the low SM background makes diphoton$+$MET a promising search to look for signals of the dark penguin.\footnote{If kinematically allowed, the outgoing photons in Fig.~\ref{fig:diphotonloop} can also be replaced by $Z$ bosons, although the smaller $\mathrm{BR}(\chi_h \to \chi_l Z)\sim s_w^2$ and the additional branching ratios for the $Z$ decays further suppress the cross section, making the search for $Z$ final states less promising.} 

\begin{figure}
\begin{center}
        \begin{subfigure}[b]{0.45\textwidth}
                \includegraphics[width=\textwidth]{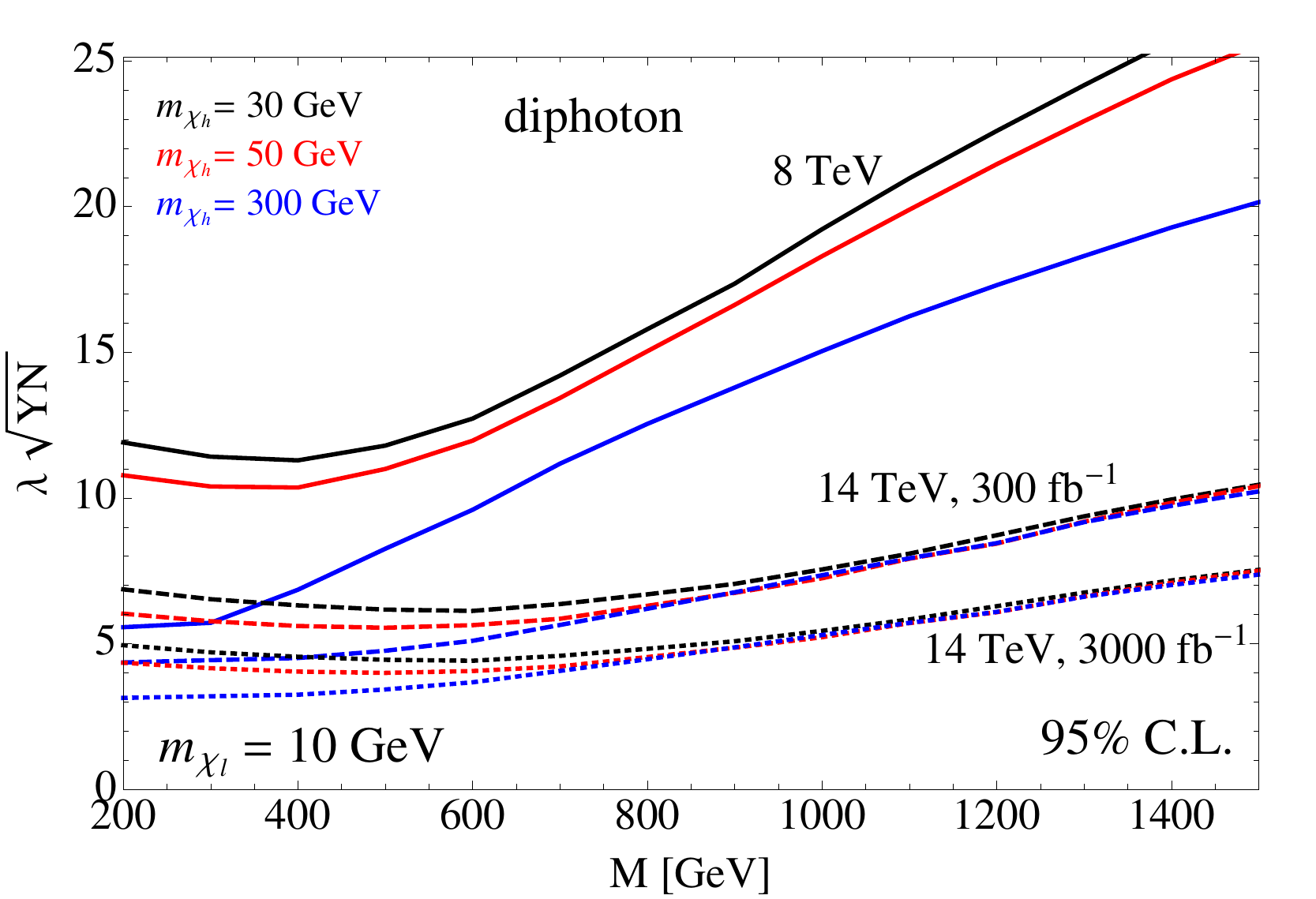}
                \caption{}
                \label{Fig:diphotonbounds}
        \end{subfigure}%
        \quad
        \begin{subfigure}[b]{0.455\textwidth}
                \includegraphics[width=\textwidth]{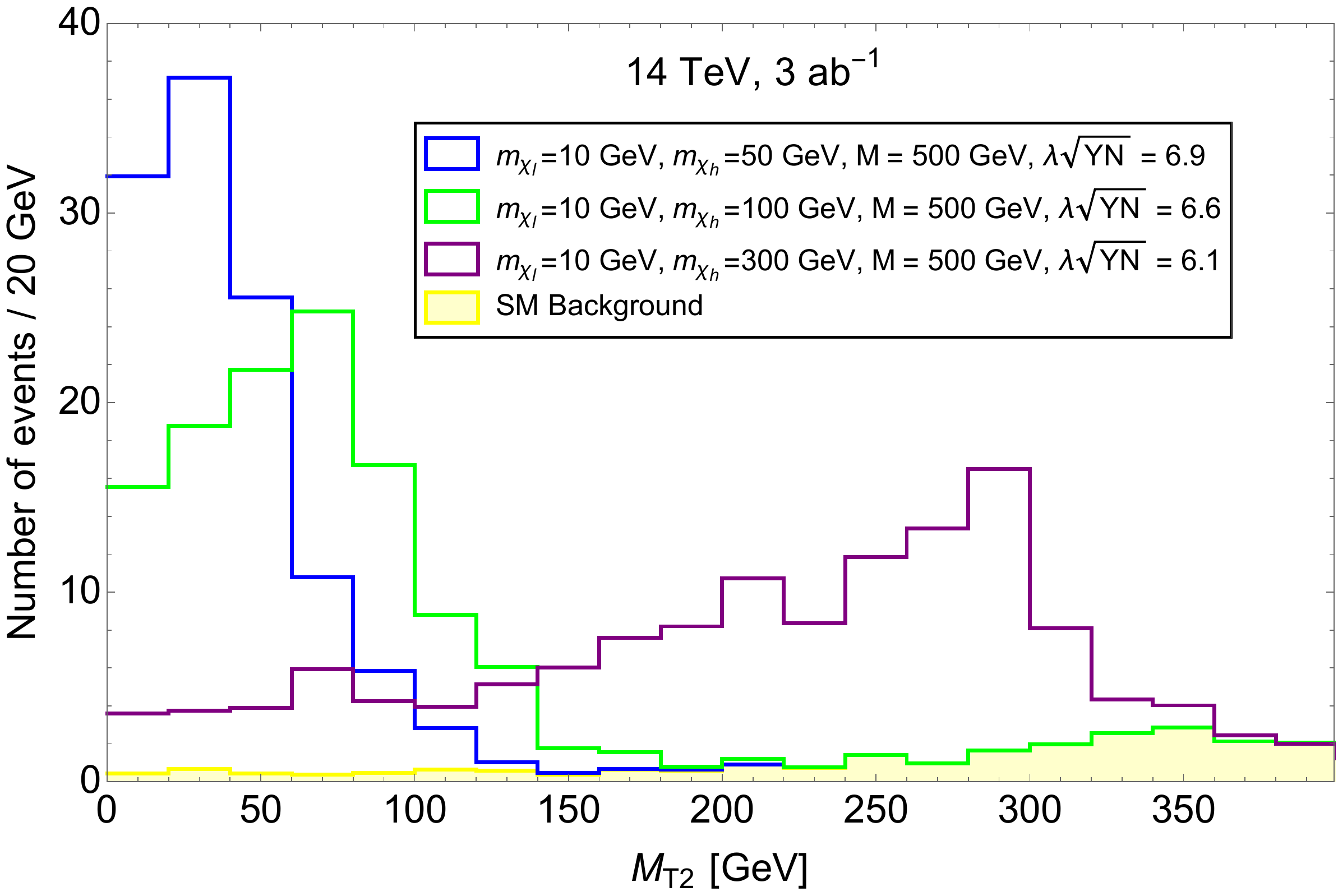}
                \caption{}
                \label{Fig:MT2}
         \end{subfigure}
\end{center}
\caption{Left panel: LHC diphoton bounds for $N_\chi = 2$. Right panel: $M_{T2}$ distributions for the diphoton signal and background at the luminosity of 3 ab$^{-1}$, for various $m_{\chi_h}$ values. $\mu_N = m_{\chi_l}$ is assumed. For each value of $m_{\chi_h}$, the coupling $\lambda\sqrt{YN}$ is chosen to correspond to a 3$\sigma$ excess with 100 fb$^{-1}$ of data. The perturbative bound on $\lambda\sqrt{YN}$ depends on the hypercharge of the mediator, $\lambda\sqrt{YN}\lsim 4\pi \sqrt{Y}$.}
\label{fig:diphoton}
\end{figure}

The three relevant signal regions for 8 TeV are defined in the first three columns of Table~\ref{Tab:diphotonATLASsignalreg}. For each point in parameter space we compute the constraint from each of the three regions, and choose the strongest one. The reach of the 14 TeV LHC is estimated using the signal region defined in the last column of Table~\ref{Tab:diphotonATLASsignalreg}. The results for both 8 and 14 TeV are shown in Fig.~\ref{Fig:diphotonbounds}.

If a signal is observed, the diphoton$+$MET channel offers the possibility to measure $m_{\chi_h}$ through the stransverse mass variable $M_{T2}$ \cite{Lester:1999tx, Barr:2003rg, Cheng:2008hk}. This variable is constructed from the two photons and the missing energy and is defined as 
\begin{equation}
M_{T2}^2(\mu_N) \equiv  \min_{{\mathbf p}_T^1+{\mathbf p}_T^2=\mathbf{\sla
    p}_T}\left[\max \{ m_T^2(\mu_N;\,{\mathbf p}_T^1,\,{\mathbf p}_T^{\gamma_1}),\,
    m_T^2(\mu_N;\,{\mathbf p}_T^2,\,{\mathbf
      p}_T^{\gamma_2})\}\right],
\end{equation}
where
\begin{equation}
m_T^2(\mu_N; \mathbf{\sla p}^i_T, \mathbf p^{\gamma_j}_T) = \mu_N^2 + 2 ( {\sla E}^i_T E^{\gamma_j}_T - \mathbf {\sla p}^i_T \cdot \mathbf p^{\gamma_j}_T).
\end{equation}
In the above equations $\mu_N$ is an unknown trial mass, $\mathbf{\sla p}_T$ is the transverse missing momentum, $\mathbf p^{\gamma_j}_T$ is the transverse momentum of photon $\gamma_j$, while the two transverse energies are defined as 
\begin{equation}
{\sla E}^i_T = \sqrt{\mu_N^2 + |\mathbf{\sla p}^i_T|^2} \qquad \text{and} \qquad E^{\gamma_j}_T = |\mathbf p^{\gamma_j}_T|.
\end{equation}
If the trial mass $\mu_N$ is chosen equal to the $\chi_l$ mass, the distribution of $M_{T2}$ has an edge at the value of $m_{\chi_h}$. In Fig.~\ref{Fig:MT2} we show the $M_{T2}$ distribution at the $14$ TeV LHC with 3 ab$^{-1}$ of data, for some illustrative choices of $m_{\chi_h}$, assuming $\mu_N = m_{\chi_l}$. For each $m_{\chi_h}$, we chose the value of the coupling $\lambda\sqrt{YN}$ such that a $3\sigma$ excess would be observed at an earlier stage of LHC running, namely with 100 fb$^{-1}$ of data. The edge at $M_{T2}\sim m_{\chi_h}$ can be seen in all cases, with some uncertainty of $O(10)$ GeV due to detector effects.

In Fig.~\ref{Fig:MT2} we assumed the mass of the DM to be known, which allowed us to set $\mu_N = m_{\chi_l}$ in the computation of $M_{T2}$. But how could $m_{\chi_l}$ be determined experimentally? References~\cite{Cho:2007qv, Barr:2007hy} showed that the value of $m_{\chi_l}$ can be estimated by observing a kink of the edge of the $M_{T2}$ distribution as a function of $\mu_N$. However, a large number of signal events is required in order for the kink to be observable, making the application of this method, and thus the determination of the DM mass directly in the diphoton process, not likely for our model. On the other hand, direct detection constraints may hint that the DM is light. As we will discuss in Sec.~\ref{sec:compare}, in the region of parameters where the LHC has sensitivity to a dark penguin signal, current direct detection searches require $m_{\chi_l} \lesssim 10$ GeV. Hence one can argue that if a diphoton$+$MET signal is observed at the LHC, and interpreted as involving as missing energy the particle that provides the dominant dark matter density in the universe, one should use the value $\mu_N \lesssim 10$ GeV. 

\begin{figure}
\begin{center}
\includegraphics[width=0.6\textwidth]{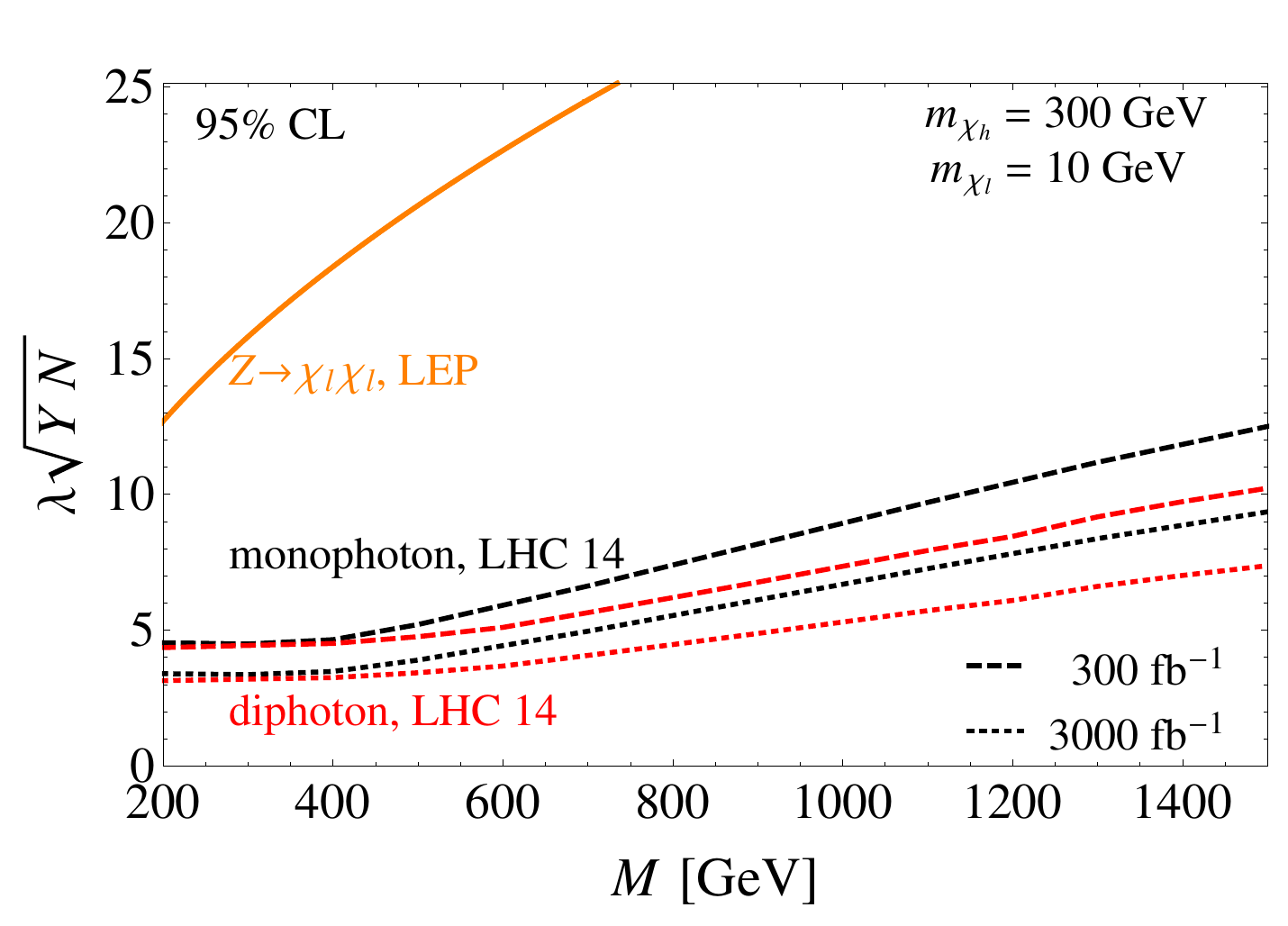}
\end{center}
\caption{A comparison between the monophoton and diphoton constraints for $N_{\chi}=2$. The perturbative bound on $\lambda\sqrt{YN}$ depends on the hypercharge of the mediator, $\lambda\sqrt{YN}\lsim 4\pi \sqrt{Y}$.}
\label{fig:mchih_300}
\end{figure}

It is interesting to compare the results from the monophoton and diphoton searches. The comparison requires some further assumption, since the monophoton search sets a constraint on the dark penguin coupling between $\chi_h$ and $\chi_l$, whereas the diphoton search constrains the coupling between $\chi_h$ and $\chi_h$. As discussed in Sec.~\ref{sec:model}, here we focus on the totally anarchic scenario, where all couplings are assumed to be of the same order. In Fig.~\ref{fig:mchih_300} we show a comparison between the $14$ TeV monophoton and diphoton searches, for the illustrative choice $(m_{\chi_l},\,m_{\chi_h})=(10,\,300)$ GeV. Diphoton gives a slightly stronger bound, but the difference is below the uncertainties associated with our analysis. Notice that, while it is useful to compare the two channels within a specific natural scenario such as the anarchic one, in general the monophoton and diphoton searches are complementary to each other.

\subsection{Constraint from rare $Z$ decays}\label{sec:Zdecay}
\begin{figure}
\begin{center}
\includegraphics[width=0.6\textwidth]{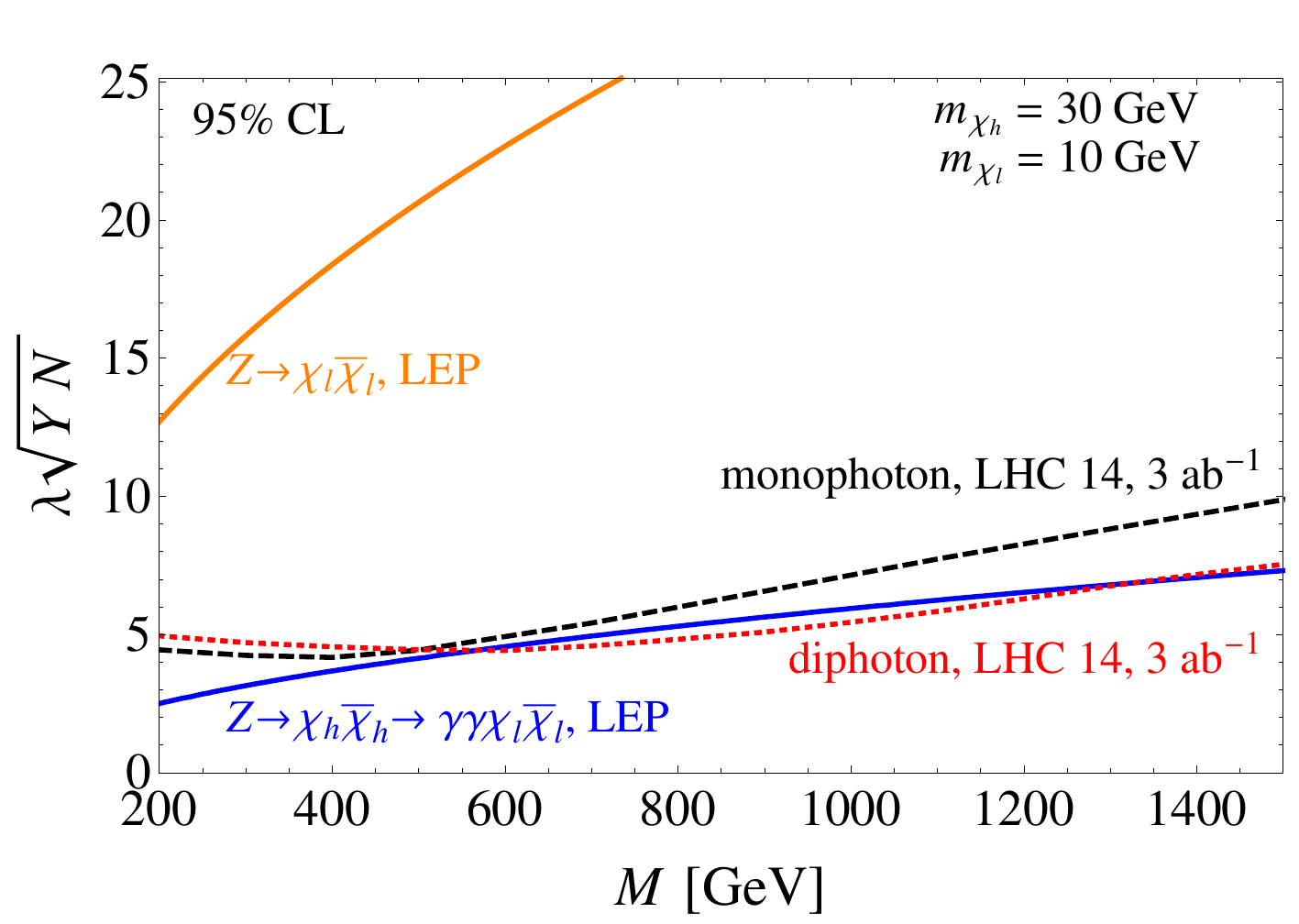}
\end{center}
\caption{Comparison of the LEP bounds from $Z\to \gamma\gamma+$MET (blue) and from the invisible $Z$ width (orange) to the projected LHC sensitivity in monophoton (black) and diphoton (red), at 14 TeV and with $3$ ab$^{-1}$ of data. The perturbative bound on $\lambda\sqrt{YN}$ depends on the hypercharge of the mediator, $\lambda\sqrt{YN}\lsim 4\pi \sqrt{Y}$.}
\label{fig:mchih_30}
\end{figure}

We now turn to the case of light $\chi_h$. If $m_{\chi_h}<m_Z/2$, the decay $Z\to\chi_h \bar{\chi}_h\to\gamma\gamma \chi_l \bar{\chi}_l$ is kinematically allowed, and is constrained by LEP1 data collected at the $Z$-pole. A search performed by OPAL \cite{Acton:1993kp} for events containing at least two photons and missing energy sets a strong constraint on this decay. The search region required exactly two photons, satisfying the cuts $E_\gamma > 1\;\mathrm{GeV}$, $\left|\eta_\gamma \right| < 1.74$, $m_{\gamma\gamma}> 10\;\mathrm{GeV}$ and $\pi-\Delta\phi_{\gamma\gamma}> 0.0873$ (acoplanarity angle). In addition, at least one of the two photons was required to have $\left|\eta_\gamma \right| < 1.10$, and additional photons with $E_\gamma > 0.5\;\mathrm{GeV}$ were vetoed. No events of this type were observed in $43\;\mathrm{pb}^{-1}$ of data, corresponding to $1.8\times 10^6$ $Z$ bosons. The expected SM background comes from $e^+ e^-\to Z(\nu\bar{\nu})\gamma\gamma$, where the two photons arise from initial state radiation, and is quoted by OPAL to amount to 0.2 events.\footnote{Simulating the process $e^+e^-\to \nu\bar{\nu} \gamma\gamma$ in the SM we find a prediction of $\simeq 0.1$ events, in rough agreement with the OPAL number.} Therefore, assuming no systematic uncertainty, the $95\%$ CL limit is of $2.8$ signal events. In order to set a bound on the parameter space, we still need to compute the efficiency of the cuts on the signal. For $m_{\chi_h}=30$ GeV the efficiency is of $\sim 70\%$, leading to the $95\%$ CL bound (the branching ratio for the decay $\chi_h \to \chi_l \gamma$ is unity, since $m_{\chi_h} < m_Z + m_{\chi_l}$)
\begin{equation}
\mathrm{BR}(Z\to \chi_h \bar{\chi}_h) \lesssim 2\times 10^{-6}\,,
\end{equation}
which under the assumption of anarchic couplings is much stronger than the other LEP constraint from the invisible $Z$ width. The corresponding excluded region is shown in Fig.~\ref{fig:mchih_30}, where it is compared to the projected sensitivity of monophoton and diphoton searches at the 14 TeV LHC. We conclude that if $m_{\chi_h} < m_Z/2$, only the diphoton search will be able to marginally improve the LEP constraint. This result motivates our focus on the scenario $m_{\chi_h}> m_Z/2$, in which case LEP data only give a mild constraint from the invisible $Z$ width.   

\section{Non-pointing photon signals}\label{sec:displacedphoton}
\begin{figure}
\begin{center}
\includegraphics[width=0.6\textwidth]{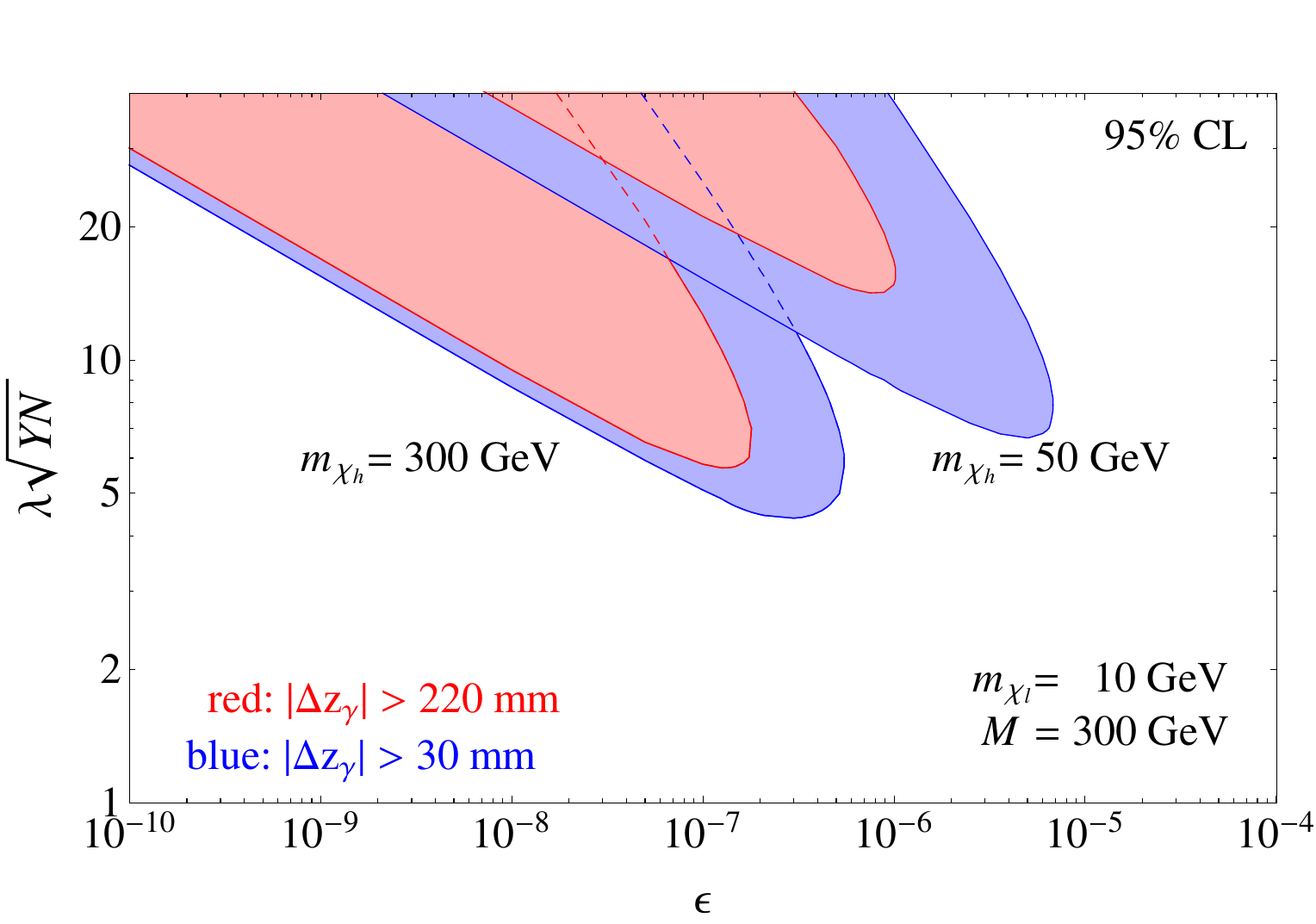}
\end{center}
\caption{Bounds in the $(\epsilon,\lambda\sqrt{YN})$ plane obtained from the $8$ TeV ATLAS search for non-pointing photons. The mass of $\chi_l$ is fixed to $10$ GeV, and the mediator mass to $300$ GeV. The perturbative bound on $\lambda\sqrt{YN}$ depends on the hypercharge of the mediator, $\lambda\sqrt{YN}\lsim 4\pi \sqrt{Y}$.}
\label{fig:displacedbound}
\end{figure}

As discussed in Sec.~\ref{sec:alignment}, when the flavor structure of the dipole operator is aligned with the mass matrix, the decay rate of a heavier fermion $\chi_h$ into the DM $\chi_l$ and a photon can be highly suppressed, while the coupling of $\gamma,Z$ to a pair of $\chi_h$'s or $\chi_l$'s is still sizable. This motivates the study of displaced photon signals from pair produced $\chi_h$'s, each decaying into $\chi_l+ \gamma$. The bounds obtained using the ATLAS non-pointing photon search \cite{Aad:2014gfa} are shown in Fig.~\ref{fig:displacedbound}, for some representative choices of the parameters. The bounds are computed by means of a simple counting experiment, by comparing the signal and background yields in two different regions, corresponding to $\left|\Delta z_\gamma \right| > 30$ mm (exclusion shaded in blue) and $\left|\Delta z_\gamma \right| > 220$ mm (shaded in red). The displacement $\Delta z_\gamma$ was defined in Fig.~\ref{fig:nonpointing}. For both signal regions, an upper bound $\left|\Delta z_\gamma \right| < 750$ mm is also imposed, since the ATLAS paper does not report the background expectation for larger values of the displacement. In the first signal region, approximately $149$ events were expected from the SM background, and $140$ were observed. In the second region, approximately $14$ events were expected and $18$ were observed.  
A few comments are in order. The first observation is that, independently of the value of $m_{\chi_h}$, the lower bound on the coupling\footnote{In the following discussion, $\lambda$ is understood as a shorthand for $\lambda\sqrt{YN}$.} $\lambda$ as a function of $\epsilon$ scales as $\lambda \propto \epsilon^{-1/4}$. This can be understood by noticing from Eq.~\eqref{eq:displdistr} that, for large $\Gamma_{\chi_h}$ (the lower bound on $\lambda$ corresponds to large displacement of the decay), the number of signal events passing the cuts is $N_{S}\propto \lambda^4 \Gamma_{\chi_{h}}$, since the production cross section scales like $\lambda^4$. Recalling that $\Gamma_{\chi_h}\propto \lambda^4 \epsilon^2$ and that in our counting experiment the exclusion bound is given by $N_S = \mathrm{constant}$, we obtain the scaling $\lambda \propto \epsilon^{-1/4}$. Second, comparing the cases $m_{\chi_h}=300$ GeV and $m_{\chi_h}=50$ GeV we note that the bound is weaker in the latter case, especially for the harder cut $\left|\Delta z_\gamma \right| > 220$ mm. The reason for this is twofold. In first place, the efficiency of the cuts on the $E_T$ of the photons and on the $\slashed E_T$ (see Sec.~\ref{sec:displacedphotonTech} for details) is smaller for lighter $\chi_h$. In second place, a lighter $\chi_h$ is more boosted, which implies that the photon from its decay is typically more collinear with $\chi_h$. This fact, combined with the upper cut of $4$ ns that we impose on the time of flight of $\chi_h$ to avoid spurious collisions due to LHC satellite bunches, implies that the typical photon displacement is smaller for lighter $\chi_h$. As a consequence, for $m_{\chi_h} = 50\;\mathrm{GeV}$ the cut $\left|\Delta z_\gamma \right| > 220$ mm significantly reduces the signal yield. 
\section{Comparison to the direct and indirect detection results}\label{sec:compare}
\begin{figure}
\begin{center}
\includegraphics[width=0.5\textwidth]{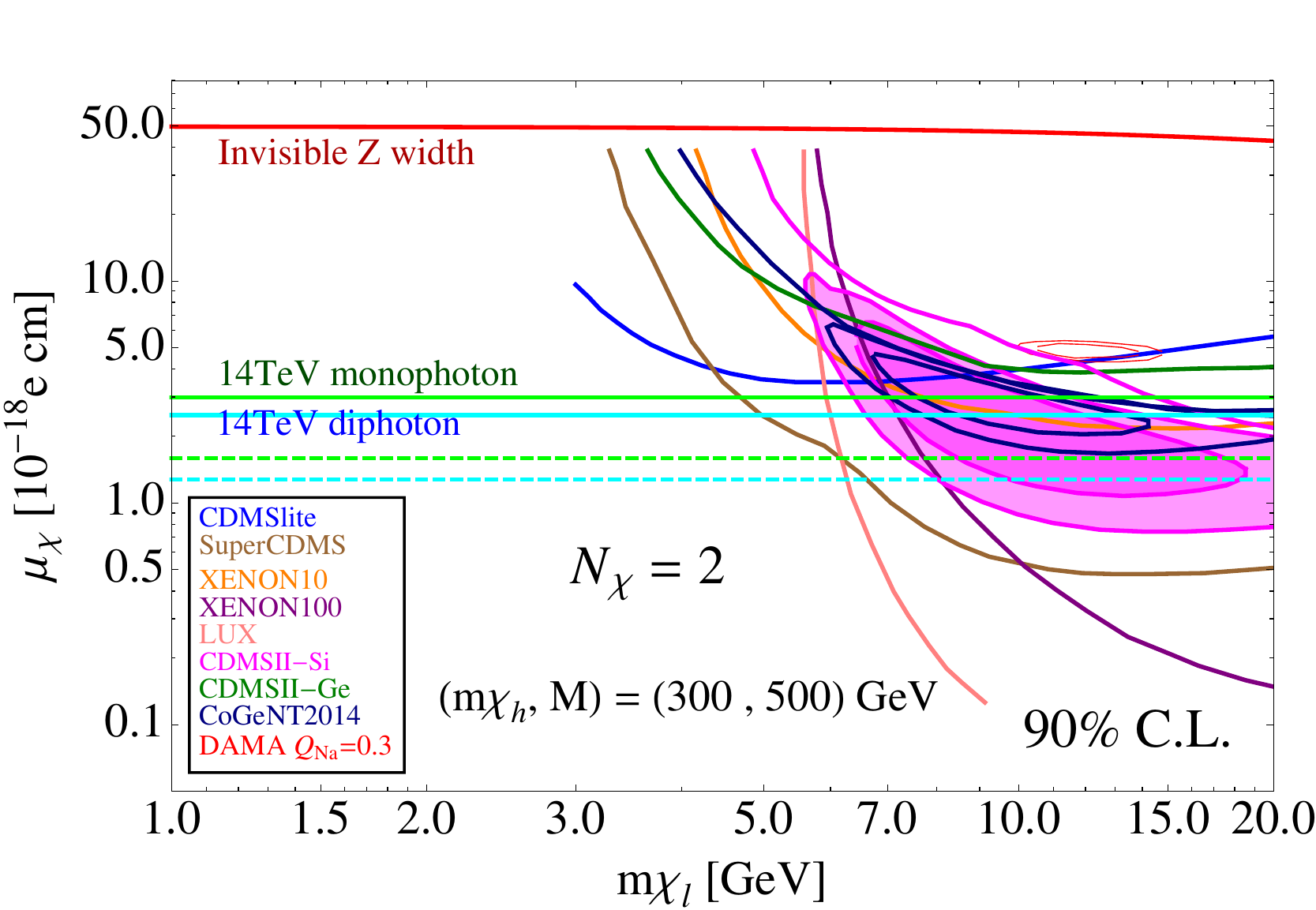}\,\,\includegraphics[width=0.5\textwidth]{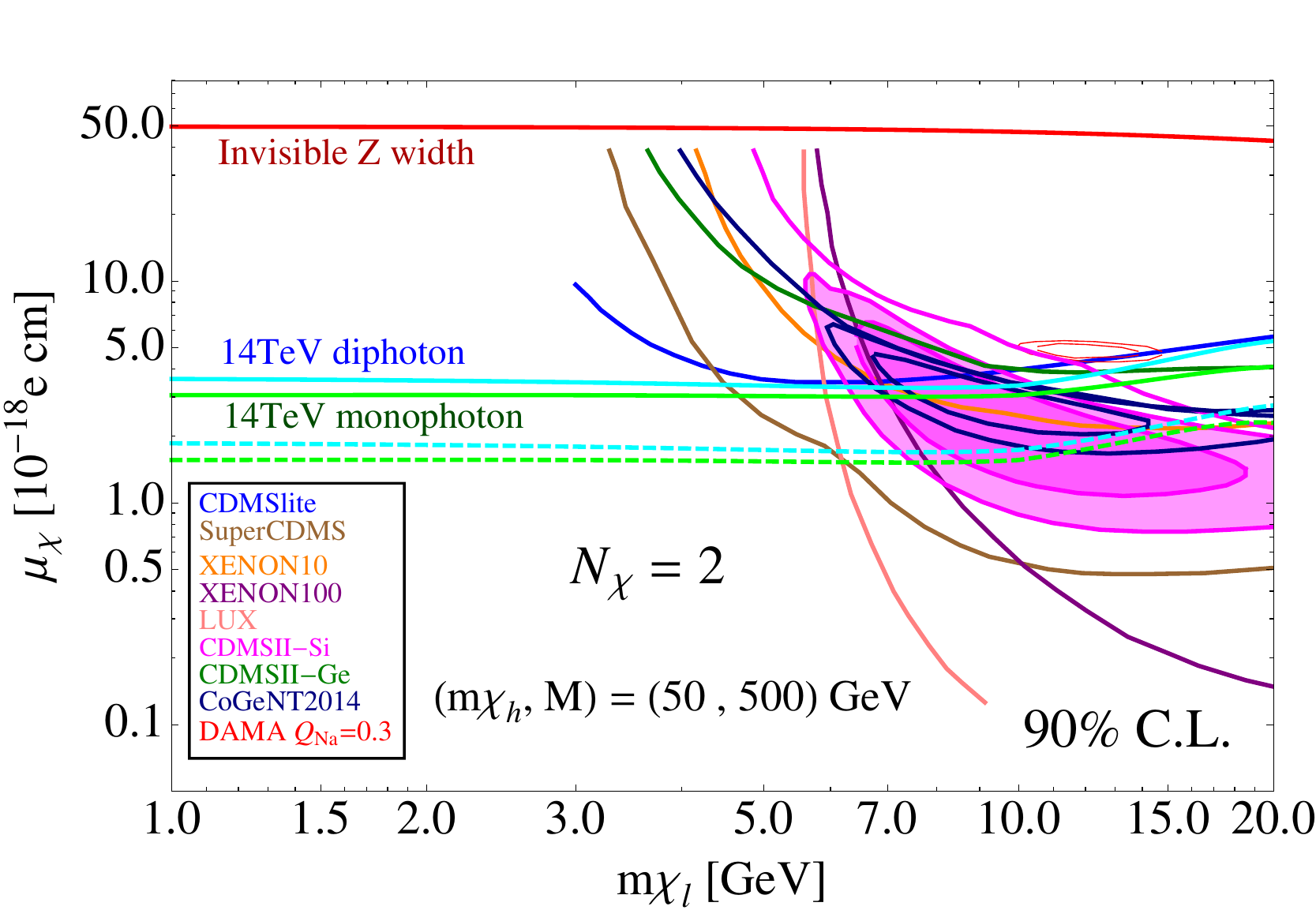}\\\includegraphics[width=0.5\textwidth]{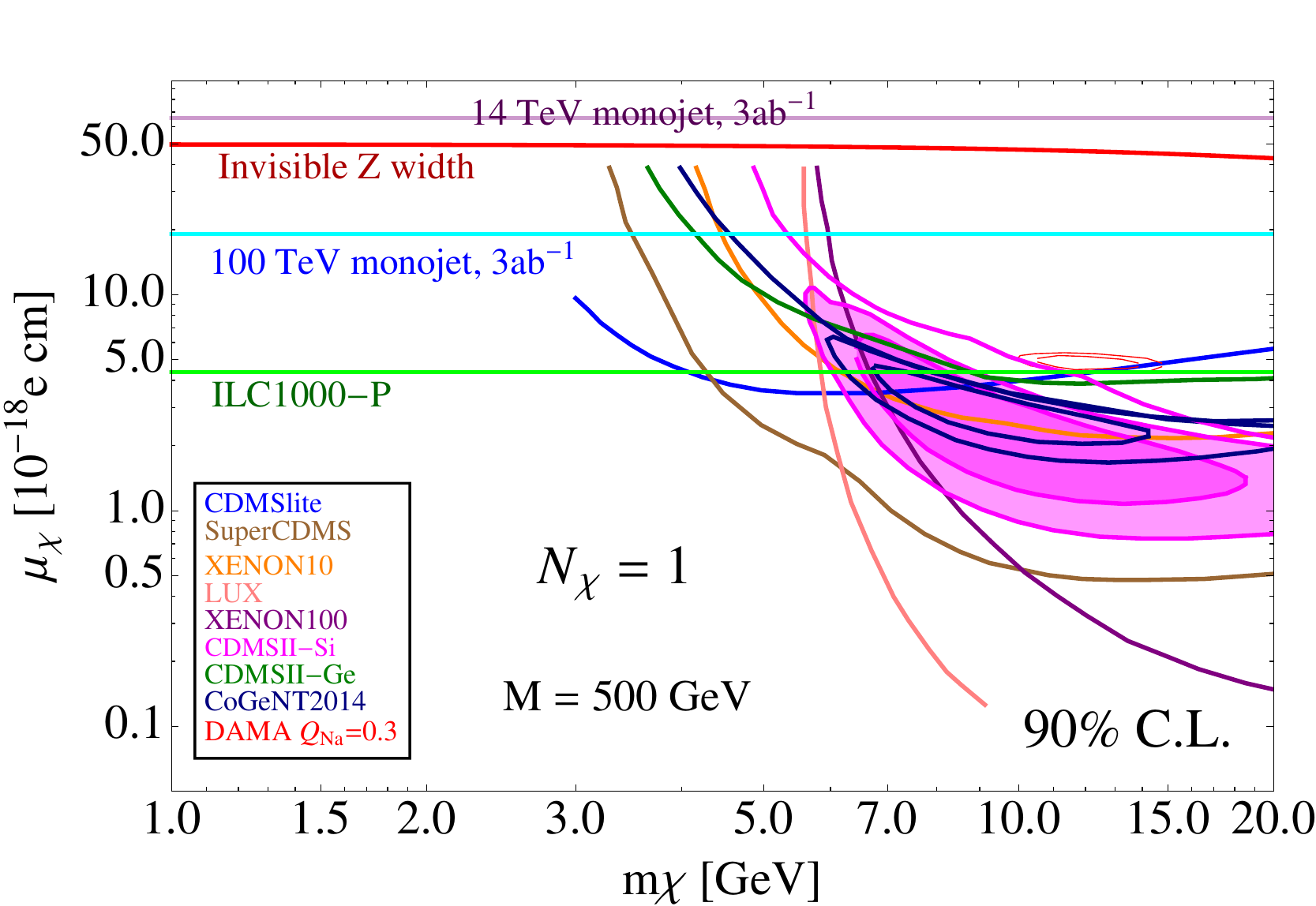}
\end{center}
\caption{Top row: comparison between the projected collider constraints and the current bounds from direct detection experiments for $N_\chi=2$, based on the assumption of anarchic couplings between the mediators and dark fermions. Here we adapt the direct detection bounds from \cite{DelNobile:2013cva,DelNobile:2014eta}, assuming a standard halo model (SHM). See the text for more details. The magnetic dipole moment is defined by the effective coupling of the DM to the photon, $(\mu_{\chi}/2)\bar{\chi}\sigma^{\mu\nu}\chi F_{\mu\nu}$. The red solid lines show the bound from the measurement of the invisible $Z$ decay width at LEP. The green (cyan) lines show bounds from the monophoton (diphoton) searches at the $14$ TeV LHC, with solid (dashed) line for $300$ ($3000$) fb$^{-1}$ of data. Bottom: projected collider constraints and current bounds from direct detection experiments for $N_\chi=1$. The lines correspond to the invisible $Z$ width at LEP (red), $14$ TeV LHC monojet (purple), $100$ TeV collider monojet (cyan) and ILC1000-P monophoton (green).}
\label{fig:directdetection}
\end{figure}

The comparison between the collider and direct detection constraints is shown in Fig.~\ref{fig:directdetection}, under the assumption of anarchic couplings between the mediators and the dark fermions. Here we follow the analysis in \cite{DelNobile:2013cva,DelNobile:2014eta}, comparing the bounds set by the two classes of experiments on the magnetic dipole moment $\mu_\chi$, defined by the effective operator
\begin{equation}\label{eq:effop}
\frac{\mu_{\chi}}{2}\bar{\chi}\sigma^{\mu\nu}\chi F_{\mu\nu}
\end{equation}
where $\chi$ is the DM, and related to the parameters of the dark penguin that couples two DM particles to the photon by
\begin{equation}\label{eq:dipolem}
\mu_{\chi}=\frac{e\lambda^2YN}{32\pi^2M}\,.
\end{equation}
Since the momentum exchange in direct detection experiments is small, we can safely integrate out the mediators and use the effective description in Eq.~\eqref{eq:effop}. Results from direct detection experiments were obtained in \cite{DelNobile:2014eta}, with the assumption of the standard halo model and velocities $v_{esc}=544\,\text{km}/\text{s}$, $v_{\odot}=232\,\text{km}/\text{s}$, $v_{0}=220\,\text{km}/\text{s}$. For the local DM density, the standard value $\rho=0.3\,\text{GeV}/c^2/\text{cm}^3$ was assumed. Figure~\ref{fig:directdetection} shows the $90\%$ CL bounds from CDMSlite~\cite{Agnese:2013jaa}, SuperCDMS~\cite{Agnese:2014aze}, the XENON10 $S_2$-only analysis~\cite{2011PhRvL.107e1301A}, XENON100~\cite{2012PhRvL.109r1301A}, LUX~\cite{Akerib:2013tjd}, CDMS-II-Si~\cite{2011PhRvL.106m1302A}, CDMS-II-Ge low threshold~\cite{2011PhRvL.106m1302A}, and CoGeNT2014 data~\cite{Aalseth:2014eft}, together with the $68\%$ and $90\%$ CL allowed regions for DAMA~\cite{2010EPJC...67...39B} (assuming quenching factor $Q_{\text{Na}}=0.30$), CoGeNT2014, and CDMS-II-Si. For XENON10 we take the result with a conservative setting of the electron yield to zero below $1.4$ keVnr, as in Ref.~\cite{2012PhRvL.109r1301A}. For LUX we adapt the limit corresponding to zero observed events. For further details on the experimental data, as well as on the assumptions made on the low energy thresholds and quenching factors, we refer the reader to Ref.~\cite{DelNobile:2014eta}.

We discuss first the $N_{\chi}=2$ case, shown in the top row of Fig.~\ref{fig:directdetection}. The red horizontal line indicates the upper bound on $\mu_{\chi}$ obtained from the invisible $Z$ decay at LEP, whereas the green and cyan horizontal lines correspond to the bounds obtained from the projected monophoton and diphoton searches at the $14$ TeV LHC, respectively. Due to the anarchic assumption, all these searches can be interpreted as effectively constraining the size of the dark penguin that couples two DM particles to the photon, and thus the size of the effective interaction in Eq.~\eqref{eq:effop} via Eq.~\eqref{eq:dipolem}. This allows us to make the comparison to direct detection experiments. 

To gain some understanding of the dependence of the bounds on the mass of $\chi_h$, we show results using two benchmark values, $50$ and $300$ GeV. On the other hand, the collider bounds are essentially independent of $m_{\chi_l}$ in the range considered $m_{\chi_l}\lesssim 20\;\mathrm{GeV}$, the only appreciable effect being that the LHC monophoton and diphoton bounds worsen slightly for $m_{\chi_l}\gtrsim 10\;\mathrm{GeV}$ in the case $m_{\chi_h}=50\;\mathrm{GeV}$. The plots in Fig.~\ref{fig:directdetection} assume the benchmark value of the mediator mass $M = 500\;\mathrm{GeV}$, but they serve as useful order of magnitude estimates also for heavier mediators. By the end of the high-luminosity LHC run, monophoton and diphoton$+$MET searches will be able to test values of the dipole interaction strength that are comparable to the current XENON10 reach for $m_{\chi_l}>10$ GeV, and will provide a much better bound than the current direct detection limits for lighter $\chi_l$. 

The $N_{\chi}=1$ case is shown in the bottom panel of Fig.~\ref{fig:directdetection}. The projected $14$ TeV LHC monojet bound (purple) is slightly worse than the current limit from the invisible $Z$ width (red), showing that to improve the constraint it will be necessary to go to future colliders. The $100$ TeV collider monojet bound is shown in cyan and corresponds to $\mu_{\chi}\lsim 2\times 10^{-17}e\,\text{cm}$, while the ILC1000-P monophoton search (green) can reach a bound $\lesssim 5\times 10^{-18}e\,\text{cm}$. For comparison, the LEP constraint derived in \cite{Fortin:2011hv} corresponds to $\mu_{\chi}\lsim 10^{-16}e$ cm. Similarly to the $N_\chi = 2$ case, collider searches play a more important role than direct detection experiments when the DM is light, $m_{\chi}\lsim 10$ GeV.  

\begin{figure}
\begin{center}
\includegraphics[width=0.47\textwidth]{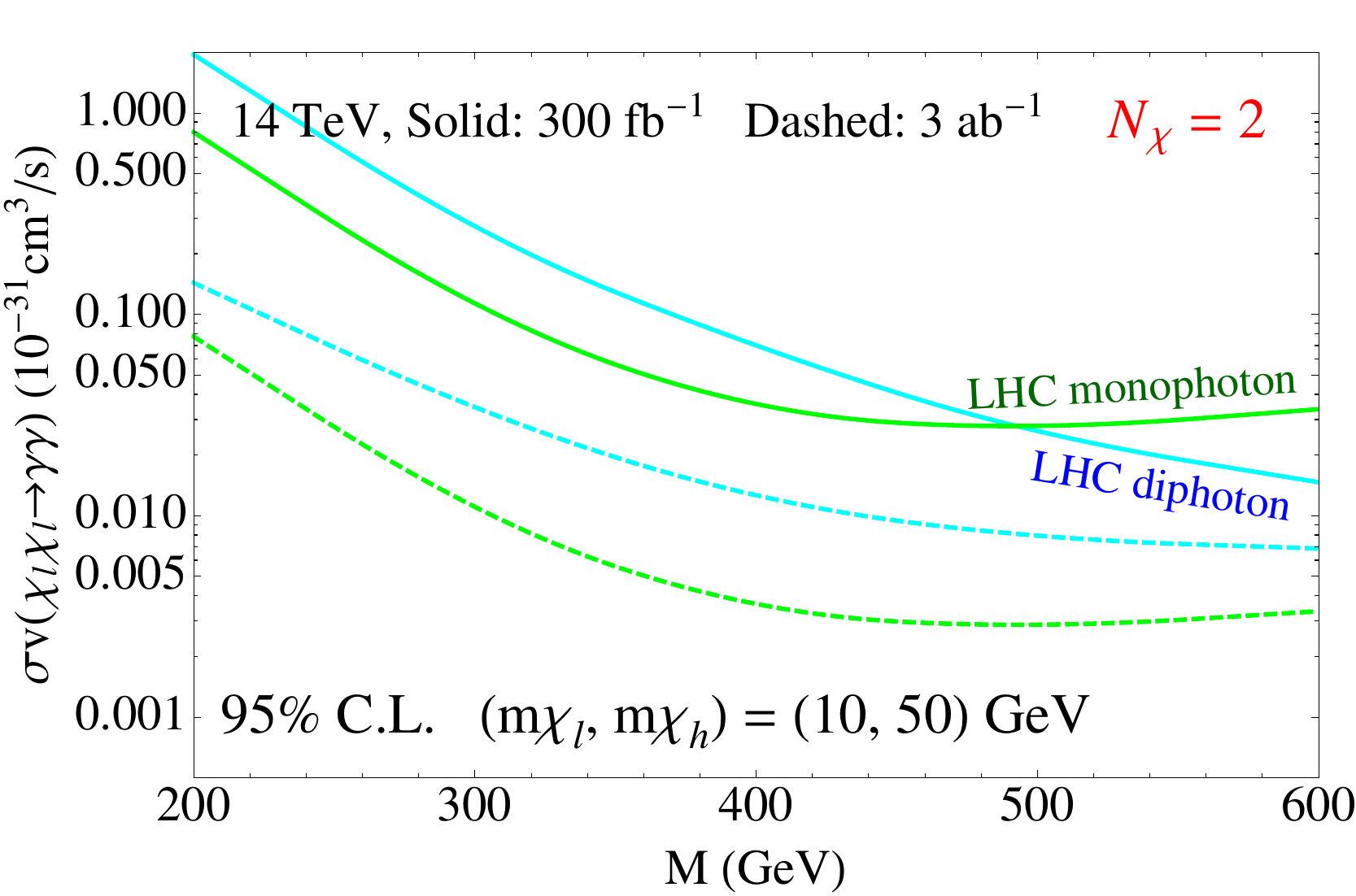}\qquad\includegraphics[width=0.47\textwidth]{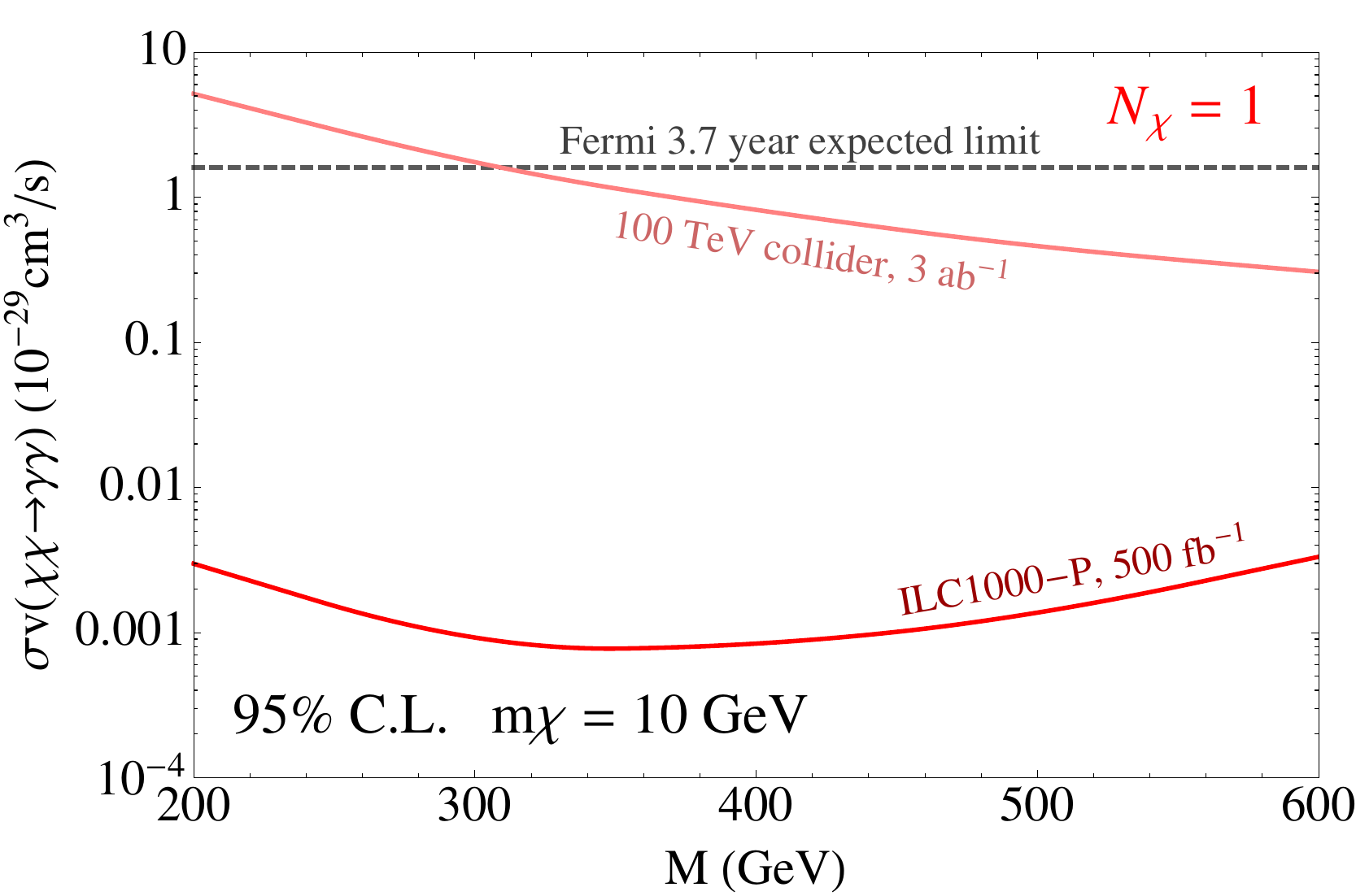}
\end{center}
\caption{Upper bounds on the annihilation of DM into photons, derived from the projected collider constraints computed in this paper. For $N_{\chi}=2$ (left) we show the projected LHC bounds, while for $N_\chi=1$ (right) the reach of future colliders is presented. In the right panel, the Fermi bound corresponds to the expected limit from the 3.7 years data, as reported in \cite{Ackermann:2013uma} (we take the most stringent limit in Ref.~\cite{Ackermann:2013uma}, obtained using the R16 Einasto DM profile).}
\label{fig:indirectdetection}
\end{figure}

The collider bounds can also be compared to the indirect detection searches, where a pair of DM particles annihilates into photons. There are two mechanism through which the annihilation can proceed in our model: either via two dipole interactions connected by a $t$-channel DM exchange, or via a Rayleigh operator. Using naive dimensional analysis, the ratio of the former amplitude to the latter \cite{Weiner:2012cb} is $\sim \lambda^2 M/(16\pi^2 m_{\chi})$. Considering the most conservative choice of parameters relevant to our discussion, namely $\lambda\sim 4$, $M\sim 200\;\mathrm{GeV}$ and $m_{\chi}\sim 10\;\mathrm{GeV}$, we find the ratio to be about $2$. We conclude that for our purposes it is a safe approximation to neglect the contribution of the Rayleigh operator, which is typically strongly subdominant compared to the diagram with $t$-channel DM exchange. The latter dominates the annihilation, with a cross section~\cite{Sigurdson:2004zp}
\begin{equation}
\sigma_{\bar{\chi}\chi\to\gamma\gamma}v = \frac{1}{2\pi}\,\mu_{\chi}^4 m_{\chi}^2\,.
\end{equation}
As shown in Fig.~\ref{fig:indirectdetection}, if more than one flavor of dark fermions are accessible at the LHC, the size of the annihilation into photons can be tested down to approximately $10^{-33}$ cm$^3/$s. In the $N_{\chi}=1$ case, the ILC1000-P can set the strongest bound on the annihilation, of order $10^{-32}$ cm$^3/$s.
\section{Conclusions}\label{sec:conclude}
The search for DM plays a central role in the physics program of current and future collider experiments. For many plausible DM-SM couplings, a correct description of the DM production processes, which is crucial to extract meaningful information from the experimental results, requires that the EFT parameterization be UV completed by a simplified model with light mediators. This is especially important when the DM-SM coupling arises at loop level. In this paper we performed the first simplified model collider study for a loop process, focusing on the \emph{Dark Penguin}, whose form factors reduce to the magnetic dipole operator at low energies. We computed bounds from monojet, monophoton, and diphoton searches at the $8$ and $14$ TeV LHC, as well as from the future ILC and $100$ TeV hadron collider. Differently from searches for EFT interactions, when light mediators are included the optimal search strategy requires the cuts on the MET to be as loose as possible, to capture the enhancement of the form factors near the threshold for production of on-shell mediators. As we showed through a detailed comparison, for light DM mass the collider bounds are complementary to those derived from direct and indirect detection experiments.

Based on general considerations, it is plausible that the dark sector may be endowed with a non-trivial flavor structure. If some of the additional states beyond the DM are kinematically accessible, collider experiments offer a unique opportunity to probe the dark flavor. By employing a simplified model with a second dark fermion in addition to the DM, we showed that, under the natural assumption of anarchic structure in the dark sector, the bounds on the DM-SM coupling set by collider searches are much stronger than in the case where the DM is the only accessible state. Collider searches not only have the capability to probe the flavor structure of the dark sector, but might even allow to \emph{measure} some of its properties. For example, the determination of the mass of the heavy dark fermion could be possible by using $M_{T2}$ in the diphoton$+$MET channel. Furthermore, if the flavor structures of the dark penguin and the dark fermion mass matrix are nearly aligned, the decay of the heavy dark fermion into DM and photon can be displaced. In this case, it is achievable to extract information about the small mixing angle in the dark sector from the search for non-pointing photons$+$MET. 

We end with an outlook to future developments. While our study was focused on the dark penguin mediating the dipole operator, our method is fully general and can be extended to any other loop-mediated interaction between the DM and the SM fields, such as for example the Rayleigh and $\bar{\chi}\chi G^{\mu\nu}G_{\mu\nu}$ operators, where $G^{\mu\nu}$ is the SM gluon field strength. It would also be interesting to analyze in detail the prospects for the ILC, which the preliminary results presented here show to be very promising. Additionally, while our first estimates for a very high energy hadron collider appear less favorable, the detailed assessment of the design requirements that would allow a substantial improvement warrants a dedicated study.   

\section*{Acknowledgments}
The authors thank Yang~Bai, Heribertus~B.~Hartanto, David~E.~Kaplan, Simon~Knapen, Tongyan~Lin, Mani~Tripathi, William~Shepherd, Ciaran~Williams, Itay~Yavin and Ning~Zhou for useful discussions, and are especially grateful to Roni~Harnik for comments about the manuscript. R.P. was supported by the NSF under grant PHY-1214000. E.S. and Y.T. were supported by the Department of Energy under grant DE-FG02-91ER40674. Y.T. thanks the Aspen Center for Physics (National Science Foundation Grant PHYS-1066293) for hospitality during the completion of this work.

\appendix

\section{Analytical expressions of the form factors}\label{append:formfactor}
In this appendix we present the expressions of the form factors that appear in Eq.~\eqref{eq:loopresult}. By applying the standard Passarino-Veltman decomposition \cite{Passarino:1978jh}, they can be written as\footnote{In this appendix, we adopt the shortened notations $m_{i,j}\equiv m_{\chi_{i,j}}$ and $q^{4,6}\equiv (q^2)^{2,3}$.}
\begin{align}
F_{q}\,=&\, \frac{f_{q}(u_{1},u_{2},u_{3}) + a_{q}C_{0}(m_{i}^{2}, m_{j}^2, q^2, M_{s}^2, M_{f}^2, M_{s}^2)+b_{q}C_{0}(m_{i}^{2}, m_{j}^2, q^2, M_{f}^2, M_{s}^2, M_{f}^2)}{[q^2-(m_{i}-m_{j})^2][q^2-(m_{i}+m_{j})^2]^2}\,, \nonumber\\
F_{\sigma}\,=&\, \frac{f_{\sigma}(u_{1},u_{2},u_{3}) + a_{\sigma}C_{0}(m_{i}^{2}, m_{j}^2, q^2, M_{s}^2, M_{f}^2, M_{s}^2)+b_{\sigma}C_{0}(m_{i}^{2}, m_{j}^2, q^2, M_{f}^2, M_{s}^2, M_{f}^2)}{[q^2-(m_{i}-m_{j})^2][q^2-(m_{i}+m_{j})^2]^2}\,. \label{ScalarIntegrals}
\end{align}
where
\begin{align}
u_{1}\,=&\, B_{0}(m_{j}^{2},M_{f}^{2},M_{s}^{2})-B_{0}(q^{2},M_{s}^{2},M_{s}^{2})\,, \nonumber\\
u_{2}\,=&\, B_{0}(m_{i}^{2},M_{f}^{2},M_{s}^{2})-B_{0}(q^{2},M_{f}^{2},M_{f}^{2})\,, \nonumber\\
u_{3}\,=&\, B_{0}(m_{j}^{2},M_{f}^{2},M_{s}^{2})-B_{0}(m_{i}^{2},M_{f}^{2},M_{s}^{2})\,.
\end{align}
For the scalar integrals, we followed the convention of LoopTools \cite{Hahn:1998yk}. The building blocks of $F_{q}$ are given by
\begin{align}
f_{q}(&u_{1},u_{2},u_{3}) \nonumber\\ =&\, -2 m_{i}^{4} (1+u_{2})-m_{i}^{2} [q^{2} (-4+u_{1}-3 u_{2}-u_{3})-4 (M_{f}^2-M_{s}^2) (u_{1}-u_{2}-u_{3})] \nonumber \\  
-&\, 4 m_{i} M_{f} q^{2} (u_{1}-u_{2}-u_{3})- 2 m_{j}^4 (1+u_{2}+u_{3})-4 m_{i}^3 M_{f} (-u_{1}+u_{2}+u_{3})\nonumber \\ 
-&\, 4 m_{j}^3 [-(m_{i}+M_{f}) (u_{1}-u_{2})+M_{f} u_{3}]-q^{2} [2 (M_{f}^2-M_{s}^2) (-u_{1}+u_{2}+u_{3})\nonumber\\ 
+&\, q^{2} (2-u_{1}+u_{2}+u_{3})]-m_{j}^2 [q^{2} (-4+u_{1}-3 u_{2}-3 u_{3})-4 (M_{f}^2-M_{s}^2) (u_{1}-u_{2}-u_{3}) \nonumber \\ 
+&\, 12 m_{i} M_{f} (-u_{1}+u_{2}+u_{3})+m_{i}^2 (-4-8 u_{1}+4 u_{2}+6 u_{3})]+4 m_{j} \{m_{i}^3 (u_{1}-u_{2}-2 u_{3}) \nonumber \\  
+&\, 3 m_{i}^2 M_{f} (u_{1}-u_{2}-u_{3})+M_{f} q^{2} (-u_{1}+u_{2}+u_{3})+m_{i} [2 M_{f}^2 (u_{1}-u_{2}-u_{3}) \nonumber \\
+&\, 2 M_{s}^2 (-u_{1}+u_{2}+u_{3})+q^{2} (2 u_{2}+u_{3})]\},
\end{align}
\begin{align}
a_{q}/2& \nonumber\\ \,=&\, m_{j}^5 (m_{i}+M_{f})+m_{j}^4 [(m_{i}+M_{f}) (2 m_{i}+M_{f})-2 M_{s}^2]+m_{i}^2 [M_{f} (m_{i}+M_{f})(m_{i}^2-2 M_{f}^2) \nonumber \\ 
-&\,2 (m_{i}-2 M_{f}) (m_{i}+M_{f}) M_{s}^2-2 M_{s}^4]-[2 m_{i}^3 M_{f}-2 m_{i} M_{f}^3+M_{f}^4-2 (m_{i}^2-m_{i} M_{f}+M_{f}^2) M_{s}^2 \nonumber \\ +&\,M_{s}^4] q^{2} 
+(m_{i}-M_{f}) M_{f} q^{4}+m_{j} (m_{i}^2-2 M_{f}^2+2 M_{s}^2-q^{2}) \{m_{i} [(m_{i}+M_{f}) (m_{i}+2 M_{f})-2 M_{s}^2]\nonumber \\-&\,M_{f} q^{2}\} 
+ m_{j}^3 [2 m_{i}^3+4 m_{i}^2 M_{f}-m_{i} q^{2}-2 M_{f} (M_{f}^2-M_{s}^2+q^{2})]+m_{j}^2 [2 m_{i}^4+4 m_{i}^3 M_{f} \nonumber \\-&\,2 (M_{f}^2-M_{s}^2)^2 
+ m_{i}^2 (-2 M_{f}^2+4 M_{s}^2-3 q^{2})+2 M_{s}^2 q^{2}+m_{i} (-6 M_{f}^3+6 M_{f} M_{s}^2-4 M_{f} q^{2})],
\end{align}
\begin{align}
b_{q}\,=&\, a_{q}+2 (m_{j}+m_{i})^2 \{-(m_{j}+m_{i}+2 M_{f}) [m_{j} m_{i} (m_{j}+m_{i})+2 (m_{j}^2-m_{j} m_{i}+m_{i}^2) M_{f}] \nonumber\\
\,+&\,4 (m_{j}^2-m_{j} m_{i}+m_{i}^2) M_{s}^2\}+4 (m_{j}+m_{i}) \{2 m_{i} M_{f} (m_{i}+M_{f})+m_{j}^2 (m_{i}+2 M_{f})-2 m_{i} M_{s}^2 \nonumber \\
\,+&\, m_{j} [m_{i}^2+2 m_{i} M_{f}+2 (M_{f}^2-M_{s}^2)]\} q^2 - 2 [m_{j} m_{i}+2 (m_{j}+m_{i}) M_{f}] q^4\,.
\end{align}
The building blocks of $F_{\sigma}$ are
\begin{align}
f_{\sigma}(&u_{1},u_{2},u_{3}) \nonumber\\\,=&\, -2 m_{j}^5-2 m_{i}^5+m_{i} q^{2} [6 (M_{f}^2-M_{s}^2) (u_{1}-u_{2}-u_{3})+q^{2} (-2+u_{1}+u_{2}-u_{3})] \nonumber \\
+&\,4 m_{i}^2 M_{f} q^{2} (u_{1}-u_{2}-u_{3})-m_{i}^3 q^{2} (-4+u_{1}+u_{2}-u_{3})+2 m_{j}^4 m_{i} (-1+u_{3}) \nonumber \\
+&\,4 M_{f} q^{4} (-u_{1}+u_{2}+u_{3})+m_{j}^3 [2 m_{i}^2 (2+u_{3})-q^{2} (-4+u_{1}+u_{2}+u_{3})]+m_{j}^2 [4 M_{f} q^{2} (u_{1} \nonumber \\-&\,u_{2}-u_{3}) - 2 m_{i}^3 (-2+u_{3})+m_{i} q^{2} (4+3 u_{1}+3 u_{2}+3 u_{3})]+m_{j} \{m_{i}^2 q^{2} (4+3 u_{1}+3 u_{2}-3 u_{3}) \nonumber \\
+&\,8 m_{i} M_{f} q^{2} (u_{1}-u_{2}-u_{3})-2 m_{i}^4 (1+u_{3})+q^{2} [6 (M_{f}^2-M_{s}^2) (u_{1}-u_{2}-u_{3}) \nonumber \\
+&\,q^{2} (-2+u_{1}+u_{2}+u_{3})]\},
\end{align}
\begin{align}
a_{\sigma}&\nonumber\\ =&\, -2 (m_{j}-m_{i})^2 (m_{j}+m_{i})^3 M_{s}^2+2 (m_{j}+m_{i}) \{(m_{j}+M_{f}) (m_{i}+M_{f}) [m_{j}^2-m_{j} m_{i}+m_{i}^2\nonumber\\ +&\,(m_{j}+m_{i}) M_{f} - 3 M_{f}^2]+2 [m_{j} (m_{i}+M_{f})+M_{f} (m_{i}+3 M_{f})] M_{s}^2-3 M_{s}^4\} q^{2}-2 [m_{j}^2 (m_{i}+2 M_{f})\nonumber \\+&\, 2 M_{f} (m_{i}^2+m_{i} M_{f}-M_{f}^2)
-(m_{i}-2 M_{f}) M_{s}^2+m_{j} (m_{i}^2+2 m_{i} M_{f}+2 M_{f}^2-M_{s}^2)] q^{4}+2 M_{f} q^{6}\,,
\end{align}
\begin{align}
b_{\sigma} \,=&\,\, a_{\sigma}-2 (m_{j}-m_{i})^2 (m_{j}+m_{i})^3 \left[M_{f} (m_{j}+m_{i}+M_{f})-M_{s}^2\right]+2 (m_{j}-m_{i})^2 (m_{j}+m_{i}) \nonumber\\ \,\times &\, \left[M_{f}(m_{j}+m_{i}-2 M_{f}) +2 M_{s}^2\right] q^2+2 (m_{j}+m_{i}) \left[M_{f} (m_{j}+m_{i}+3 M_{f})-3 M_{s}^2\right] q^4-2 M_{f} q^6\,.
\end{align}
The above expressions can be easily implemented in a numerical code, and provide stable results for all values of $q^2$, in particular for $q^2 > 4M^2$ (where $M$ denotes generically the mediator mass), where the form factors develop an imaginary part as a consequence of the virtual states going on-shell. For completeness, we also report the expressions of the form factors in terms of integrals over Feynman parameters, which are more suitable for analytical expansions. These read
\begin{align}
F_q \,=& \, \frac{2}{m_{i}-m_{j}}\int d^3 x \left(\frac{c_q}{\Delta_f} + \frac{d_q}{\Delta_s}\right), \nonumber \\
F_\sigma \,=&\, 2\, \int d^3 x \left(\frac{c_\sigma}{\Delta_f} + \frac{d_\sigma}{\Delta_s}\right), \label{FeynmanParam}
\end{align}
where
\begin{align}
\Delta_{f} \,\,=& \,\, -xy q^{2}- zx m_{i}^{2}-yz m_{j}^{2}+ (x+y) M_{f}^{2} + z M_{s}^{2}\,, \nonumber\\
\Delta_{s} \,\,=& \,\, \Delta_f(M_f \leftrightarrow M_s)\,, \nonumber\\
c_q\,\,=&\,\,  y (2 y+z-2)m_{j} -  (y+z-1) (2 y+z) m_{i} -  (2 y+z-1)M_{f}\,, \nonumber \\
d_q\,\,=&\,\, y(2 y+z-1)m_{j}  -  (y+z-1)(2y+z-1) m_{i} +(2y+z-1)M_{f}\,, \nonumber \\
c_\sigma\,\,=&\,\, -y z m_{j} + z(y+z-1) m_{i} +  (z-1)M_{f}\,, \nonumber \\
d_\sigma\,\,=&\,\, - y z m_{j} + z(y+z-1) m_{i} - z M_{f}\,,
\end{align}
and we defined $\int d^{3}x\equiv \int_0^1 dx dy dz \,\delta(x+y+z-1)$. It is straightforward to check numerically that the Eqs.~\eqref{FeynmanParam} agree exactly with Eqs.~\eqref{ScalarIntegrals} for $q^2 < 4M^2$, whereas for $q^2 > 4M^2$ the Feynman parameter integrals can become numerically unstable, and only the expression in terms of scalar integrals should be used.\footnote{We have cross-checked our results against Ref.~\cite{Weiner:2012gm}, where the form factors were computed for the special case $m_i = m_j$, $Y=1/2$, $N=2$. We find agreement, except for a few small differences that we believe are due to typos in App.~A of \cite{Weiner:2012gm}: 1) in their Eq.~(A7), a factor $z$ should multiply the second term on the right-hand side (RHS), and an overall factor $(-1)$ should multiply the RHS; 2) in their Eq.~(A8), an overall factor $(-2)$ should multiply the RHS.}

By using Eqs.~\eqref{FeynmanParam}, one can derive the expressions of the form factors for large mediator mass, $M_f = M_s = M \gg m_{i},m_{j},\sqrt{q^2}$,
\begin{equation}
F_q \,\to \, \frac{1}{6M^2}\,, \qquad F_\sigma \,\to \, -\frac{1}{M}\,. \label{FeynmanParamEFT}
\end{equation}
In this limit, the amplitude in Eq.~\eqref{eq:loopresult} can be seen as generated by the effective Lagrangian
\begin{equation}\label{eq:eftcoupling}
\frac{1}{1+\delta_{ij}}\frac{(\lambda^2)_{ij}\, g^\prime Y N}{192\pi^2 M^2}\overline{\chi}_i \gamma^\mu \chi_j \partial^\nu B_{\mu\nu} + \mathrm{h.c.} + \frac{1}{1+\delta_{ij}}\frac{(\lambda^2)_{ij}\, g^\prime Y N}{64\pi^2 M}\overline{\chi}_i \sigma^{\mu\nu} \chi_j B_{\mu\nu} + \mathrm{h.c.}\,. 
\end{equation}
%
\section{Statistics}\label{append:statistics}
In this appendix we briefly describe our procedure for setting limits on the DM parameter space. Given the expected number of background events $N_B$, we exclude a signal model yielding $N_S$ events if 
\begin{equation}
\int_0^{N_{obs}}\frac{dP}{dx}(x;N_B + N_S) dx < p\,,
\end{equation}
where $dP/dx\,(x;N_B+N_S)$ is the normalized probability distribution function for signal plus background, $N_{obs}$ is the number of events observed, and $p$ is the chosen probability. For example, for a $95\%$ CL exclusion, $p=0.05$. Notice that we are setting a `one-sided' limit. Throughout our analysis we neglect systematic uncertainties on the signal, because they are subleading to those associated with the background. 

Under the assumption that signal plus background follows a Poisson distribution with mean $N_B + N_S$, and neglecting all systematic uncertainties, the exclusion limit is given by
\begin{equation} \label{PoissonLimit}
\frac{\Gamma(N_{obs}+1, N_B + N_S)}{\Gamma(N_{obs}+1,0)} = p\,,
\end{equation}
where $\Gamma(s,q) = \int_q^\infty t^{s-1}e^{-t}dt\,.$ For example, for $p=0.05$, we find for $N_{B}=N_{obs}=0$ that $N_S \simeq 3.0$.

If we assume that signal plus background follows a Gaussian distribution with mean $N_B + N_S$ and standard deviation $\sqrt{N_S + (\delta N_B)^2}$, where $\delta N_B = \sqrt{N_B +(\epsilon N_B)^2}$, with $\epsilon$ the relative systematic uncertainty on the background, we find the exclusion limit
\begin{equation} \label{GaussLimit}
\frac{N_B+N_S-N_{obs}}{\sqrt{N_S+(\delta N_B)^2}}=\sqrt{2}\,\mathrm{erf}^{-1}\,(1-2p)\,.
\end{equation}
For example, for $p=0.05$ the right-hand side is equal to $\simeq 1.645$. 

Equation~\eqref{PoissonLimit} was used to set the limit from prompt diphoton and from the rare $Z$ decay $Z\to \gamma\gamma+$MET, whereas all the remaining limits in this paper were computed using Eq.~\eqref{GaussLimit}.

\bibliographystyle{utphys}
\bibliography{./DMref}

\end{document}